\documentclass[preprint,showpacs,preprintnumbers,amsmath,amssymb,floatfix,fleqn]{revtex4}
\usepackage{bm}
\usepackage{graphicx}
\usepackage{dcolumn}

\begin{document}

\title{
Improved $\alpha^4$ Term of the Electron Anomalous Magnetic Moment}

\author{Toichiro Kinoshita }
\email{tk@hepth.cornell.edu}
\affiliation{Laboratory for Elementary Particle Physics  \\  
Cornell University,  Ithaca, New York, 14853  }

\author{M. Nio}
\email{nio@riken.jp}
\affiliation{Theoretical Physics Laboratory,
RIKEN, Wako, Saitama, Japan 351-0198 }

\date{\today}

\begin{abstract}
We report a new value 
of electron $g-2$, or $a_e$, from 891 Feynman diagrams
of order $\alpha^4$. The FORTRAN codes of 373
diagrams containing closed electron loops have been verified by
at least two independent formulations. For the remaining 518 diagrams, 
which have no closed lepton loop, 
verification by a second formulation is not yet attempted
because of the enormous amount of additional work required.  
However, these integrals have structures that allow 
extensive cross-checking as well as
detailed comparison with lower-order diagrams
through the renormalization procedure.  
No algebraic error has been uncovered for them.  
The numerical evaluation of the entire $\alpha^4$ term
by the integration routine VEGAS
gives $-1.7283 (35) (\alpha/\pi)^4$, 
where the uncertainty is obtained
by careful examination of error estimates by VEGAS.
This leads to 
$a_e = 1~159~652~175.86~(0.10)~(0.26)~(8.48) \times 10^{-12}$,
where the uncertainties come from the $\alpha^4$ term,
the estimated uncertainty of $\alpha^5$ term, and the inverse 
fine structure constant, $\alpha^{-1} = 137.036~000~3~(10)$,
measured by atom interferometry 
combined with a frequency comb technique, respectively.
The inverse fine structure constant $\alpha^{-1} (a_e)$
derived from the theory and the Seattle measurement of $a_e$ is
$137.035~998~83~(51)$.
\end{abstract}


\maketitle

\section{Introduction and summary}
\label{sec:intro}

It was more than 18 years ago that an 
amazingly precise measurement of 
$g-2$ values of the electron and positron
 in a Penning trap was reported \cite{vandyck}:
\begin{eqnarray}   
 a_e^- ({\rm exp}) 
  &=& 1~159~652~188.4~(4.3) \times 10^{-12}~~~~[3.7~{\rm ppb}] , 
  \nonumber  \\
 a_e^+ ({\rm exp}) 
  &=&1~159~652~187.9~(4.3) \times 10^{-12}~~~~[3.7~{\rm ppb}] .
\label{ae_exp}
\end{eqnarray}   
Numerals 4.3 in parentheses represent
uncertainties in the last digits of measurements,
and 3.7 ppb is the fractional precision (1 ppb = $1 \times 10^{-9}$).
At present a new measurement of $a_e$, which 
is based on a cylindrical cavity whose property has been studied analytically
\cite{brown}, is in progress.  It is expected to
improve the precision substantially \cite{gabrielse}.
Unlike the muon $g-2$, which is sensitive to short distance physics,
the electron $g-2$ is rather insensitive to short distance physics.
It is therefore calculable to very high precision from the QED
or, more generally, the Standard Model.
High precision measurements of $a_e$ 
thus enables us to subject the QED to a most stringent test.
If disagreement is found between theory and experiment of $a_e$,
it would have a very profound impact on the validity of QED,
or modern quantum field theory in general.
This is why high precision measurement of $a_e$ is of such
an importance.

The QED contribution to $a_e$ can be written in general as
\begin{equation}
a_e ({\rm QED}) = A_1 
+ A_2 (m_e /m_\mu) 
+ A_2 (m_e /m_\tau) 
+ A_3 (m_e /m_\mu, m_e /m_\tau).
\end{equation}
$A_1$, $A_2$, $A_3$ can be expanded as power series in $\alpha /\pi$:
\begin{equation}
A_i = 
A_i^{(2)}
\left (\frac {\alpha}{\pi} \right )
+ A_i^{(4)}
\left (\frac {\alpha}{\pi} \right )^2
+ A_i^{(6)}
\left (\frac {\alpha}{\pi} \right )^3
+ \ldots ,~~~~~~~ i=1,2,3.
\end{equation}
$A_1^{(2)}$, $A_1^{(4)}$, and $A_1^{(6)}$ are known 
analytically \cite{schwinger,som-pet,lap-rem} whose numerical values are
\begin{eqnarray}
A_1^{(2)} &=&~~0.5,
  \nonumber  \\
A_1^{(4)} &=&-0.328~478~965~579~\ldots ,
  \nonumber  \\
A_1^{(6)} &=&~~1.181~241~456~\ldots .
\label{analytic}
\end{eqnarray}
Actually, the analytic result $A_1^{(6)}$ is 
the culmination of a long sequence of analytic work
that started around 1969 \cite{a6analytic}.
It took more than 25 years of hard work before it was
finally completed by \cite{lap-rem}.

From the vantage of 1960 $-$ 1970, analytic solution of
the sixth-order muon g-2 diagrams containing
an (external) light-by-light scattering subdiagram 
seemed to be extremely difficult.
Thus, one of the authors (T.K.) and Aldins developed a numerical approach
based on the technique of Feynman-parametric integral \cite{masssing},
calculating the integrand by hand.
The subtle problem of renormalizing the
light-by-light scattering subdiagram 
was bypassed taking advantage of gauge invariance.
Meanwhile, Brodsky and Dufner were working on the same problem.
We decided to join force and write the report together \cite{aldins}.
The resulting 7-dimensional integral was evaluated by an adaptive-iterative
Monte-Carlo routine \cite{czyz}.
Sixth-order diagrams containing vacuum-polarization loops
were also evaluated in a similar manner \cite{bk}.

This method was then extended to general
diagrams that require explicit renormalization \cite{ck1,ck2}.
It consists of three steps:

(1) Conversion of momentum space integrals
into integrals over Feynman parameters.
This step is fully analytic and 
carried out by an algebraic manipulation software,
such as SCHOONSCHIP \cite{veltman} and FORM \cite{vermas}.

(2) Renormalization by a subtraction term
that subtracts divergence {\it point-by-point} throughout 
the domain of integration.
The renormalization term is constructed in such a way that  
it is analytically factorisable into
products of a divergent part of renormalization constant and
a g-2 term of lower order.

(3) Numerical integration by VEGAS \cite{lepage}, 
an improved version of adaptive-iterative
Monte-Carlo integration routine \cite{czyz}.

We applied this scheme to two different formulations, both
in Feynman gauge.
One is to apply it to
each vertex individually and add them up.
Another starts by combining a set of vertices into
one with the help of the Ward-Takahashi identity
\begin{equation}
q_\mu \Lambda^{\mu} (p,q) = - \Sigma (p+\frac{q}{2}) + \Sigma (p-\frac{q}{2}),
\label{wtid}
\end{equation}
where $\Lambda^{\mu} (p,q)$
is the sum of vertices obtained by inserting the external
magnetic field in fermion lines of a self-energy diagram
$\Sigma (p)$.  $p \pm q/2$ is outgoing (incoming) electron momentum.
Differentiating both sides of (\ref{wtid})
with respect to $q_\nu$ one obtains
\vspace{2mm}
\begin{equation}
\Lambda^{\nu} (p,q) \simeq - q^{\mu} \left [ \frac{\partial \Lambda_\mu (p,q)}{\partial q_\nu} \right ]_{q=0} - \frac{\partial \Sigma (p)}{\partial p_\nu}.
\label{wtid2}
\end{equation}
The magnetic projection of right-hand-side (RHS)
provides the starting point of an independent formulation.
This will be called {\it Version A} \cite{kn1}.
The approach starting from the left-hand-side will be called {\it Version B}.
Although {\it Version A} requires additional algebraic work
it results in fewer and more compact integrals than {\it Version B}
and ensures significant economy of computing time.

The entire $\alpha^3$ term was coded 
in both {\it Version A} and {\it Version B}.
By 1974 this led to a crude estimate of $A_1^{(6)}$ \cite{ck2}.
To achieve a higher precision in this approach, however,
very extensive computer work was required,
which was not easily available at that time.
It was only in 1995 that we were able to obtain 
a sufficiently precise result \cite{kino1}
\begin{equation}
A_1^{(6)}({\rm num.}) =1.181~259~(40).
\label{numerical6}
\end{equation}
Shortly afterwards this value was confirmed by the analytic 
result \cite{lap-rem} to more than 5 digits.

Although analytic technique developed for integrating $A_1^{(6)}$,
in particular that of integration by part, have been useful
for analytic study of eighth-order term $A_1^{(8)}$,
further development seems to be needed
for a complete analytic integration of $A_1^{(8)}$ \cite{anal_dev}.
Only a small number of eighth-order diagrams without
closed lepton loop have been
integrated analytically thus far \cite{caffo}.

On the other hand, no theoretical difficulty has been encountered
in the Feynman-parametric approach to $A_1^{(8)}$.
The renormalization scheme developed in \cite{ck1} for the sixth-order
term could be readily extended to eighth-order.
However, the enormous size of integrals of Group IV (see Sec. \ref{sec:group4})
and Group V (see Sec. \ref{sec:group5}) of eighth-order term
and a large number of renormalization terms required 
presented a tremendous challenge to both algebraic construction of
integrands and their evaluation by numerical means.
For these reasons numerical evaluation of these diagrams
were carried out initially in {\it Version A} only, which required
smaller codes than {\it Version B}.

The first result of $A_1^{(8)}$ obtained by VEGAS \cite{lepage}
\begin{equation}
A_1^{(8)} = -1.434 (138) 
\label{firsta8}
\end{equation}
was reported by 1990 \cite{kl1,qedbook}.
This was a very preliminary result 
based on about $10^7$ sampling points per iteration
and about 30 iterations.
The estimated uncertainty of numerical evaluation was of the same order
of magnitude as that of the experimental value \cite{vandyck}.

In order to improve $A_1^{(8)}$ beyond (\ref{firsta8}) by an order
of magnitude or more
and move from a {\it qualitative} to {\it quantitative} calculation,
it is necessary to evaluate the integrals
using more than two orders of magnitude of sampling points.
It was estimated that this would require computation of
up to ten years on high speed computers then available.

Because of this lengthy time scale and numerous 
requests for information on
the status of calculation, preliminary values
have been reported at several Conferences and Meetings and some of them
have been printed in Proceedings and books:
\begin{equation}
\begin{array}{lclr}
A_1^{(8)} &=&  
-1.557~~(70)~~~~~~~~~~~~~~~~~~~~&\mbox{\rm \cite{prelim1}}, \\
A_1^{(8)} &=&  
-1.4092~(384),~~~~~~~~~~~~~~~~~ & \mbox{\rm \cite{prelim2}},  \\
A_1^{(8)} &=&  
-1.5098~(384),~~~~~~~~~~~~~& \mbox{\rm \cite{prelim3,hk}},  \\
A_1^{(8)} &=&  
-1.7366~(60),~~~~~~~~~~~~~~~~~~~ & \mbox{ \rm \cite{prelim4}},  \\
A_1^{(8)} &=&  
-1.7260~(50),~~~~~~~~~~~~~~~& \mbox{\rm \cite{prelim5,prelim6}}.  
\end{array}
\label{snapshots}
\end{equation}
%
Note that these values are not entirely independent.
This is due to the fact that complete evaluation of each integral
took typically three to six months of intense runs on fast computer.
At any particular moment only a few of these integrals are being upgraded.
The values in (\ref{snapshots}) are snapshots of this continuously
evolving numerical work.

The difficulty encountered in improving numerical
precision of $A_1^{(8)}$ turned out to be far greater than that of $A_1^{(6)}$.
It was partly due to 
the enormous size of integrands,
which made accumulation of good sampling of integrand difficult.
However, the most serious problem we have encountered is that our
method is very sensitive to the round-off error 
(called digit-deficiency or $d$-$d$ error)
which is present in any computer calculation based
on finite digit arithmetic, but affects our calculation
particularly severely.
(See Appendix B of Ref. \cite{kn2} for details.)
Because of these problems, some values given in (\ref{snapshots}) 
differ from each other considerably.
For instance, the results of \cite{prelim3} and \cite{prelim2}
differ by more than 2.6 s. d.
Actually, error estimate of \cite{prelim3} generated by
VEGAS was considerably smaller than that of \cite{prelim2}.
However, unstable and erratic behavior of some integrals, in particular $M_{40}$
discussed in detail in Sec. \ref{sec:group5}, indicated that
the error estimate generated by VEGAS could not be trusted
at face value.
This is why the error assigned to \cite{prelim3}
was enlarged to that of \cite{prelim2}.

More recently, 
we discovered an algebraic error in the integrands
formulated in {\it Version A} \cite{hk} 
while reevaluating 18 integrals of Group IV(d)
by an independent formulation in {\it Version B}.
Correction of this error is the cause of most of 
the jump from \cite{prelim3,hk} to \cite{prelim4}.

The discovery of this error forced us to re-examine other diagrams, too.
For this purpose, we have re-evaluated all diagrams of Groups IV(b) and IV(c)
by an independent formulation in {\it Version B}.
The result is that FORTRAN codes of all eighth-order $g-2$ diagrams 
containing closed electron loops
(373 diagrams) have now been verified by more than one
 independent formulations \cite{kn1,kn2}.
Remaining 518 diagrams belonging to Group V, which have no closed lepton loops, 
have thus far been formulated only in {\it Version A},
because the amount of work required for {\it Version B} would 
be several times larger than that of {\it Version A}.

On the other hand, Group V diagrams are subject to very tight
algebraic constraints.
Namely, these diagrams have numerous
renormalization terms which are constructed 
systematically by a power-counting rule that enables us to carry out an
extensive cross-checking among themselves and with lower-order
diagrams which are known exactly \cite{qedbook}.
Having examined this procedure once again,
we are convinced
that they are completely free from algebraic error,
although we found several typographical errors in
the intermediate renormalization scheme described in Appendix C and
Appendix D of \cite{qedbook}.
These errors are corrected in Appendix \ref{renormalization} of this paper.

The really time-consuming task is to make sure that 
the result of numerical integration by VEGAS \cite{lepage}
can be trusted.
This required a very good control of $d$-$d$ error.
(See Appendix B of Ref. \cite{kn2} for details.)
This work has been carried out on a number of
supercomputers over the period of more than 15 years.
The latest results of this on-going effort 
are given in Eqs. (\ref{a8I}), (\ref{a8II}), (\ref{a8III}), 
(\ref{a8IV}), and (\ref{a8V}).
Summing them up we obtain the total eighth-order term
\begin{equation}
A_1^{(8)} = -1.7283 (35), 
\label{newa8}
\end{equation}
where the uncertainty is determined by careful analysis
of errors estimated by VEGAS.
This is an order of magnitude improvement over the previous value \cite{prelim3,hk}.

At the level of precision we are interested in we
cannot totally ignore the tenth-order term $A_1^{(10)}$,
which is not known at present.
In view of the gently increasing trend of coefficients
of $(\alpha/\pi )^n$ as $n$ increases from 1 to 4,
however, it is unlikely that $A_1^{(10)}$ becomes suddenly very large.
Following the argument of \cite{mohr}, let us assume, until proved otherwise,
that
\begin{equation}
A_1^{(10)} = 0.0 ~(3.8).
\label{a10}
\end{equation}
In order to find out whether this is a justifiable assumption
we have embarked on a very extensive project to evaluate $A_1^{(10)}$ \cite{kn3,ahkn}.

The electron anomaly receives  also small mass-dependent contributions \cite{massdepcorr}
\begin{eqnarray}
A_2^{(4)} (m_e/m_\mu ) &=& 5.197~386~70~(27) \times 10^{-7},
  \nonumber  \\
A_2^{(4)} (m_e/m_\tau ) &=& 1.837~63~(60) \times 10^{-9},
  \nonumber  \\
A_2^{(6)} (m_e/m_\mu ) &=& -7.373~941~58~(28) \times 10^{-6},
  \nonumber  \\
A_2^{(6)} (m_e/m_\tau ) &=& -6.5819~(19) \times 10^{-8},
  \nonumber  \\
A_3^{(6)} (m_e/m_\mu, m_e/m_\tau ) &=& 0.190~95~(63)~ \times 10^{-12},
\label{massdep}
\end{eqnarray}
where the uncertainties 
are due to measurement uncertainties of $m_e/m_\mu$
and $m_e/m_\tau$ only.
Evaluation of $A_2^{(8)} (m_e/m_\mu )$,
$A_2^{(8)} (m_e/m_\tau )$, and
$A_3^{(8)} (m_e/m_\mu, m_e/m_\tau )$
are not difficult since we already have their codes.
But, numerically, they are totally insignificant at present.

To evaluate the theoretical value of $a_e$ we must
also include non-QED contributions of the Standard Model 
\cite{mohr,nonQED,czarnecki}:
\begin{eqnarray}
a_e ({\rm hadron}) &=& 1.671~(19) \times 10^{-12} ,
  \nonumber  \\
a_e ({\rm weak}) &=& 0.030~(1) \times 10^{-12} .
\end{eqnarray}
Although they are adequate for the moment, updates of these values
would be desirable.

In order to compare theory with experiment we need a value of
$\alpha$, and the most precise one for that.
Only recently such an $\alpha$ has become available by
the progress of atom interferometry \cite{wicht}.
Combined with the measurement of the cesium $D1$ line \cite{udem},
this leads to
\begin{equation}
\alpha^{-1} (h/{M_{Cs}}) = 137.036~000~3~(10)~~~~~~[7.4~{\rm ppb}].
\label{atominter}
\end{equation}
Using this $\alpha$ and  $a_e$
calculated from (\ref{newa8}), (\ref{massdep}), and (\ref{a10}) we obtain
\begin{equation}
 a_e (h/M_{Cs}) = 1~159~652~175.86~(0.10)~(0.26)~(8.48)~ \times 10^{-12} ,
\label{newae}
\end{equation}
where the first and second uncertainties are from (\ref{newa8})
and (\ref{a10}), and the third is from (\ref{atominter}).
This leads to
\begin{equation}
 a_e (exp) - a_e (h/M_{Cs}) = 12.4~(4.3)~(8.5) \times 10^{-12} ,
\label{diff}
\end{equation}
where first and second errors 
are from measurements of $a_e$ and
$\alpha (h/{M_{Cs}})$, respectively.

It is noted in \cite{wicht} that 
the uncertainty 7.4 ppb in (\ref{atominter}) may be brought down to 3.1 ppb 
when the  nature of systematic
error in the $h/M_{Cs}$ measurement is understood better.
This is very exciting:
The discrepancy (\ref{diff}) between theory and experiment would then
become 2.2 s. d., which may provide the first serious challenge 
to the theory and measurement of $a_e$.
Of course further work is needed to find out whether
this is a real discrepancy or something else.
The result of new measurement of $a_e$ \cite{gabrielse}
is anxiously waited for.

An alternate and better test of QED
is to compare $\alpha$ from (\ref{atominter}) with $\alpha$ obtained from 
theory and measurement of $a_e$:
\begin{eqnarray}
\alpha^{-1} (a_e) &=& 137.035~998~834~(12)~(31)~(502)~\nonumber   \\
                  &=& 137.035~998~834~(503)~~~~~~[3.7~{\rm ppb}],
\label{alpha_ae}
\end{eqnarray}
where, in the first line, 
12 and 31 are uncertainties in theoretical values (\ref{newa8}) and (\ref{a10})
and 502 is that of experimental value 
(\ref{ae_exp}).
This shows clearly that, if the measurement of $a_e$ is improved
by an order of magnitude, the precision of $\alpha (a_e)$
will be enhanced to about 0.3 ppb, putting it an order of magnitude
ahead of $\alpha (h/M_{Cs})$ of (\ref{atominter}).
The current theoretical uncertainty in $(\alpha/\pi )^4 A_1^{(8)}$
is $0.10 \times 10^{-12}$, which is smaller than the uncertainty
assumed for the $(\alpha /\pi )^5$ term.
This means that 
more accurate evaluation of $A_1^{(8)}$ will not
improve the QED prediction significantly until
a better estimate of $A_1^{(10)}$ is obtained.
Complete evaluation of $A_1^{(10)}$,
which has contributions from 12672 Feynman diagrams,
each of which may occupy FORTRAN code up to two orders
of magnitudes larger than those of eighth-order terms,
clearly requires an enormous amount of work \cite{kn3,ahkn}.

Appendix \ref{sec:elimination} describes how 
algebraic errors of Group V integrals are eliminated.
Two-step renormalization scheme of Group V diagrams in Version A
is described in Appendix \ref{renormalization}.

\section{Classification of  diagrams contributing to $A_1^{(8)}$}
\label{sec:qedcontribution}

$A_1^{(8)}$ has contributions from 891 eighth-order vertex diagrams.
Feynman integrals for these diagrams
consist of twelve propagators integrated over 4 four-dimensional
loop momenta.
They may have subdiagrams of
vacuum-polarization ({\it v-p}) type and/or  
light-by-light scattering ({\it l-l}) type.
They fall into five gauge-invariant groups according to 
the type of proper vertex diagrams from which they are derived
and the kind of 
closed electron loops they contain.
The {\it v-p} diagrams found in $A_1^{(8)} $ are as follows:

\noindent
$\Pi_2$, which consists of one closed lepton loop of second-order.

\noindent
$\Pi_4$, which consists of three proper
closed lepton loops of fourth-order.

\noindent
$\Pi_{4(2)}$, which consists of three lepton loops 
of type $\Pi_4$ whose internal photon line has a $\Pi_2$ insertion.

\noindent
$\Pi_6$, which consists of 15 proper
closed lepton loops of sixth-order.

The {\it l-l} diagrams we need are:

\noindent
$\Lambda_4$, which consists of six proper
closed lepton loops of fourth-order, with four photon
lines attached to them.

\noindent
$\Lambda_4^{(2)}$, which consists of 60 diagrams
in which lepton lines and vertices of $\Lambda_4$
are modified by second-order radiative corrections.

We are now ready to classify the diagrams
into five (gauge-invariant) groups:

\noindent
{\bf Group I.}  Second-order electron vertex diagrams containing lepton 
{\it v-p} loops $\Pi_2$, $\Pi_4$, $\Pi_{4(2)}$ and/or $\Pi_6$.
This group consists of 25 diagrams. 

\noindent
{\bf Group II.}  Fourth-order proper vertex diagrams containing lepton 
{\it v-p} loops $\Pi_2$ and/or $\Pi_4$. 
This group consists of 54 diagrams. 

\noindent
{\bf Group III.}  Sixth-order proper vertex diagrams containing a 
{\it v-p} loop $\Pi_2$.  This group consists of 150 diagrams.  

\noindent
{\bf  Group IV}.  Vertex diagrams containing a light-by-light 
scattering subdiagram $\Lambda_4^{(2)}$ 
or $\Lambda_4$ with further radiative 
corrections of various kinds.
This group has 144 diagrams, which can be divided
further into subgroups IV(a), IV(b), IV(c), and IV(d). 
 
\noindent
{\bf  Group V}.  Vertex diagrams with no closed electron loop.
It consists of 518 diagrams.

Diagrams of Groups I, II, III 
and IV(a) are relatively easy to deal with
since their structures are closely related to the second-,
fourth-, and sixth-order diagrams, which are known exactly.
They have been checked by several independent calculations \cite{kn2}. 

In order to deal with the remaining diagrams in general, we need
a full power of algebraic formulation originally developed
for sixth-order diagrams \cite{ck1}
and extended to the eighth-order diagrams
in \cite{kl1}.
But, for diagrams of Groups IV(b), IV(c), IV(d), as well as Group V,
the {\it Version B} codes 
are so huge that we chose initially to work only with {\it Version A}. 
Since we fixed the error in Group IV(d) with the help of {\it Version B}
calculation, however, we decided to 
evaluate Groups IV(b) and IV(c) in {\it Version B}, too.
The numerical results are in excellent agreement with {\it Version A} \cite{kn2}.

This leaves Group V as the only group which has
been evaluated in {\it Version A} only.
Until it is evaluated in {\it Version B} or other methods,
we have to rely on the very tight algebraic construction 
as the guarantor of the FORTRAN code of Group V.

\section{Group I Diagrams}
\label{sec:group1}

\begin{figure}[ht]
\resizebox{9.0cm}{!}{\includegraphics{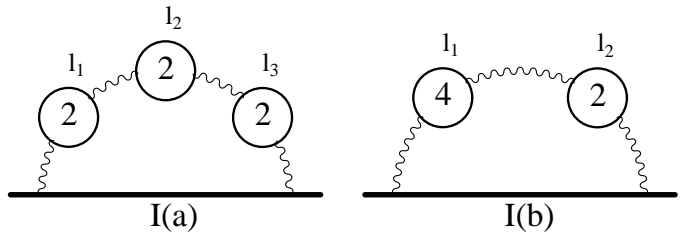}}
\caption{\label{vertex1} (a) Diagrams contributing to subgroup I(a).
(b) Diagrams contributing to subgroup I(b). 
Solid horizontal lines represent the electron in external magnetic field.
Numerals $``$2", $``$4" within solid circles refer to the 
proper renormalized {\it v-p} diagrams $\Pi_2$ and 
$\Pi_4$ , respectively.
One and six Feynman diagrams contribute to I(a) and I(b),
respectively.}
\end{figure}

Group I diagrams can be classified further into
four gauge-invariant subgroups:

{\em Subgroup I(a)}.  
Diagrams obtained by inserting three $\Pi_2$'s
in a second-order vertex.  
This subgroup consists of one Feynman diagram.
See Fig. \ref{vertex1}(a).  

{\em Subgroup I(b)}.  
Diagrams obtained by inserting a $\Pi_2$ and a $\Pi_4$
in a second-order vertex.  
Six Feynman diagrams belong to this subgroup.
See Fig. \ref{vertex1}(b).  

\begin{figure}[ht]
\vspace*{1cm}
\resizebox{9.0cm}{!}{\includegraphics{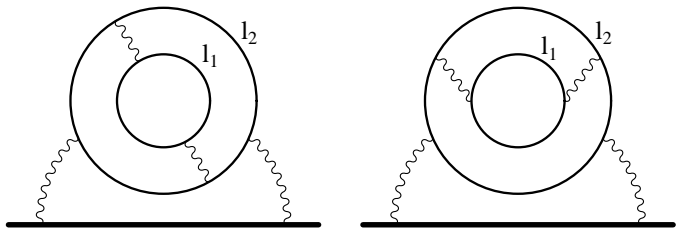}}
\caption{\label{vertex2} Diagrams contributing to subgroup I(c). $(l_1 , l_2 ) = (e, e)$.
See FIG. 1 for notation.}  
\end{figure}

{\em Subgroup I(c)}.  
Diagrams containing $\Pi_{4(2)}$.
There are three Feynman diagrams that belong to this subgroup. 
See Fig. \ref{vertex2}.  

\begin{figure}[ht]
\vspace*{1cm}
\resizebox{10.0cm}{!}{\includegraphics{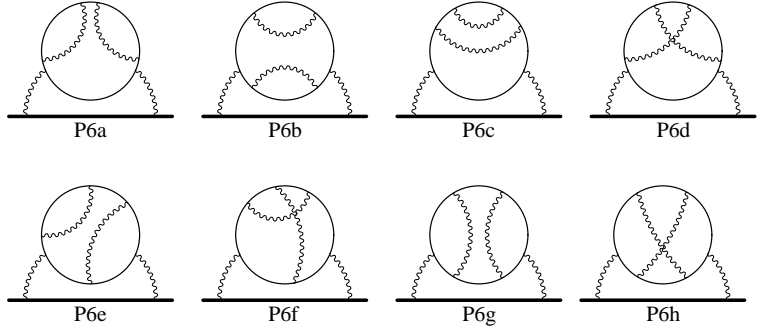}}
\caption{\label{vertex3} Eighth-order vertices of subgroup I(d) 
obtained by insertion of sixth-order (single electron loop) 
{\it v-p} diagram $\Pi_6$ in a second-order electron vertex.}  
\end{figure}

{\em Subgroup I(d)}.  
Diagrams obtained by insertion of $\Pi_6$
in a second-order electron vertex.  Fifteen
Feynman diagrams belong to this subgroup.  
Eight are shown in Fig. \ref{vertex3}.  
Diagrams $a, c,
d, e, f$ and the time-reversed diagram of $e$ have charge-conjugated
counterparts.

The evaluation of subgroups I(a) and I(b) is greatly
facilitated by the analytic formulas available for the second- and fourth-order 
spectral representations of the renormalized photon
propagators \cite{Kallen}.
The contribution to 
$a_e$ from the diagram obtained by sequential insertion of $m$ {\em k}-th
order electron and $n$ {\em l}-th order electron {\it v-p} loops into a
second-order electron vertex is reduced to a simple formula
\begin{equation}
a= \int_{0}^{1} dy(1-y) \left [ \int_{0}^{1} ds~{\frac{\rho_k (s)}{1~+~{\displaystyle {\frac{4}{1-s^2 }
{\frac{1-y}{y^2}} \left ( \frac{m_e}{m_{\mu} } \right )^2}} }} \right ]^m
\left [ \int_{0}^{1} dt~{\frac{\rho_l (t)}{1~+~{\displaystyle {\frac{4}{1-t^2} {\frac{1-y}{y^2}}}}}} \right ]^n ,           \label{a2mn}
\end{equation}
where $\rho_k$ is the {\em k}-th order photon spectral function.  
Exact $\rho_2$ and $\rho_4$ can be found in Ref. \cite{Kallen,kl1}.
An exact spectral function for $\Pi_{4(2)}$ and
an approximate one for $\Pi_6$ are also available \cite{hoang,broadhurst}.

The contribution of diagrams of Fig. \ref{vertex1} can be 
obtained by choosing $(k=2, m=3, n=0), (k=2, m=2,l=2, n=1), 
(k=2, m=1,l=2, n=2)$.
The latest numerical values obtained by
evaluating these integrals using VEGAS \cite{lepage} 
are listed in Table \ref{table1}, where the number of sampling points 
per iteration and the number of iterations are also listed.

Note that these diagrams need no additional renormalization.
Thus the renormalized amplitude  
$a_{2,p2:3}$ is given by
\begin{equation}
a_{2,p2:3} = M_{2,p2:3}.
\end{equation}

From Table \ref{table1}, we obtain the contribution of subgroup I(a) 
\begin{equation}
a_{I(a)}^{(8)}~ =~0.000~876~911~(40),             \label{a8Ia}
\end{equation}
which is 5 times more precise than the earlier result \cite{km},
and is in excellent agreement with the analytic result \cite{crt}:
\begin{equation}
a_{I(a)}^{(8)} (anal.)~ =~0.000~876~865~\ldots.            \label{a8Ia_asym}
\end{equation}

The contributions of Fig. \ref{vertex1}(b) for $(l_1 , l_2 ) = (e, e)$ 
can be written down in a similar fashion.  
The most recent result of numerical integration
by VEGAS is listed in the second row of Table \ref{table1}:
\begin{equation}
a_{I(b)}^{(8)}~ =~0.015~325~6~(6).             \label{a8Ib}
\end{equation}

\begin{table}
\renewcommand{\arraystretch}{0.80}
\begin{center}
\caption{Contributions of diagrams of Figs. \ref{vertex1}(a), \ref{vertex1}(b),
and \ref{vertex3}.
$n_F$ is the number of Feynman diagrams represented by the integral.
Evaluation was carried out on $\alpha$ workstations in 1997.
\\
\label{table1}
} 
\begin{tabular}{lcrrr}
\hline
\hline
~Integral~~& ~~~$n_F$~~~  &~Value (Error)~~~~~&~Sampling~per~& ~~~No. of~~~ \\ [.1cm]  
& &~~including $n_F$~&~iteration~~~~~& ~~iterations~  \\ [.1cm]   \hline
\\
$M_{2,P2:3}$ &1&~$0.876~911~(40)\times 10^{-3}$~&~~$1 \times 10^7$ \hspace{4mm}~~&55\hspace{4mm}~~\\ [.1cm]
\\
$M_{2,P2,P4}$ &6&~0.015~325~6~(6)~&~~$1 \times 10^7$ \hspace{4mm}~~&55\hspace{4mm}~~\\ [.1cm]
\\
$\Delta M_{2,P4a(P2)}$&1 &~0.011~403~1~(8)~&~~~~~~~$1 \times 10^7$ \hspace{4mm}~~&64\hspace{4mm}~~ \\ [.1cm]
$\Delta M_{2,P4b(P2)}$&2 &~0.001~716~8~(4)~&~~~~~~~$1 \times 10^7$ \hspace{4mm}~~&55\hspace{4mm}~~ \\ [.1cm]
\\
$\Delta M_{2,P6a}$&2 &~~0.044~448~(3)~&~~~~~$1 \times 10^8$ \hspace{4mm}~~&70\hspace{4mm}~~   \\ [.1cm]
$\Delta M_{2,P6b}$&1 &~~0.028~594~(3)~&~~~~~$1 \times 10^8$ \hspace{4mm}~~&70\hspace{4mm}~~   \\ [.1cm]
$\Delta M_{2,P6c}$&2 &~-0.038~374~(2)~&~~~~~$1 \times 10^8$ \hspace{4mm}~~&70\hspace{4mm}~~   \\ [.1cm]
$\Delta M_{2,P6d}$&2 &~-0.027~507~(2)~&~~~~~$1 \times 10^8$ \hspace{4mm}~~&70\hspace{4mm}~~   \\ [.1cm]
$\Delta M_{2,P6e}$&4 &~ 0.179~335~(3)~&~~~~~$4 \times 10^8$ \hspace{4mm}~~&70\hspace{4mm}~~   \\ [.1cm]
$\Delta M_{2,P6f}$&2 &~-0.062~004~(2)~&~~~~~$1 \times 10^8$ \hspace{4mm}~~&70\hspace{4mm}~~   \\ [.1cm]
$\Delta M_{2,P6g}$&1 &~ 0.038~879~(2)~&~~~~~$1 \times 10^8$ \hspace{4mm}~~&70\hspace{4mm}~~   \\ [.1cm]
$\Delta M_{2,P6h}$&1 &~~0.023~675~(2)~&~~~~~$1 \times 10^8$ \hspace{4mm}~~&70\hspace{4mm}~~   \\  [.1cm]
\\
\hline
\hline
\end{tabular}
\end{center}
\end{table}
\renewcommand{\arraystretch}{1}

In evaluating the contribution to $a_e$ from the 
9 Feynman diagrams of subgroup I(c) shown in FIG. \ref{vertex2}, 
our approach is to make use of the parametric integral
representation of the {\it v-p} term $\Pi_{4(2)}$.
Following the two-step renormalization procedure,
these contributions can be written in the form
\cite{seeref} 
\begin{equation}
a_{2,P4(P2)}~=~ 
\Delta {M_{2,P4a(P2)}}
~+~  \Delta {M_{2,P4b(P2)}}          
~-~ 2 \Delta B_{2,P2} M_{2,P2} ,  
\label{a2p4p2}
\end{equation}
where each term is
finite integral obtained by the 
$ {\bf K}_S $ renormalization procedure
described in Ref. \cite{kl1,kn2}.  The 
suffix $P2$ stands for the second-order {\it v-p} diagram
$\Pi_2$, $ P4 $ for the fourth-order {\it v-p}
diagram $\Pi_4$, while 
$P4 (P2)$ represents the diagram $\Pi_{4(2)}$.
$P4$ receives contributions from $ P_{4a} $ 
(vertex correction) and $ P_{4b} $ (two lepton self-energy insertions),
$P4=P_{4a}+ P_{4b}$.

The results of numerical evaluation of (\ref{a2p4p2}),
obtained by VEGAS,
are listed in Table \ref{table1}.  
Note that the value listed in column 3 for
$ \Delta {M_{2,P4b(P2)}}$ includes the factor $n_F$.          
Numerical values of 
lower-order Feynman integrals, in terms of which the residual renormalization
terms are expressed, are given in Table \ref{table3aux}.  From 
these Tables we obtain
\begin{equation}
a_{Ic}^{(8)} =~0.011~130~8~(9)   .          
\label{a8Ic_ee}
\end{equation}

We obtained an independent check of (\ref{a8Ic_ee}) using 
a two-dimensional integral form of
$\alpha^3$ spectral function for $\Pi_{4(2)}$ of Fig. \ref{vertex2},
which was derived from the QCD spectral function obtained in \cite{hoang}.
Numerical integration using this spectral function gives
\begin{equation}
a_{Ic}^{(8)} =~0.011~131~2~(12) ,          
\label{a8Ic_exact}
\end{equation}
for 10 million sampling points iterated 20 times in quadruple precision.
This is in agreement with (\ref{a8Ic_ee}) to the fifth decimal points
although their approaches are quite different.

\renewcommand{\arraystretch}{0.80}
\begin{table}
\caption{ Auxiliary integrals for Group I.
Some integrals are known exactly. 
Remaining integrals are obtained numerically by VEGAS.
\label{table3aux}
} 
\begin{tabular}{llll}
\hline
\hline
~~~Integral~~~  &~~~Value~(Error)~~~~& ~~~Integral~~~~   &~~~Value~(Error)~~~~~~  \\ [.1cm]   \hline
~$M_{2 ,P2}$~~~&~~0.015~687~421~$\cdots$ ~& ~$M_{2 ,P2^*}$~~~&~-0.012~702~383~$\cdots$ ~\\[.1cm]
~$\Delta M_{2 ,P4}$~~~&~~0.076~401~785~$\cdots$ ~& ~~&~~\\[.1cm]
~$\Delta B_{2}$~~~&~~0.75~~~~&~~$\Delta B_{2,P2}$~~&~~0.063~399~266~$\cdots$ \\[.1cm]
~$\Delta L_4$~~~&~~0.465~024~(17)~~~~&~~$\Delta B_4$~~&~-0.437~094~(21)~~\\
[.1cm]
~$\Delta \delta m_{4}$~~~&~~1.906~340~(21)~~~~&~~~&~~~\\
[.1cm]
\hline
\hline
\end{tabular}
\end{table}
\renewcommand{\arraystretch}{1}

\noindent
The new result (\ref{a8Ic_ee})
confirms the old result \cite{qedbook} but with a much higher precision.

The contribution to $a_e$ from 15 diagrams of subgroup I(d) (see Fig. \ref{vertex3})
can be written as
\begin{equation}
a_{2,P6i}~=~\Delta M_{2,P6i}~+~{\rm {residual~renormalization~terms}}~, 
~~~~~(i~=~a,~ .~.~.~,~h).     \label{groupId}
\end{equation}
Divergence-free integrals $\Delta M_{2,P6i}$ are defined by (4.13)
of Ref. \cite{kl1}.  
Their numerical values (summed over the diagrams related by time-reversal and
charge-conjugation symmetries) 
are evaluated numerically by VEGAS
and listed in the third column of Table \ref{table1}.

Summing up the contributions of diagrams $P6a$ to $P6h$ of Fig. \ref{vertex3},
we obtain the following expression:
\begin{align}   
a_{I(d)}^{(8)} &=~\sum_{i=a}^h  \Delta M_{2,P6i}
~-~4 \Delta B_2 \Delta M_{2,P4}    \nonumber  \\
&+~5( \Delta B_2 )^2 M_{2,P2}
~-~2( \Delta L_4~+~\Delta B_4) M_{2,P2}   \nonumber  \\
&-~2 \Delta \delta m_4 M_{2,P2^*} ,  
\label{a8Idform}
\end{align}
where the value of $\Delta M_{2,P6i}$ in column 3 
of Table \ref{table1} is the sum of $n_F$ vertex diagrams and
\begin{align} 
\Delta B_2  ~&=~{ \Delta}^{'} B_2 ~+~
{\Delta}^{'} L_2 ~=~ {\frac{3}{4}} ,    \nonumber  \\
{\Delta} M_{2,P4}~&=~\Delta M_{2,P4a}~+~
2 \Delta M_{2,P4b} ,    \nonumber  \\
\Delta L_4 ~&=~\Delta L_{4x} ~+~2 \Delta L_{4c} +~\Delta L_{4l}
~+~2 \Delta L_{4s} ,    \nonumber  \\
\Delta B_4 ~&=~\Delta B_{4a} +~\Delta B_{4b} ,    \nonumber  \\
\Delta \delta m_4 ~&=~\Delta \delta m_{4a} +~\Delta \delta m_{4b} . 
\label{aux}
\end{align}
The quantities listed in (\ref{aux}) are defined in Ref. \cite{kl1}.  Their
numerical values are listed in Table \ref{table3aux}.  
The 1998 results of numerical integration of $\Delta M_{2,P6i}$
are listed in Table \ref{table1}.  From 
the numerical values in Tables \ref{table1} and \ref{table3aux} we obtain
\begin{equation}
a_{I(d)}^{(8)} =~0.049~516~1~(63) .            
\label{a8Id}
\end{equation}

We obtained an independent check of (\ref{a8Id}) using the $Pad\acute{e}$
approximant of the sixth-order photon spectral function \cite{broadhurst} 
\begin{equation}
a_{I(d)}^{(8)}(Pad\acute{e}) = ~0.049~519~5~(38) ,            
\label{a8IdPade}
\end{equation}
which is in excellent agreement with (\ref{a8Id}).
Calculation was carried out in real*16 for 10 million sampling points
iterated 100 times.

Collecting the results (\ref{a8Ia_asym}), (\ref{a8Ib}), 
(\ref{a8Ic_ee}) and (\ref{a8Id}), we find the best value of the
contribution to the electron anomaly from the 25 diagrams of group I to be 
\begin{equation}
A_1^{(8)} [I]~=~0.076~849~(7) .             
\label{a8I}
\end{equation}

\section{Group II Diagrams}
\label{sec:group2}

Diagrams of this group are generated by inserting $\Pi_2$
and $\Pi_4$ in the photon lines of
fourth-order vertex diagrams.
Use of analytic expressions for the second- and fourth-order spectral
functions for the photon propagators
and time-reversal symmetry cuts down the number of 
independent integrals in {\it Version A}
from 54 to 8.

\begin{figure}[ht]
\vspace*{1cm}
\resizebox{10.0cm}{!}{\includegraphics{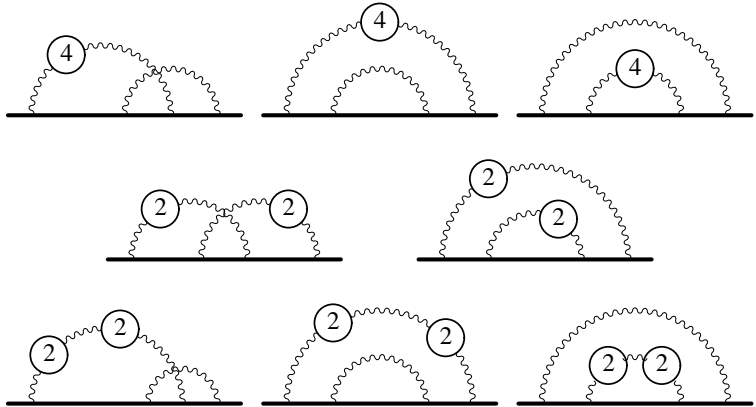}}
\vspace{0.5cm}
\caption{\label{vertex5} Eighth-order diagrams obtained from 
the fourth-order vertex diagrams by inserting vacuum-polarization loops
$\Pi_2$ and $\Pi_4$.
}
\end{figure}

The contribution to $a_e$ arising from the set of vertex diagrams
represented by the $``$self-energy" diagrams of Fig. \ref{vertex5}
can be written in the form
\begin{equation}
a_{4,P_{\alpha}}~ =~\Delta M_{4,P_{\alpha}} +
~{\rm {residual ~renormalization~ terms}}  ,
\end{equation}
where $\Delta M_{4,P_{\alpha}}$ are finite integrals
obtained in the intermediate step of two-step
renormalization \cite{qedbook}.  Their 
numerical values, obtained by VEGAS 
are listed in Table \ref{table4}.   The values of auxiliary integrals
needed to calculate the total contribution of group II diagrams are given in
Tables \ref{table3aux} and \ref{table4aux}.

Summing the contributions of diagrams of the first, second, 
and third rows of Fig. \ref{vertex5}, 
one obtains (see Ref. \cite{qedbook} for notation)
\begin{align} 
a_{4,P4} ~&=~ \Delta M_{4a,P4} +~
\Delta M_{4b,P {1^{'}} : 4} ~+~\Delta M_{4b,P0:4}~~~~   \nonumber  \\
 ~&-~{\Delta}  B_2 M_{2,P4}
~-~{\Delta} B_{2,P4} M_2 ,   
\label{P4form}
\end{align}
\begin{equation} 
a_{4,P2,P2} ~=~\Delta M_{4a,P2,P2}~+~
\Delta M_{4b,P1^{'} : 2 ,P0:2}   
~-~ {\Delta} B_{2,P2}~M_{2,P2}  ,
\label{P2P2form}
\end{equation}
\begin{eqnarray} 
a_{4,P2:2}~&=~ \Delta M_{4a,P2:2}~+~
\Delta M_{4b,P {1^{'}} :2:2}+~
\Delta M_{4b,P0:2:2}   \nonumber  \\
~&-~\Delta B_2 M_{2,P2:2}~-~
{\Delta} B_{2,P2:2}~M_2~ ,
\label{P22form}
\end{eqnarray}
respectively, 
where $M_{2,P4}$ is equal to 
$\Delta M_{2,P4} - 2 \Delta B_2 M_{2,P2}$.  
$M_{4a,..}$ corresponds to the left-most diagrams
of Fig. \ref{vertex5} and $M_{4b,..}$ ones on the right.
$\Delta M_{4a,...}$ is the sum of $n_F$ vertex diagrams.
On the other hand, the auxiliary integrals listed in Tables \ref{table3aux}
and \ref{table4aux} do not include multiplicity.

\renewcommand{\arraystretch}{0.80}
\begin{table}
\caption{ Contributions of diagrams of Fig. \ref{vertex5}.
$n_F$ is the number of Feynman diagrams represented by the integral.
Suffixes $P0$ and $P1'$ 
are remnants of notations in \cite{qedbook}
and refer to two virtual photons in which v-p insertions are made.
Numerical evaluation was carried out on $\alpha$ workstations in 2001.
\\
\label{table4}
} 
\begin{tabular}{lclrr}
\hline
\hline
~Integral~~& ~~~$n_F$~~~  &~Value (Error)~~~~~&~Sampling~per~& ~~~No. of~~~ \\ [.1cm]  
& &~~including $n_F$~&~iteration~~~~~& ~~iterations~  \\ [.1cm]   \hline
$\Delta M_{4a,P4}$ &~18&~~0.131~298~(8)&$~~2 \times 10^9$ \hspace{8mm}~~&100\hspace{4mm}~~ \\ [.1cm]
$\Delta M_{4b,P0:4}$&&&&   \\ [.1mm]
~~~+$\Delta M_{4b,P1':4}$
&18&$-$0.420~295~(8)&$~~4 \times 10^8$ \hspace{8mm}~~&100\hspace{4mm}~~   \\ [.1cm]
$\Delta M_{4a,P2,P2}$ &3&~~-0.004~351~(2)&$~~1 \times 10^8$ \hspace{8mm}~~&70\hspace{4mm}~~  \\ [.1cm]
$\Delta M_{4b,P1':2,P0:2}$
&3&$-0.022~332~(1)$&$~~1 \times 10^8$ \hspace{8mm}~~&70\hspace{4mm}~~ \\ [.1cm]
$\Delta M_{4a,P2:2}$ &6&$~~0.007~875~(8)$&$~~2 \times 10^8$ \hspace{8mm}~~&70\hspace{4mm}~~  \\ [.1cm]
$\Delta M_{4b,P0:2:2}$ &&&&   \\ [.1mm]
~~~+$\Delta M_{4b,P1':2:2}$
&6&$-0.065~458~(4)$&$~~1 \times 10^8$ \hspace{8mm}~~&70\hspace{4mm}~~ \\ [.1cm]
\hline
\hline
\end{tabular}
\end{table}
\renewcommand{\arraystretch}{1}

\renewcommand{\arraystretch}{0.80}
\begin{table}
\caption{ Auxiliary integrals for Group II.
Some integrals are known exactly. 
Remaining integrals are obtained by VEGAS integration, with
total sampling points of order $10^{11}$.
\\
\label{table4aux}
} 
\begin{tabular}{llll}
\hline
\hline
~~~Integral~~~  &~~~Value~(Error)~~~~& ~~~Integral~~~~   &~~~Value~(Error)~~~~~~  \\ [.1cm]   \hline
~$M_{2}$~~~&~~0.5~~~~&~~$M_{2,P4}$~~&~~0.052~870~652~$\cdots$\\[.1cm]
~$ M_{2 ,P2:2}$~~~&~~0.002~558~524~$\cdots$~~~~&~~~&~~\\[.1cm]
~$\Delta B_{2,P4}$~~~&~~0.183~666~5~(18)~&~~$\Delta B_{2,P2:2}$~~&~~0.027~902~3~(4)~~~\\[.1cm]
\hline
\hline
\end{tabular}
\end{table}
\renewcommand{\arraystretch}{1}

Substituting the data from Tables \ref{table3aux} and \ref{table4}
 into (\ref{P4form}), 
(\ref{P2P2form}), and (\ref{P22form}) we obtain
\begin{align} 
a_{4,P4} ~&=~-0.420~483~(12) ,      \nonumber  \\
a_{4,P2,P2} ~&=~-0.027~677~(~2)  ,  \nonumber  \\
a_{4,P2:2} &=~-0.073~452~(~9) .     
\label{a8IInum}
\end{align}
Note that values listed in column 3 of Table \ref{table4}
include the factor $n_F$.

The sum of contributions of 54
diagrams of group II is
\begin{equation}
a_{~II}^{(8)} ~=~-0.521~613~(15)  .  ~~    
\label{a8II}
\end{equation}

\section{Group III Diagrams}
\label{sec:group3}

\begin{figure}[ht]
\vspace*{1cm}
\resizebox{12.0cm}{!}{\includegraphics{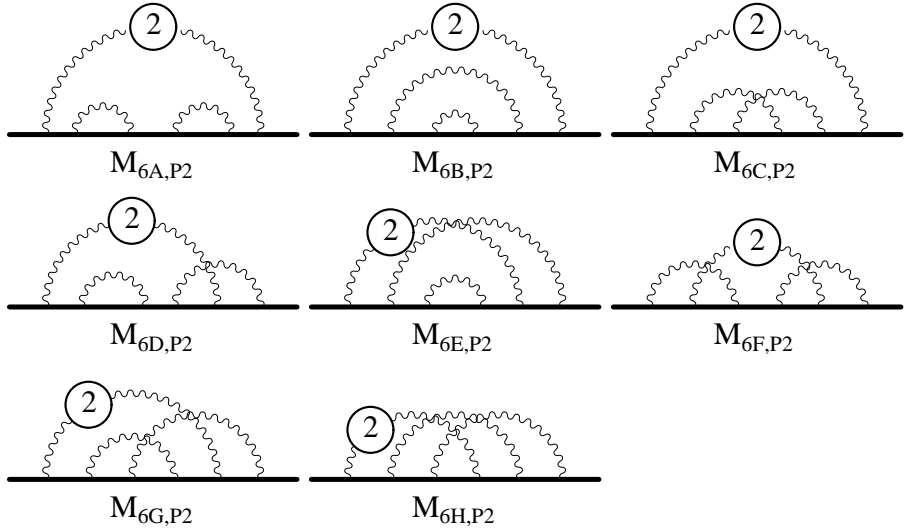}}
\vspace{0.5cm}
\caption{\label{vertex6} 
Typical eighth-order diagrams obtained by insertion of
a vacuum-polarization loop $\Pi_2$ in electron 
diagrams of the three-photon-exchange type.  Altogether there are
150 diagrams of this type.
}
\end{figure}

Diagrams belonging to this group are generated 
by inserting a second-order
vacuum-polarization loop  $\Pi_2$ in
the photon lines of sixth-order electron vertex
diagrams of the three-photon-exchange type.
Time-reversal invariance and use of the function 
${\rho}_2$ (see (\ref{a2mn})) for the
photon spectral function 
reduce the
number of independent integrals in {\it Version A} from 150 to 8.  
Some of these integrals are represented by the 
$``$self-energy" diagrams of Fig. \ref{vertex6}.

Let $M_{6 \alpha ,P2}$ be the magnetic moment 
projection in {\it Version A}
of the set of 150 diagrams generated from a self-energy
diagram $ \alpha $ (=A through H) of Fig. \ref{vertex6} 
by insertion of $\Pi_2$ and
an external vertex.
The renormalized
contribution due to the group III diagrams can then 
be written as 
\begin{equation}
a_{III}^{(8)} ~=~\sum_{\alpha = A}^H  a_{6 \alpha ,P2} ,
\end{equation}
where
\begin{equation}
a_{6 \alpha ,P2} ~=~\Delta M_{6 \alpha ,P2} ~+~ {\rm {residual~renormalization~terms}} .
\end{equation}
$\Delta M_{6 \alpha ,P2}$ is UV- and IR-finite: 
all divergences of $M_{6 \alpha ,P2}$ 
have been projected out by ${\bf K}_S$ and
${\bf I}_R$ operations.  (See Ref. \cite{qedbook}.)  

The latest numerical values
of Group III integrals are summarized in 
Table \ref{table5}.  
Numerical values of auxiliary integrals
needed in the renormalization scheme are listed in Tables \ref{table3aux},
\ref{table4aux} and \ref{table5aux}.
For comparison, the
results of
old calculation \cite{km} carried out in double precision
are listed in the last column of Table \ref{table5}. 
This is to examine the effect of {\it digit-deficiency} error.
In this case the effect is relatively mild because
the introduction of a vacuum-polarization loop tends to make
the integrand less sensitive to the singularity.

When summed over all the diagrams of group III, the UV- and IR-divergent pieces cancel
out and the total contribution to $a_e$ can be written as a sum of
finite pieces (see Ref. \cite{qedbook} for notation):
\begin{align} 
a_{III}^{(8)}~&=~ \sum_{ \alpha = A}^H
\Delta M_{6 \alpha ,P2}        \nonumber  \\
~&-~3 \Delta B_{2,P2} \Delta M_{4} -3 \Delta B_2  \Delta M_{4,P2}              \nonumber  \\
~&+~( M_{2^{*} ,P2} [I] - M_{2^* , P2} )
\Delta \delta m_4
~+~(M_{ 2^{*}} [I] -M_{2^*} )  \Delta \delta m_{4,P2}
             \nonumber  \\
~&-~M_{2,P2} [ \Delta B_4 + 2 \Delta L_4 ~-~ 2( \Delta B_2 )^2 ]                \nonumber  \\
~&-~M_2 ( \Delta B_{4, P2} + 2 \Delta L_{4,P2} ~
~-~4 \Delta B_2 \Delta B_{2,P2} ) ,   
\label{aIII}
\end{align}
where $\Delta M_{6 \alpha ,P2}$ listed in Table \ref{table5} 
is the sum of $n_F$ vertex diagrams.
$M_{2^*}$ is the magnetic moment obtained from $M_2$ by insertion of a two-point
vertex in the electron lines. $M_{2^*}[I]$ is the part of $M_{2^*}$ which may be
IR-divergent according to the IR power counting. An analytic evaluation
of $M_{2^*}[I]$ turns out to be finite and is listed in Table \ref{table5aux}. 

\renewcommand{\arraystretch}{0.80}
\begin{table}
\caption{ Contributions of diagrams of Fig. \ref{vertex6}.
$n_F$ is the number of Feynman diagrams represented by the integral.
Integrals were evaluated in double precision except for
a small part of $\Delta M_{6G,P2}$ which was evaluated
in quadruple precision.
\\
\label{table5}
}
\begin{tabular}{lcrrrr}
\hline
\hline
~Integral~~& ~~~$n_F$~~~  &~Value (Error)~~~~~&~Sampling~per~& ~~~No. of~~~&~~~~Data from~~~ \\ [.1cm]  
& &~~including $n_F$~~~~~&~iteration~~~~~& ~~iterations &~~Ref. \cite{km}~~~~ \\ [.1cm]   \hline
$\Delta M_{6A,P2}$&5& $-$0.440~307~(56)~&~~$1 \times 10^8$ \hspace{4mm}~~&97\hspace{4mm}~&$-$0.440~3~(3) \\[.1mm]
$\Delta M_{6B,P2}$&5&~~0.689~660~(84)~&~~$2 \times 10^8$ \hspace{4mm}~~&100\hspace{4mm}~&0.690~7~(7) \\[.1mm]
$\Delta M_{6C,P2}$&5&~~~0.638~503~(80)~&~~$2 \times 10^8$ \hspace{4mm}~~&100\hspace{4mm}~&0.638~2~(7) \\[.1mm]
$\Delta M_{6D,P2}$
&10&~~0.440~999~(77)~&~~$2 \times 10^8$ \hspace{4mm}~~&100\hspace{4mm}~&0.442~0~(7) \\[.1mm]
$\Delta M_{6E,P2}$
&5&~~0.431~541~(55)~&~~$1 \times 10^8$ \hspace{4mm}~~&100\hspace{4mm}~&0.432~1~(4) \\[.1mm]
$\Delta M_{6F,P2}$
&5&~~~0.442~163~(83)~&~~$2 \times 10^8$ \hspace{4mm}~~&100\hspace{4mm}~&0.441~9~(8) \\[.1mm]
$\Delta M_{6G,P2}$ &10&$~~~~~~1.140~811~(76)$&~&~~&1.140~4~(10) \\[.1mm]
~~~~{\it d-part}&&$~~1.054~635~(60)$&$2 \times 10^{8} \hspace{4mm} $ &100 \hspace{4mm}  &  \\
~~~~{\it q-part}&&$~~0.086~176~(47)$&$2 \times 10^7 \hspace{4mm} $&100 \hspace{4mm}  &  \\
$\Delta M_{6H,P2}$
&5&$-$0.797~941~(83)~&~~$6 \times 10^8$ \hspace{4mm}~~&100\hspace{4mm}~&$-$0.797~5~(12) \\ [.1cm]
\hline
\hline
\end{tabular}
\end{table}
\renewcommand{\arraystretch}{1}

\renewcommand{\arraystretch}{0.80}
\begin{table}
\caption{ Auxiliary integrals for Group III.
Some integrals are known exactly. 
Remaining integrals are obtained by VEGAS integration, with
total sampling points of order $10^{11}$.
\\
\label{table5aux}
} 
\begin{tabular}{llll}
\hline
\hline
~~~Integral~~~  &~~~Value~(Error)~~~~& ~~~Integral~~~~   &~~~Value~(Error)~~~~~~  \\ [.1cm]   \hline
~$M_{2^*}$~~~&~~1.0~~~~&~ ~$M_{2^*}[I]$~~~&~~$-$1.0~~~\\[.1cm]
~$M_{2^*,P2}$~~~&~~0.044~077~4~(3)~~&~ ~$M_{2^*,P2}[I]$~~~&~~0.010~255~3~(11)~~\\[.1cm]
~$\Delta M_4$~~~&~~0.030~833~612~$\cdots$~~~~&~~$\Delta M_{4,P2}$~~~&~~$-$0.106~707~082~$\cdots$~~\\[.1cm]
~$\Delta \delta m_4$~~~&~~1.906~340~(21)~~~~&~~$\Delta \delta m_{4,P2}$~~&~~0.679~769~(15)\\[.1cm]
~$\Delta B_{4,P2}+2 \Delta L_{4,P2}$~~~&~~0.085~865~(30)~~~~&~~~~&~~ \\[.1cm]
\hline
\hline
\end{tabular}
\end{table}
\renewcommand{\arraystretch}{1}

Plugging the values listed in Tables 
\ref{table3aux}, \ref{table4aux} and \ref{table5aux}
in (\ref{aIII}), we obtain
\begin{equation}
a_{III}^{(8)}~=~1.417~722~(214)  .     
  \label{a8III}
\end{equation}

\section{Group IV Diagrams}
\label{sec:group4}

Diagrams of this group can be divided into four subgroups:
IV(a), IV(b), IV(c), and IV(d).  Each subgroup
consists of two equivalent sets of diagrams related by charge conjugation 
(reversal of the direction of momentum flow in the loop of the light-by-light
scattering subdiagram).
Diagrams of subgroups IV(a), IV(b), and IV(c) are obtained
by modifying the sixth-order diagram $M_{6LL}$ 
which contains the 
light-by-light scattering subdiagram $\Lambda_4$, one of
whose external photon line represents the magnetic field. 
The magnetic moment contribution $M_{6LL}$
is known analytically \cite{laporta2}, whose numerical value is
\begin{equation}
M_{6LL} =~0.371~005~292 ... .
\label{a6LL}
\end{equation}

\begin{figure}[ht]
\vspace*{1cm}
\resizebox{10.0cm}{!}{\includegraphics{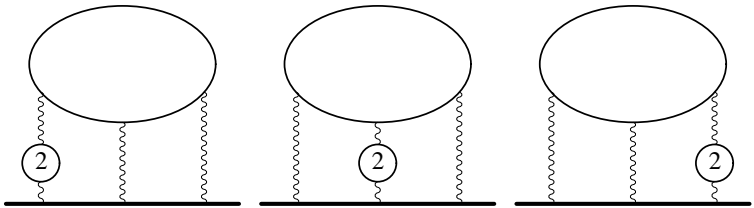}}
\vspace{.6cm}
\caption{\label{vertex7} Electron self-energy-like diagrams 
representing the external-vertex-summed integrals of subgroup IV(a). 
$(l_1 , l_2) = (e,e)$. }  
\end{figure}

{\em Subgroup IV(a).}
Diagrams obtained by inserting a second-order
vacuum-polarization loop $\Pi_2$ in $M_{6LL}$.
This subgroup consists of 18 vertex diagrams. 
In {\it Version A} they are represented by the
self-energy-like diagrams shown in Fig. \ref{vertex7}.
Insertion of the magnetic vertex in these figures generates
diagrams of {\it Version B}.
Since $M_{6LL}$ is known analytically and since insertion of $\Pi_2$
is straightforward, this subgroup is known exactly.

\begin{figure}[ht]
\vspace*{1cm}
\resizebox{10.0cm}{!}{\includegraphics{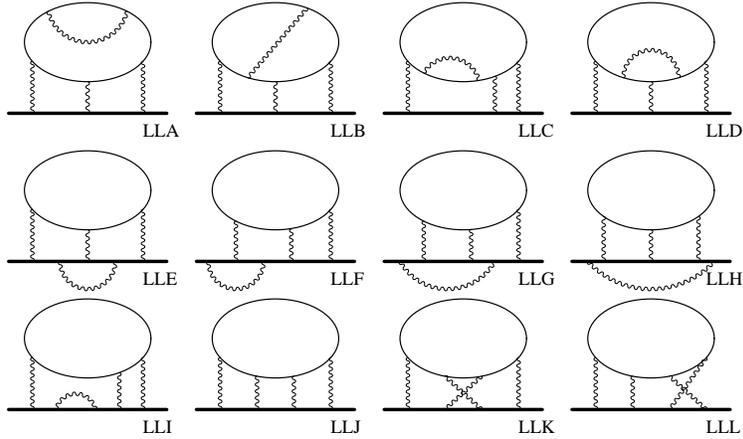}}
\vspace{.6cm}
\caption{\label{vertex8} Electron self-energy-like diagrams representing 
(external-vertex-summed) integrals of subgroup IV(b), IV(c), and IV(d).}  
\end{figure}

{\em Subgroup IV(b).}
Diagrams containing sixth-order light-by-light
scattering subdiagrams $\Lambda_{4(2)}$.  
Altogether, there are 60 diagrams of this
type.  Charge-conjugation and time-reversal 
symmetries and summation over external vertex insertions reduce
the number of independent integrals to 4 in {\it Version A},
represented by the self-energy-like diagrams LLA, LLB, LLC and
LLD of Fig. \ref{vertex8}.
They can be generated from a common template by permutation of
tensor indices of virtual photons attached to the $\Lambda_4^{(2)}$ loop.
Similarly, all diagrams of {\it Version B} can be generated from
a common template.

{\em Subgroup IV(c).}
Diagrams obtained by including 
second-order radiative corrections on
the open electron line of $M_{6LL}$.
There are 48 diagrams that belong to this subgroup.  
Summation
over external vertex insertions and use of the interrelations
available due to charge-conjugation and time-reversal symmetries leave five
independent integrals in {\it Version A},  
represented by the self-energy-like diagrams 
LLE, LLF, LLG, LLH and LLI of Fig. \ref{vertex8}.
They can be generated from a common template by permutation of
tensor indices of virtual photons attached to the open electron line.
Similarly, all diagrams of {\it Version B} can be generated from
a common template.

{\em Subgroup IV(d).}
Diagrams generated by inserting $\Lambda_4$
{\it internally} in fourth-order vertex
diagrams.  Diagrams of this type appear
for the first time in the eighth order.  
Charge-conjugation invariance and
summation over the external vertex insertion 
with the help of the Ward-Takahashi
identity lead us to three independent integrals in {\it Version A},  
represented by the diagrams LLJ, LLK and LLL of Fig. \ref{vertex8}.
They can be generated from a common template by permutation of
tensor indices of virtual photons attached to the $\Lambda_4$ loop.
Similarly, all diagrams of {\it Version B} can be generated from
a common template.

In subgroups IV(a), IV(b), and IV(c) UV-divergences 
arising from the light-by-light scattering subdiagram
$\Lambda_4$, or more explicitly
$ {\Pi}^{ \nu \alpha \beta \gamma } (q, k_i , k_j , k_l )$,
can be taken care of by making use of the identity:
\begin{equation}
{\Pi}^{ \nu \alpha \beta \gamma } ( q, k_i , k_j , k_l ) ~=~
-~q_{\mu} \left [{\frac{\partial}{{ \partial q}_{\nu}} } {\Pi}^{ \mu \alpha \beta  
\gamma } ( q , k_i , k_j , k_l ) \right ] ,     \label{WTid}
\end{equation}
which follows from the Ward-Takahashi identity.
Namely, no explicit UV renormalization is needed
if one uses the RHS of (\ref{WTid}) instead of LHS
 and the fact that $\Sigma (p)$ of (\ref{wtid2})
vanishes by Furry's theorem.  
On the other hand, $\Sigma (p)$ is nonzero for subgroup IV(d) 
and the UV-divergence associated with the light-by-light
scattering subdiagram  $\Lambda_4$ must be regularized, e. g., 
by dimensional regularization.   For
these diagrams it is necessary to carry out explicit renormalization
of $\Lambda_4$ as well as that of the two sixth-order 
vertex subdiagrams containing $\Lambda_4$.  
See \cite{kn1} for a detailed discussion of renormalization
based on a combination of dimensional regularization and
Pauli-Villars regularization.

Many diagrams of Groups IV(b) and IV(c) have UV divergences associated with
the second-order vertex or self-energy parts.
Their renormalization is achieved by a subtraction term which can be
factorized analytically into a product of UV-divergent part
of the renormalization constant and the sixth-order amplitude
$M_{6LL}$ (in {\it Version A}) or its component (in {\it Version B}).
This enables us to verify the analytic structure 
of many terms of the integrand.
A similar argument applies to diagrams with IR divergence.
Since different diagrams generated by the same template
have different divergences, this procedure explores the validity 
of template itself exhaustively.

On the other hand, Group IV(d) diagrams LLJ, LLK, and LLL
have no UV-divergent subdiagram of second-order.
Thus they could not be verified by comparison with known lower-order
structure.
Indeed, this is why the errors in 18 integrals 
of Group IV(d) remained undetected
until the integrand was constructed
in {\it Version B} \cite{kn1}.
The error was in the template of Group IV(d)
where some non-divergent terms were left out inadvertently.

The argument presented above suggests that the validity of codes for
Groups IV(b) and IV(c)
can be established within {\it Version A}. 
However, as a further check, we have  also evaluated 
them in {\it Version B}.
In the following let us consider 
{\it Version A} and {\it Version B} separately
since renormalization is handled slightly differently in two cases.

\subsection{Version A}
\label{subsec:versa} 

The calculation of group IV(a) contribution is particularly simple.
This is because $M_{6LL}$ has been fully tested by comparison
with the analytic result \cite{laporta2}, and insertion
of the vacuum polarization term is straightforward.
As a consequence the integral $M_{6LL,P2}$ 
is finite, which means

\begin{equation}
a_{6LL,P2}~=~
M_{6LL,P2}~=~
\Delta M_{6LL,P2} .
\label{a8IVa}
\end{equation}
The result of numerical calculation is listed in Table \ref{table5aux}.

Let us denote magnetic projections of subgroups IV(b) and IV(c)
as $M_{8LL\alpha}$ where $\alpha = A,..,I$.
Relating the IR- and UV-divergent $M_{8LL \alpha }$ to the finite, numerically
calculable piece $\Delta M_{8LL \alpha}$ defined by the procedure of two-step
renormalization of Ref. \cite{KL4}, one can write 
the contributions of the diagrams of subgroups IV(b) and IV(c) as
\begin{equation}
a_{IV(b)}^{(8)} ~=~ \sum_{ \alpha = A }^D
\Delta M_{8LL \alpha } ~-~
3 \Delta B_2 M_{6LL} ,       \label{a8IVb}
\end{equation}
and
\begin{equation}
a_{IV(c)}^{(8)} =~\sum_{ \alpha = E }^I
\Delta M_{8LL \alpha } ~-~
2 \Delta B_2 M_{6LL} ,          \label{a8IVc}
\end{equation}
where the six-order light-by-light contribution to the
anomaly is given by (\ref{a6LL}),
and $\Delta M_{8LL \alpha } $ listed in Table \ref{table6}
represents the sum of $n_F$ diagrams.

\renewcommand{\arraystretch}{0.75}
\begin{table}
\caption{Contributions of diagrams of Fig. \ref{vertex7}
and Fig. \ref{vertex8}
excluding LLJ, LLK, and LLL which were evaluated  separately in Ref. \cite{kn1}.
$n_F$ is the number of Feynman diagrams represented by the integral.
Some integrals are split
into two parts: $d$-$part$ is evaluated in real*8 and $q$-$part$
is evaluated in real*16. $a$-$part$ refers to the adjustable precision
method developed by \cite{sinkovits}.
The superscript *  indicates that indicated contributions 
were obtained by extrapolation from calculations
in which the edges of integration domain were
chopped off by 1.d-10.
See Appendix B of Ref. \cite{kn2} for details. 
Numerical work was carried out on SP3, velocity cluster, SP2, 
Condor cluster, and $\alpha$ workstations over several years.
The table lists only the latest of results obtained by various means.
\\
\label{table6}
}
\begin{tabular}{lcrrr}
\hline
\hline
~Integral~~& ~~~$n_F$~~~  &~Value (Error)~~~~~&~Sampling~per~& ~~~No. of~~~ \\ [.1cm]  
& &~~including $n_F$~&~iteration~~~~~& ~~iterations~  \\ [.1cm]   \hline
$\Delta M_{6LL,P2}$&6&$~0.598~948~(137)$&~~$4 \times 10^{8}$ &100 \\
$\Delta M_{8LLA}$ &10&$~~~~~~~0.171~527~(292)$&~&~~  \\
~~~~{\it d-part}&&$~~0.148~729~(268)$&$2 \times 10^{8}$ &100   \\
~~~~{\it q-part}&&$~~0.022~798~(116)$&$2 \times 10^5$&380   \\
$\Delta M_{8LLB}$ &20&$~~~~~-1.296~116~(481)$&$2 \times 10^8$ &99   \\
$\Delta M_{8LLC}$ &20&$~~~~~~~2.840~029~(513)$&~&~~  \\
~~~~{\it d-part}&&$~~2.705~399~(345)$&$1 \times 10^{9}$ &60   \\
~~~~{\it q-part1}&&$~~0.132~081~(375)$&$4 \times 10^6$&63   \\
~~~~{\it q-part2}&&$~~0.002~549~(~53)$&$1 \times 10^6$&100   \\
$\Delta M_{8LLD}$ &10&$~~~~~-0.058~890~(325)$&~$2 \times 10^{8}$ &100 \\
$\Delta M_{8LLE}$ &6&$~~~~~-0.265~962~(201)$&$1 \times 10^{8}$ &80 \\
$\Delta M_{8LLF}$ &12&$~~~~~~~-0.764~679~(413)$&~&~~  \\
~~~~{\it d-part}&&$~~-0.618~362~(346)$&$2 \times 10^{8}$ &80   \\
~~~~{\it q-part1}&&$~~-0.142~801~(221)$&$2 \times 10^7$&120   \\
~~~~{\it q-part2}&&$~~-0.003~516~(~36)$&$1 \times 10^6$&400   \\
$\Delta M_{8LLG}$ &12&$~~-0.550~968~(670)$&$4 \times 10^{8}$ &80   \\
$\Delta M_{8LLH}$ &6&$~~~~~~~0.190~770~(682)$&~&~~  \\
~~~~{\it d-part}&&$~~0.069~275~(379)$&$2 \times 10^{8}$ &100   \\
~~~~{\it q-part}&&$~~0.121~495~(566)$&$2 \times 10^7$&100   \\
$\Delta M_{8LLI}$ &12&$~~~~~~0.808~638~(482)$&~$4 \times 10^{8}$ &80   \\
\\
\hline
\hline
\end{tabular}
\end{table}
\renewcommand{\arraystretch}{1}

Numerical integration of all terms 
contributing to $a_{IV}^{(8)}$ has been carried out using
VEGAS \cite{lepage}.  
The latest results 
for Groups IV(b) and IV(c) are listed in Table \ref{table6}.  
The result for Group IV(d) had been handled separately \cite{kn1}.
In general, the major difficulty
in dealing with the diagrams of Groups IV(b) and IV(c)
 arises from the enormous size of
integrands (up to 5000 terms and 240 kilobytes of FORTRAN source code per integral) and the large number of integration
variables (up to 10).

Unlike $A_2^{(8)} (m_\mu/m_e)$ for muon $g-2$, which has
a singular surface just outside of the integration domain (unit cube),
the singularity of the mass-independent $A_1^{(8)}$ is far away from
the domain of integration.
This makes evaluation of electron $g-2$ less sensitive to the
{\it digit-deficiency} problem.
Nevertheless, our point-wise renormalization scheme is susceptible
to the $d$-$d$ problem. 

In most cases the first step is to make the integrand smoother
by {\it stretching} (see Appendix B of Ref. \cite{kn2}),
which is repeated several times until the integrand behaves
more gently.

Although {\it chopping} (see Appendix B of Ref. \cite{kn2}) was
handy to obtain a rough estimate
quickly, we had to abandon it in the end
because extrapolation to $\delta = 0$ turned out to be too unreliable
in order to reach the desired precision.

Most integrals were then evaluated by {\it splitting} them into
two parts, one evaluated in real*8 and the other in real*16.
In some cases, however, even the part evaluated in real*16
suffered from severe $d$-$d$ problem, preventing us from
collecting large enough samplings for high statistics.
Analyzing this problem closely, we found that it is possible
to evaluate these integrals by the following procedure:
First try several iterations with a positive 
rescale parameter $\beta$ (typically $\beta = 0.5$)
until VEGAS begins to show sign of blowing up
due to the $d$-$d$ problem.
Then {\it  freeze} $\beta$ to 0 
(see Appendix B of Ref. \cite{kn2}).
This may solve the problem in most cases.  If not,
try several iterations and see how rapidly the calculation runs into
the $d$-$d$ problem.
It turns out that it takes place very early 
if we chose too many sampling points $N_S$ per iteration.
This is because choosing large $N_S$ increases the chance of
hitting random numbers too close to the singularity within one iteration.
As a consequence the $d$-$d$ problem is likely to dominate
each iteration and makes it very difficult to collect
large enough number of good samplings.
A better strategy would be to reduce the size of $N_S$
to a moderate value and, instead, increase the number of iterations
$N_I$ substantially.
This is acceptable since, for $\beta = 0$
which means that the distribution function $\rho$ is
no longer changed from iteration to iteration, the final error
generated by VEGAS depends only on the product $N_S N_I$.

This strategy has been applied in particular to
the diagrams LLA and LLC, as is seen from Table \ref{table6}.
Entries in Table \ref{table6} are only the best of results obtained
by various methods discussed above.
They are consistent with each other despite their
diverse approaches.

One obtains from Table \ref{table6} 
the {\it Version A}  contributions
of subgroups IV(b) and IV(c):
\begin{eqnarray}   
a_{IV(b)}^{(8)} ~&=&~ 0.821~79~(~83) ,  
  \nonumber  \\
a_{IV(c)}^{(8)} ~&=&-1.138~71~(117) . 
\label{a8IV_A_num}
\end{eqnarray}

\subsection{Version B}
\label{subsec:vb}

In {\it Version B} the magnetic moment projection 
is evaluated for each vertex diagram on
the LHS of (\ref{wtid2}).
It is convenient to denote these diagrams 
in terms of self-energy-like diagrams
of Fig. \ref{vertex8}, by attaching suffix $i$ to indicate
the lepton line in which an external magnetic field vertex
is inserted.
For instance, we obtain vertex diagrams LLA1, LLA2, ..., LLA5 from
the diagram LLA.

We will not discuss subgroup IV(a) here since  its {\it Version A}
has already been fully tested.
For subgroup IV(b) we find 
\begin{equation}
a_{IV(b)}^{(8)} ~=~ \sum_{ \alpha = A }^D 
\sum_{i=1}^5
\Delta M_{8LL \alpha i} ~-~
4 \Delta B_2 M_{6LL} ,       
\label{a8IVbB}
\end{equation}
instead of (\ref{a8IVb}).
Note that the last term of (\ref{a8IVbB}) is different
from that of (\ref{a8IVb}).
This is not an error.
It arises from difference in the definition
of $\Delta M$ terms.
Similarly, for subgroup IV(c) we obtain
\begin{equation}
a_{IV(c)}^{(8)} =~\sum_{ \alpha = E }^I \sum_{i=1}^3
 \Delta M_{8LL \alpha i} ~-~
2 \Delta B_2 M_{6LL} .          \label{a8IVcB}
\end{equation}

Numerical evaluation of $\Delta M_{8LL \alpha i}$ are carried out with
$10^9$ sampling points per iteration.
The results  are listed in Table \ref{tablexx}.
Precision of these calculations is somewhat higher than that of
{\it Version A}.
See the last column of Table \ref{tablexx} for comparison
of two Versions. 
Although the values of $\Delta M_{8LLA}$ and $\Delta M_{8LLC}$
in two versions differ by several s. d., 
this is due to the influence of $d$-$d$ problem.
The general agreement
between {\it Version A} and {\it Version B} leaves no doubt that
their codes are free from any error.

One obtains from Table \ref{tablexx} the
values of $a_{IV(b)}^{(8)}$ and $a_{IV(c)}^{(8)}$
\begin{eqnarray}   
a_{IV(b)}^{(8)} ~&=&~ 0.822~57~(30) ,  
  \nonumber  \\
a_{IV(c)}^{(8)} ~&=&-1.138~93~(37) , 
\label{a8IV_numB}
\end{eqnarray}   
which are consistent with those given in (\ref{a8IV_A_num}).

\begin{table}
\caption{Contribution of Group IV(b)
and Group IV(c) diagrams of Fig. \ref{vertex8}
evaluated in {\it Version B}.
Double precision is used for all calculations
which are carried out on RSCC at RIKEN.   
Finite renormalization  terms $\Delta B_2 M_{6LLi}, i=1, 2, 3$, are 
needed for LLA and LLC, respectively, in order to compare  
them with the calculations in  {\it Version A}.   $M_{6LL(2+3)}
\equiv M_{6LL2} + M_{6LL3}$ is obtained
subtracting $M_{6LL1}$ from the known value of  
$M_{6LL}(\equiv M_{6LL1}+ M_{6LL2}+ M_{6LL3})$ given in (\ref{a6LL}).
\\
\label{tablexx}
}
\begin{tabular}{lcrrrr}
\hline
\hline
~Integral~~& ~~~$n_F$~~~  &~Value (Error)~~~~~&~Sampling~per~& ~~~No. of~~~
&~~~Difference ~~~ \\ [.1cm]  
& &~~including $n_F$~&~iteration~~~~~& ~~iterations~
&~~~ {\it Ver.A} - {\it Ver.B}   \\ [.1cm]   \hline
$\Delta M_{8LLA}$ &10&     0.171~789~(131)     &&&-0.000~262~(320)  \\
~~$\sum_{i=1}^5 {\eta}_A \Delta M_{8LLAi}$ 
& &   -0.096~030~(128) & $1 \times 10^{9}$  & 300~~~ &  \\
~~~~$-$$\Delta B_2 M_{6LL2}$ & &0.267~819~(~27)  & $1 \times 10^{9}$  & 100 ~~~&  \\ 
$\Delta M_{8LLB}$ &20& $-$1.296~748~(161) & $1 \times 10^{9}$  & 350~~~ &$+$0.000~631~(507)  \\
$\Delta M_{8LLC}$ &20&    2.840~956~(196)  &&    &$-$0.000~927~(549)    \\  
~~$\sum_{i=1}^5 {\eta}_C \Delta M_{8LLCi}$ 
 & &    3.387~028~(194)  & $1 \times 10^{9}$  & 516 ~~~& \\
~~~~$-$$\Delta B_2 M_{6LL(1+3)}$ & &$-$0.546~072~(~27)      &$1 \times 10^{9}$ &  &      \\ 
$\Delta M_{8LLD}$ &10&    $-$0.058~661~(~72)     & $1 \times 10^{9}$  & 420~~~ &$-$0.000~229~(332) \\
$\Delta M_{8LLE}$ &6&   $-$0.265~908~(~84)    &$1 \times 10^{9}$ &420~~~  &$-$0.000~054~(218)    \\
$\Delta M_{8LLF}$ &12&    $-$0.764~424~(213)   &$1 \times 10^{9}$ &520~~~  &$-$0.000~255~(465)   \\
$\Delta M_{8LLG}$ &12&    $-$0.551~233~(148)   &$1 \times 10^{9}$ &400~~~  &$$0.000~265~(686)   \\
$\Delta M_{8LLH}$ &6&   0.190~023~(111)  &$1 \times 10^{9}$ &400~~~  &$+$0.000~747~(691)    \\
$\Delta M_{8LLI}$ &12&   0.809~122~(217)  &$1 \times 10^{9}$ &520~~~  &$-$0.000~484~(529)   \\
\hline 
\hline
\end{tabular}
\end{table}

\subsection{Total contribution of Group IV}
\label{subsec:grp4contrib}

The contribution of subgroup IV(a) is listed only in 
(\ref{a8IV_A_num}) since it was not evaluated in {\it Version B}.
Previously only subgroup IV(d) was evaluated in both versions
with comparable statistical weights \cite{kn1}.
Now that subgroups IV(b) and IV(c) have been evaluated
in two independent ways, we may combine their results
statistically.
This leads to our best estimate for the
gauge-invariant subgroups of group IV:
\begin{eqnarray}   
a_{IV(a)}^{(8)} ~&=&~~0.598~95~(14) ,   
  \nonumber  \\
a_{IV(b)}^{(8)} ~&=&~~0.822~49~(28) ,  
  \nonumber  \\
a_{IV(c)}^{(8)} ~&=&-1.138~91~(35) , 
  \nonumber  \\
a_{IV(d)}^{(8)} ~&=&-0.990~72~(10) . 
\label{a8IV_num}
\end{eqnarray}   
Summing up these terms one find that the contribution from
all 180 diagrams of group IV is given by
\begin{equation}
a_{IV}^{(8)} ~=-0.708~19~(48).
\label{a8IV}
\end{equation}
%

\section{Group V Diagrams}
\label{sec:group5}

\begin{figure}[ht]
\vspace*{1cm}
\resizebox{12.0cm}{!}{\includegraphics{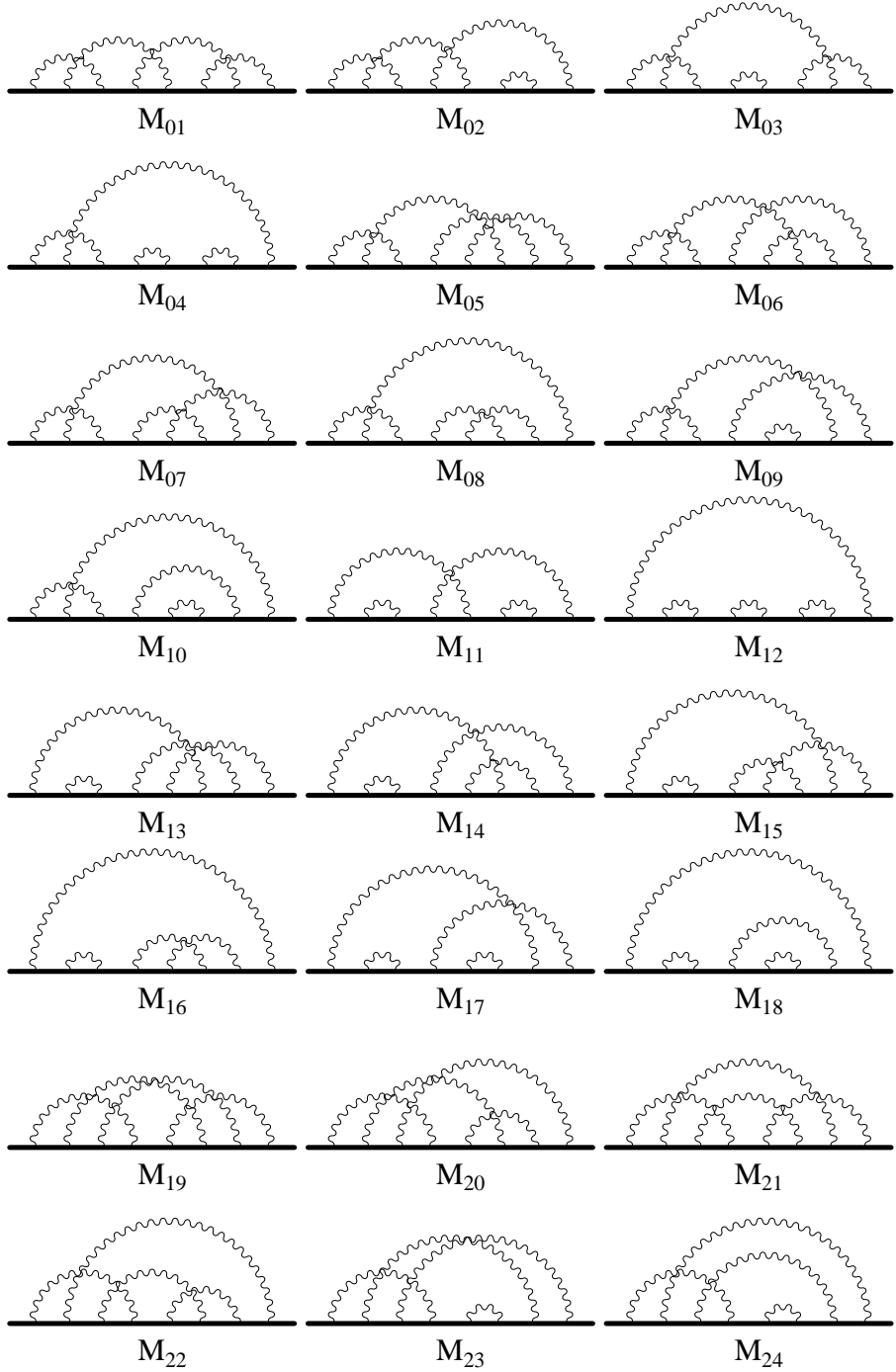}}
\vspace{.6cm}
\caption{\label{vertex9} Part of the electron self-energy-like diagrams 
$M_{01}$ - $M_{24}$ representing 
(external-vertex-summed) integrals of subgroup V.}  
\end{figure}

\begin{figure}[ht]
\vspace*{1cm}
\resizebox{12.0cm}{!}{\includegraphics{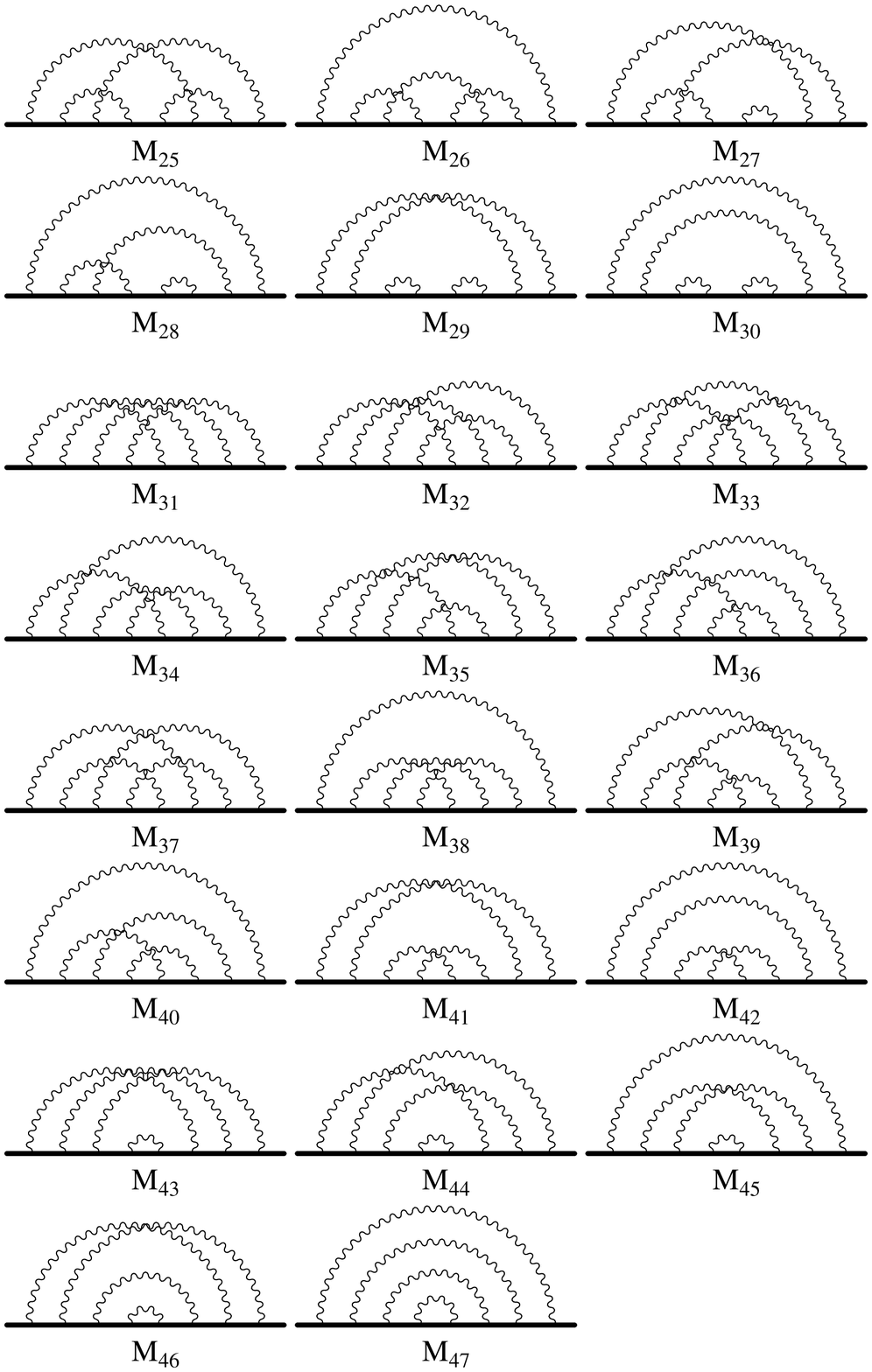}}
\vspace{.6cm}
\caption{\label{vertex10} Part of the electron self-energy-like diagrams 
$M_{25}$ - $M_{47}$ representing 
(external-vertex-summed) integrals of subgroup V.}  
\end{figure}

There are 518 vertex diagrams without closed electron loops
that contribute to Group V.
We have treated them only in {\it Version A} thus far,
in which these vertex diagrams are represented by 74
self-energy-like diagrams.  Since some of them are time-reversal
of others, the number of independent integrals is reduced to 47,
which are shown in Figs. \ref{vertex9} and \ref{vertex10}.

Carrying out momentum space integration analytically
by SCOONSCHIP and, more recently, by FORM, we obtain
the magnetic moment projection of the Ward-Takahashi-summed
amplitude as an integral over Feynman-parametric space of the form
\begin{eqnarray}   
M_{8\alpha} &=& \frac{3!}{256} \int (dz) \left( \frac{E_0 + C_0}{3U^2V^3}+
\frac{E_1 + C_1}{3U^3V^2}+ \frac{E_2 + C_2}{3U^4V} \right.   \nonumber  \\
&+& \left. \frac{N_0 + Z_0}{U^2V^4}+ \frac{N_1 + Z_1}{U^3V^3}+ \frac{N_2 + Z_2}{U^4V^2}+ \frac{N_3 + Z_3}{U^5V} \right),
\label{M8alpha}
\end{eqnarray}   
where $\alpha = 1, \ldots, 47.$
$E_0$ through $Z_3$ are polynomials of scalar currents
$B_{ij}$ and $A_i$.
See \cite{kn2,qedbook} for notation and definition.

Standard subtractive renormalization relates the $M_{8\alpha}$
to the renormalized contribution by
\begin{equation}
a_V^{(8)} ~= \sum_{\alpha=01}^{47} a_{8\alpha},
\label{renormalizeda8}
\end{equation}
where
\begin{equation}
 a_{8\alpha} = M_{8\alpha} - {\rm renormalization~terms.}
\end{equation}
$M_{8\alpha}$ is the sum of $n_F$ vertex diagrams where $n_F$
is given in the second column of Tables \ref{table7a}, \ref{table7b},
\ref{table7c}, and \ref{table7d}.
For simplicity we denote these amplitudes as $M_{\alpha}$
instead of $M_{8\alpha}$ in the following.
To avoid possible confusion with $M_2,~M_4, ...,$ defined elsewhere,
however, let us denote $M_1,~ M_2,\ldots,~ M_9$ as $M_{01},~ M_{02},\ldots,~ M_{09}$.

The two-step renormalization of eighth-order terms is in principle
no more difficult than that of lower-order terms,
except that it requires many more subtraction terms.
UV and IR subtraction terms 
of two-step renormalization scheme in {\it Version A} were
originally derived manually.  They were listed
in Appendix D of Ref. \cite{qedbook}.
In view of the fact that the two-step renormalization scheme plays
the central role in the success of our numerical integration
approach, we have re-examined its validity by a FORM program.
The FORM program is then translated into
LATEX format\cite{aoyama}.  The result is listed in Appendix \ref{renormalization}
together with the corresponding standard renormalization.
We corrected some typographical errors in Ref. \cite{qedbook}.

Collecting contributions of all terms of {\it Version A}
we obtain
\begin{align}   
a_V^{(8)} &= \Delta M^{(8)} -5 \Delta M^{(6)} \Delta B_2  \nonumber   \\
&- \Delta M^{(4)} [ 4 \Delta L^{(4)} + 3 \Delta B^{(4)} -9 (\Delta B_2)^2]
- \Delta M^{(4^*)} \Delta \delta m^{(4)}   \nonumber   \\
&-~M_2 [ 2 \Delta L^{(6)} + \Delta B^{(6)} - \Delta \delta m^{(4)}
(4 \Delta L_{2^*} + \Delta B_{2^*} - B_{2^*} [I])]  \nonumber   \\
&-(M_{2^*} - M_{2^*} [I]) [\Delta \delta m^{(6)} -\Delta \delta m^{(4)} (5 \Delta B_2 + \Delta \delta m_{2^*} )]   \nonumber   \\
&+ M_2 \Delta B_2 [10 \Delta L^{(4)} + 6 \Delta B^{(4)} - 5 (\Delta B_2 )^2 ],
\label{a8form}
\end{align}
where (see Ref. \cite{qedbook} for notation)
\begin{eqnarray}   
\Delta M^{(8)} &=& \sum_{i=01}^{47}  \Delta M_i,~~~
\Delta M^{(6)} = \sum_{\alpha=A}^{H} \Delta M_{6\alpha},~~~\nonumber  \\
\Delta M^{(4^*)} &=& 2 \Delta M_{4a(1^*)} + \Delta M_{4a(2^*)}
 + 2 \Delta M_{4b(1^*)} + \Delta M_{4b(2^*)},   \nonumber   \\
\Delta L^{(6)} &=& \sum_{\alpha = A}^{H}  \sum_{i=1}^5 \eta_\alpha \Delta L_{6\alpha i},
~~~\Delta B^{(6)} = \sum_{\alpha = A}^{H} \eta_\alpha \Delta B_{6\alpha},
\nonumber   \\
\Delta \delta m^{(6)} &=& \sum_{\alpha = A}^{H} \eta_\alpha  \Delta \delta m_{6\alpha},
\label{auxiliaryterms}
\end{eqnarray}
where $\Delta M_i$ represents the contribution of $n_F$ vertices.
$\Delta M_{6\alpha}$ is also the sum of several vertex diagrams.
Meanwhile, $\eta_{\alpha}$ in front of 
$\Delta L_{6\alpha i}$, $\Delta B_{6\alpha }$, and $\Delta \delta m_{6\alpha}$
is their multiplicity.

Numerical evaluation of Group V integrals 
has been carried out by VEGAS.
Tables \ref{table7a}, \ref{table7b}, \ref{table7c}, and \ref{table7d}
list only the latest of long sequence of work
carried out over more than 17 years.
Each integral was evaluated 5 to 10 times, 
starting with a relatively small number 
(typically $\sim 10^7$) of sampling points $N_S$,
increasing it gradually to more than $10^9$ as more and more computing power
became available.
Although these early data are not included in the final statistics,
they have played important roles in providing consistency check
among different runs.
In several cases two statistically independent runs of similar high precision
have been combined statistically, as is  indicated in the Tables.

It was seldom that VEGAS gave good result in the first try.
In most cases we had to search for alternate approach 
and/or transform the integrand by
various means, such as remapping, higher precision arithmetic, 
stretching, splitting, freezing, and their combinations.
(See Appendix B of Ref. \cite{kn2} for terminology.)

As is seen from Tables \ref{table7a} - \ref{table7d},
only ten integrals gave good
results with real*8 algebra.
The remaining 37 diagrams required real*16 in some ways.
In most cases {\it splitting} worked reasonably well.
Integrals $M_{04}$, $M_{16}$, $M_{18}$, 
$M_{26}$, $M_{28}$, $M_{30}$, $M_{42}$, and $M_{47}$
were evaluated in real*16 without relying on splitting.
$M_{04}$, $M_{16}$, and $M_{18}$ were evaluated 
also in the adjustable precision \cite{sinkovits}.

Evaluation of $M_{40}$ in real*16 ran into a 
rather puzzling problem:
For relatively small values of $N_S$ this integral 
seemed to behave reasonably well.
As we increased $N_S$, however, the error did not decrease
as expected.  For larger values of $N_S$, say $N_S \sim 10^9$,
for which each iteration now has small enough error bars,
several independent sequences of runs gave results which are different from
each other by much more than the error estimates generated by VEGAS.
We suspected that this was caused by a particularly severe
$d$-$d$ problem which made it difficult to choose a
reasonable mapping.
To pinpoint the cause of this annoying problem we decided to run $M_{40}$
using real*32 arithmetic, which would reduce
the $d$-$d$ problem to a negligible level.
Running $M_{40}$ with $N_S = 8 \times 10^9$
so that intrinsic resolution of individual iteration 
generated by VEGAS is $\sim 0.003$,
we discovered that this integral fluctuates between two peaks which
are far apart ($\sim 0.04$) from each other.
Once we discovered this problem, we went back to the beginning
and chose a different mapping which
does not have the double-peaking problem.
Afterwards $M_{40}$ behaved reasonably well.
The value of $M_{40}$ in Table \ref{table7d} was obtained
entirely in real*32.

In comparison with $M_{40}$ other diagrams were
less troublesome in finding a reasonable mapping.
However, diagram $M_{12}$ showed  sign of considerable
$d$-$d$ problem.  We therefore evaluated $M_{12}$
in real*32, too.
The result is listed in Table \ref{table7a}.

Combining the data from
Tables \ref{table7a}, \ref{table7b}, \ref{table7c}, and \ref{table7d}
as well as auxiliary integrals 
defined in (\ref{auxiliaryterms}) and
listed in Table \ref{table7aux},
one can evaluate $a_V^{(8)}$ from (\ref{a8form}). This leads to
\begin{equation}
a_{V}^{(8)} ~=-1.993~06~(343).
\label{a8V}
\end{equation}

The precision we have reported here is more than adequate for comparison with
experiment for the time being.  However, to prepare for the confrontation
with future measurements of $a_e$,
further improvement of precision of some diagrams 
would be highly desirable.

\renewcommand{\arraystretch}{0.75}
\begin{table}
\caption{Contributions of diagrams  $M_{01} - M_{13}$ of Fig. \ref{vertex9}. 
Some integrals are split
into two parts: $d$-$part$ is evaluated in real*8 and $q$-$part$
is evaluated in real*16. $a$-$part$ refers to the adjustable precision
method developed by \cite{sinkovits}.
The superscript ``a'' means that the adjustable precision method
is used in the whole domain.
The superscript ``q-d'' on $\Delta M_{12}$ indicates that
it is run with real*32.
Integrals without such qualification are evaluated in double precision.
Only the latest results are listed.
$\Delta M_{04}$ is a combination of two independent computations
with comparable statistical weights.
\\
\label{table7a}
}
\begin{tabular}{lcrcc}
\hline
\hline
~Integral~~& ~~~$n_F$~~~  &~Value (Error)~~~~~&~Sampling~per~& ~~~No. of~~~ \\ [.1cm]  
& &~~including $n_F$~&~iteration~~~~~& ~~iterations~  \\ [.1cm]   \hline
\\
$\Delta M_{01}$ &7&$~~~~~~-0.216~765~(522)$&~&~~  \\
~~~~{\it d-part}&&$~~-0.263~113~(397)$&$8 \times 10^{9}$ &139   \\
~~~~{\it q-part}&&$~~0.046~348~(338)$&$2 \times 10^8$&226   \\
$\Delta M_{02}$ &14&$~~~~~~-0.274~402~(573)$&~&~~  \\
~~~~{\it d-part}&&$~~-0.223~509~(347)$&$8 \times 10^{9}$ &464   \\
~~~~{\it q-part}&&$~~-0.050~893~(455)$&$2 \times 10^8$&320   \\
$\Delta M_{03}$ &7&$~~~~~~-0.681~633~(518)$&~&~~  \\
~~~~{\it d-part}&&$~~-0.601~353~(358)$&$6 \times 10^{9}$ &332   \\
~~~~{\it q-part}&&$~~-0.080~280~(374)$&$2 \times 10^8$&143   \\
$\Delta M_{04}^{(a,q)}$ &14&$~~~~~~4.560~949~(838)$&$~~~3.5 \times 10^9,4 \times 10^9$~&~233,209  \\
$\Delta M_{05}$ &14&$~~~~~~2.397~618~(266)$&$4 \times 10^9$~&~296  \\
$\Delta M_{06}$ &14&$~~~~~-1.447~248~(256)$&$4 \times 10^9$~&~286  \\
$\Delta M_{07}$ &14&$~~~~~0.065~257~(282)$&$4 \times 10^9$~&~279  \\
$\Delta M_{08}$ &14&$~~~~~-6.576~977~(472)$&~&~  \\
~~~~{\it d-part}&&$~~-6.724~530~(343)$&$1.6 \times 10^{10}$~&~391  \\
~~~~{\it q-part}&&$~~ 0.147~554~(323)$&$3 \times 10^8$&125   \\
$\Delta M_{09}$ &14&$~~~~~0.937~503~(447)$&$8 \times 10^{9}$~&~409  \\
$\Delta M_{10}$ &14&$~~~~17.859~588~(600)$&~&~ \\
~~~~{\it d-part}&&$~~16.823~753~(495)$&$2 \times 10^{10}$~&~403  \\
~~~~{\it q-part}&&$~~ 1.035~835~(338)$&$4 \times 10^8$&189   \\
$\Delta M_{11}$ &7&$~~~~~2.433~695~(336)$&~&~  \\
~~~~{\it d-part}&&$~~2.372~864~(168)$&$1 \times 10^{10}$~&~378  \\
~~~~{\it q-part}&&$~~ 0.060~831~(291)$&$1 \times 10^8$&100   \\
$\Delta M_{12}^{(q-d)}$ &7&$~~~~~-4.096~653~(871)$&$1 \times 10^9$~&~316  \\
\\
\hline
\hline
\end{tabular}
\end{table}
\renewcommand{\arraystretch}{1}

\renewcommand{\arraystretch}{0.75}
\begin{table}
\caption{Contributions of $M_{14} - M_{24}$ of Fig. \ref{vertex9}. 
See the caption of Table \ref{table7a} for notation.
Only latest results are listed.
$\Delta M_{18}$  and
$q-part_1$ of $\Delta M_{16}$ are statistical combinations of
two independent runs.
\\
\label{table7b}
}
\begin{tabular}{lcrcc}
\hline
\hline
~Integral~~& ~~~$n_F$~~~  &~Value (Error)~~~~~&~Sampling~per~& ~~~No. of~~~ \\ [.1cm]  
& &~~including $n_F$~&~iteration~~~~~& ~~iterations~  \\ [.1cm]   \hline
\\
$\Delta M_{13}$ &14&$~~~~-6.658~475~(277)$&$8 \times 10^{9}$~&~280  \\
$\Delta M_{14}$ &14&$~~~~~2.681~266~(404)$&~&~  \\
~~~~{\it d-part}&&$~~2.653~346~(289)$&$7 \times 10^{9}$~&~269  \\
~~~~{\it q-part}&&$~~ 0.027~920~(282)$&$2 \times 10^8$&114   \\
$\Delta M_{15}$ &14&$~~~~~1.100~419~(473)$&~&~  \\
~~~~{\it d-part}&&$~~1.199~683~(298)$&$5 \times 10^{9}$~&~237  \\
~~~~{\it q-part}&&$~~-0.099~264~(367)$&$2 \times 10^8$&236   \\
$\Delta M_{16}$ &14&$~~~~~2.972~899~(758)$&~&~  \\
~~~~{\it $q-part_1$}&&$~~2.821~674~(680)$&$~~~1 \times 10^{10}$,$2 \times 10^9$~&~150,799  \\
~~~~{\it $q-part_2$}&&$~~0.151~225~(334)$&$4 \times 10^8$&536   \\
$\Delta M_{17}$ &14&$~~~~~3.157~998~(415)$&~&~  \\
~~~~{\it d-part}&&$~~3.042~688~(253)$&$2.4 \times 10^{10}$~&~189  \\
~~~~{\it q-part}&&$~~0.115~310~(329)$&$1.2 \times 10^8$&326   \\
$\Delta M_{18}^{(a,q)}$ &14&$~~~~~12.705~790~(695)$&$4 \times 10^9,4 \times 10^9$~&~188,145  \\
$\Delta M_{19}$ &7&$~~~~-0.834~502~(266)$&~&~  \\
~~~~{\it d-part}&&$~~-0.753~570~(170)$&$4 \times 10^{8}$~&~221  \\
~~~~{\it q-part}&&$~~-0.080~932~(204)$&$2 \times 10^7$&99   \\
$\Delta M_{20}$ &14&$~~~~0.689~701~(320)$&~&~  \\
~~~~{\it d-part}&&$~~0.604~560~(299)$&$1 \times 10^{9}$~&~102  \\
~~~~{\it q-part}&&$~~0.085~141~(113)$&$2 \times 10^7$&105   \\
$\Delta M_{21}$ &7&$~~~~0.209~216~(281)$&~&~  \\
~~~~{\it d-part}&&$~~0.191~592~(245)$&$4 \times 10^{8}$~&~99  \\
~~~~{\it q-part}&&$~~0.017~624~(137)$&$2 \times 10^7$&100   \\
$\Delta M_{22}$ &14&$~~~-0.281~374~(454)$&~&~  \\
~~~~{\it d-part}&&$~~-0.226~185~(410)$&$1 \times 10^{9}$~&~110  \\
~~~~{\it q-part}&&$~~-0.055~189~(194)$&$2 \times 10^7$&100   \\
$\Delta M_{23}$ &14&$~~~-4.431~906~(473)$&~&~  \\
~~~~{\it d-part}&&$~~-4.139~207~(389)$&$3 \times 10^{9}$~&~289  \\
~~~~{\it q-part}&&$~~-0.292~699~(268)$&$4 \times 10^7$&80   \\
\\
\hline
\hline
\end{tabular}
\end{table}
\renewcommand{\arraystretch}{1}

\renewcommand{\arraystretch}{0.75}
\begin{table}
\caption{Contributions of diagrams  $M_{25} - M_{39}$ of Fig. \ref{vertex10}. 
See the caption of Table \ref{table7a} for notation.
Only the latest results are listed.
\\
\label{table7c}
}
\begin{tabular}{lcrcc}
\hline
\hline
~Integral~~& ~~~$n_F$~~~  &~Value (Error)~~~~~&~Sampling~per~& ~~~No. of~~~ \\ [.1cm]  
& &~~including $n_F$~&~iteration~~~~~& ~~iterations~  \\ [.1cm]   \hline
\\
$\Delta M_{24}$ &14&$~~~2.206~446~(404)$&~&~  \\
~~~~{\it d-part}&&$~~2.066~148~(263)$&$8 \times 10^{9}$~&~209  \\
~~~~{\it q-part}&&$~~0.140~276~(314)$&$2 \times 10^8$&193   \\
$\Delta M_{25}$ &7&$~~~0.080~642~(432)$&~&~  \\
~~~~{\it a-part}&&$~~0.047~651~(167)$&$1 \times 10^9$&100   \\
~~~~{\it a-part}&&$~~0.032~991~(398)$&$4 \times 10^{7}$~&~106  \\
$\Delta M_{26}^{(q)}$ &7&$~~~~~1.829~535~(545)$&$9 \times 10^8, 1 \times 10^9$~&~173,222  \\
$\Delta M_{27}$ &14&$~~~0.988~319~(383)$&~&~  \\
~~~~{\it d-part}&&$~~0.838~417~(256)$&$4 \times 10^9$&145   \\
~~~~{\it a-part}&&$~~0.149~902~(285)$&$1.2 \times 10^{8}$~&~311  \\
$\Delta M_{28}^{(q,q)}$ &14&$~~~~-4.950~479~(690)$&$2.6 \times 10^9,4 \times 10^9$~&~343,289  \\
$\Delta M_{29}$ &7&$~~~3.083~789~(245)$&~&~  \\
~~~~{\it d-part}&&$~~3.023~232~(130)$&$6 \times 10^9$&339   \\
~~~~{\it a-part}&&$~~0.060~557~(207)$&$4 \times 10^{7}$~&~234  \\
$\Delta M_{30}^{(q)}$ &7&$~~~~-3.721~347~(630)$&$2 \times 10^9, 2 \times 10^9$~&~159,206  \\
$\Delta M_{31}$ &7&$~~~~3.012~077~(179)$&$2 \times 10^{9}$~&~290  \\
$\Delta M_{32}$ &14&$~~~-2.377~888~(323)$&~&~  \\
~~~~{\it d-part}&&$~~-2.343~943~(284)$&$4 \times 10^8$&100   \\
~~~~{\it a-part}&&$~~-0.033~945~(153)$&$2 \times 10^{7}$~&~100  \\
$\Delta M_{33}$ &7&$~~~-1.194~130~(250)$&~&~  \\
~~~~{\it d-part}&&$~~-1.247~252~(243)$&$4 \times 10^8$&98   \\
~~~~{\it q-part}&&$~~ 0.053~122~(57)$&$2 \times 10^{7}$~&~52  \\
$\Delta M_{34}$ &14&$~~~1.341~091~(299)$&~&~  \\
~~~~{\it d-part}&&$~~1.237~527~(149)$&$5 \times 10^9$&411   \\
~~~~{\it q-part}&&$~~ 0.103~564~(259)$&$4 \times 10^{7}$~&~80  \\
$\Delta M_{35}$ &14&$~~~-0.539~732~(384)$&~&~  \\
~~~~{\it d-part}&&$~~-0.560~805~(376)$&$1 \times 10^9$&100   \\
~~~~{\it q-part}&&$~~ 0.021~073~(75)$&$2 \times 10^{7}$~&~100  \\
$\Delta M_{36}$ &14&$~~~-0.302~434~(240)$&~&~  \\
~~~~{\it d-part}&&$~~-0.252~129~(113)$&$2 \times 10^{10}$&275   \\
~~~~{\it q-part}&&$~~-0.050~305~(211)$&$4 \times 10^{7}$~&~101  \\
\\
\hline
\hline
\end{tabular}
\end{table}
\renewcommand{\arraystretch}{1}

\renewcommand{\arraystretch}{0.75}
\begin{table}
\caption{Contributions of diagrams  $M_{40} - M_{47}$ of Fig. \ref{vertex10}. 
See the caption of Table \ref{table7a} for notation.
The superscript ``q-d'' on $\Delta M_{40}$ indicates that it is run with real*32.
Only the latest results are listed.
\\
\label{table7d}
}
\begin{tabular}{lcrcc}
\hline
\hline
~Integral~~& ~~~$n_F$~~~  &~Value (Error)~~~~~&~Sampling~per~& ~~~No. of~~~ \\ [.1cm]  
& &~~including $n_F$~&~iteration~~~~~& ~~iterations~  \\ [.1cm]   \hline
\\
$\Delta M_{37}$ &7&$~~~~0.487~701~(305)$&$4 \times 10^{8}$~&~90  \\
$\Delta M_{38}$ &7&$~~~-3.094~936~(311)$&$8 \times 10^{9}$~&~147  \\
$\Delta M_{39}$ &14&$~~~-0.660~746~(290)$&$4 \times 10^{9}$~&~115  \\
$\Delta M_{40}^{(q-d)}$ &14&$~~~5.252~164~(977)$&$1 \times 10^{9}$~&~235  \\
$\Delta M_{41}$ &7&$~~~-1.416~949~(376)$&~&~  \\
~~~~{\it d-part}&&$~~-1.107~727~(256)$&$6 \times 10^{9}$&129   \\
~~~~{\it q-part}&&$~~-0.309~222~(275)$&$1.5 \times 10^{8}$~&~245  \\
$\Delta M_{42}^{(q)}$ &7&$~~~-4.859~600~(678)$&$2 \times 10^{9},2 \times 10^{9}$~&~168,226  \\
$\Delta M_{43}$ &7&$~~~-2.385~383~(221)$&$1 \times 10^{9}$~&~431  \\
$\Delta M_{44}$ &14&$~~~3.468~870~(249)$&~&~  \\
~~~~{\it d-part}&&$~~3.288~775~(78)$&$2 \times 10^{10}$&441   \\
~~~~{\it q-part}&&$~~0.180~091~(236)$&$3 \times 10^{8}$~&~232  \\
$\Delta M_{45}$ &7&$~~~0.711~911~(411)$&~&~  \\
~~~~{\it d-part}&&$~~1.427~913~(217)$&$4 \times 10^{9}$&351   \\
~~~~{\it q-part}&&$~~-0.716~002~(349)$&$3 \times 10^{8}$~&~179  \\
$\Delta M_{46}$ &7&$~~~-6.766~038~(242)$&~&~  \\
~~~~{\it d-part}&&$~~-6.121~804~(86)$&$3 \times 10^{10}$&216   \\
~~~~{\it q-part}&&$~~-0.644~234~(226)$&$6 \times 10^{7}$~&~130  \\
$\Delta M_{47}^{(q)}$ &7&$~~~8.352~501~(585)$&$2.4 \times 10^{9}, 2 \times 10^9$~&~205,179  \\
\\
\hline
\hline
\end{tabular}
\end{table}
\renewcommand{\arraystretch}{1}

\renewcommand{\arraystretch}{0.80}
\begin{table}
\caption{ Auxiliary integrals for Group V.
In this table we denote
$\Delta M_{4a^*} \equiv 2 \Delta M_{4a(1^*)} + \Delta M_{4a(2^*)}$.
Similarly for $\Delta M_{4b^*}$.
$A_{LBD} \equiv \sum_{i=1}^5 \Delta L_{6Ai} + \frac{1}{2} \Delta B_{6A}
+2 \Delta \delta m_{6A}$.  Similarly for $B_{LBD}$, etc.
\label{table7aux}
} 
\begin{tabular}{llll}
\hline
\hline
~~~Integral~~~  &~~~Value~(Error)~~~~& ~~~Integral~~~~   &~~~Value~(Error)~~~~~~  \\ [.1cm]   \hline
$\Delta B_{2^*}$  &   1.5         & $\Delta L_{2^*}$      &  -0.75 ~\\[.1cm]
$B_{2^*}[I]$      &     -0.5      &$\Delta \delta m_{2^*}$&  -0.75 ~\\[.1cm]
$\Delta M_{4a^*}$ & 3.636~02~(11)   & $\Delta M_{4b^*}$     &10.147~24~(14)\\[.1cm]
$\Delta L_{4}$ &  0.465~024~(17)  & $\Delta B_{4}$     &  -0.437~094~(21)              \\[.1cm]
$A_{LBD}$         & 1.370~25~(52)   & $B_{LBD}$             & 3.571~37~(26)\\[.1cm]
$C_{LBD}$         &-6.778~82~(39)  & $2 \times D_{LBD}$            &-8.783~15~(40)\\[.1cm]
$E_{LBD}$         &-1.592~21~(26)   & $F_{LBD}$             & 2.652~93~(36)\\[.1cm]
$2\times G_{LBD}$         & 3.890~15~(33)  & $H_{LBD}$             &-0.232~26~(13)\\[.1cm]
$\Delta M^{(6)}$  & 4.471~013~(45)&  & \\[.1cm]
\hline
\hline
\end{tabular}
\end{table}
\renewcommand{\arraystretch}{1}

\begin{acknowledgments}

The part of material presented by T. K. is based on work supported by the
National Science Foundation under Grant No. PHY-0098631.
M. N. is supported by the Ministry of Education, Science , Sports,
and Culture, Grant-in-Aid for Scientific Research (c),
15540303.  T. K. thanks the Eminent Scientist Invitation Program of
RIKEN, Japan, for the hospitality extended to him where a  part of this
work was carried out.
Thanks are due to J. Zollweg and R. Sinkovits for assistance in various
phases of computation.

The numerical work has been carried out over more than 17 years on a 
number of computers.
A part of work was conducted at the Cornell Theory Center using the resources 
of the Cornell University, New York State, the National Center for Research Resources of the National Institute of Health, the National Science Foundation,
the Defense Department Modernization Program, the United States Department
of Agriculture, and the corporate partners.
Another part of numerical work was supported by NSF Cooperative agreement
ACI-9619020 through computing resources provided by the National Partnership for
Advanced Computational Infrastructure at the San Diego Super Computer Center,
which enabled us to have an access to the Blue Horizon at the
San Diego Supercomputer Center, the IBM SP at the University of Michigan, and
the Condor Flock at the University of Wisconsin.
M. N. thanks various computational resources provided by
the Computer Center of Nara Women's University, RIKEN Super Computer System, 
and Matsuo Foundation.
\end{acknowledgments}

\appendix

\section{Elimination of Algebraic Error of Group V}
\label{sec:elimination} 

\vspace{2mm}
The number of independent
integrals of Group V is 47 in {\it Version A}.
They are denoted as $M_{01}$, $M_{02}$,$\cdots$, $M_{47}$ \cite{qedbook}.
All integrals of Group V diagrams can be constructed from 
just one ``template"
by permutation of tensor indices of photons.
This implies that any error in one of the integrals
will propagate to other integrals.
On the other hand, verification of one integral will
verify all other integrals of Group V.
All integrands are created by the following procedures:

\vspace{2mm}
{\it a.}   
Momentum-space integration is carried out analytically 
by an algebraic manipulation program FORM \cite{vermas}.
The result is expressed in terms of symbols $A_i$ and $B_{ij}$ (called
{\it scalar currents} \cite{qedbook} 
because they obey an analogue of Kirchhoff's laws
for electric current in which Feynman parameters 
play the role of resistance \cite{landau}).

\vspace{2mm}
{\it b.}
Diagrams of Group V belong to five {\it topologically distinct types}
\cite{qedbook}.  Correspondingly,
scalar currents $A_i$ and $B_{ij}$
can be written explicitly as rational functions of Feynman
parameters using one of the five ``templates" for $B_{ij}$.
Scalar currents must satisfy eight junction laws
and four loop laws for each diagram.
This imposes a very strict constraint on scalar currents
and provides powerful defense against trivial programming errors.

\begin{figure}[ht]
\vspace*{1cm}
\resizebox{12.0cm}{!}{\includegraphics{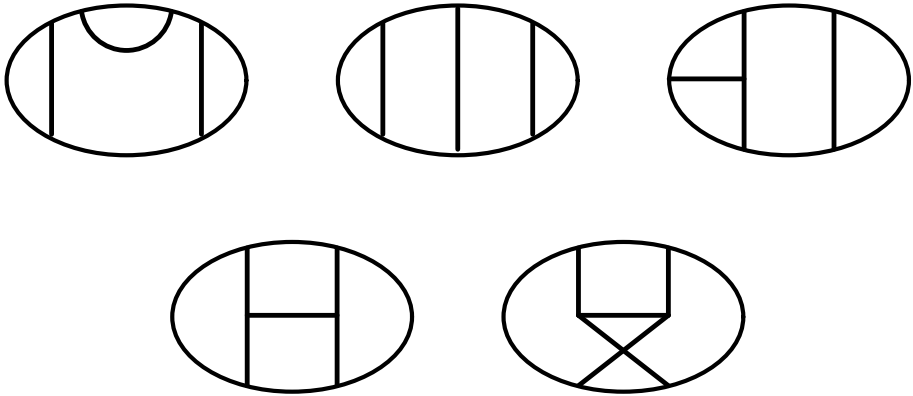}}
\vspace{.6cm}
\caption{\label{chaindgrm} Chain diagrams obtained from
eighth-order diagrams by omitting external lines and
disregarding the distinction between electron and photon
lines. The topological structure of chain diagram
is crucial for determining the structure of ``scalar currents".} 
\end{figure}

\vspace{2mm}
{\it c.}
Feynman-parametric integrals thus obtained
are ultraviolet (UV) and/or infrared (IR) divergent in general.
By a power counting rule
designed for each divergence
we can identify term by term which one is divergent \cite{qedbook}.
Collection of these terms 
(in which only leading terms are kept within each scalar current)
forms a term required for renormalization.
Its structure can be examined analytically
by comparison with a known lower-order expression
which is written as a Feynman-parametric integral over a function of
{\it its own scalar currents}.
These two integrals have different forms in general
and their identity can be established
only after some non-trivial analytic transformation.
Comparison of these integrals
provides an extremely powerful check of the subtraction term
as well as major parts of mother integrand. 

\vspace{2mm}
{\it d.}
The only remaining task is to show that scalar currents
themselves have no error.  This is not difficult 
since the number of scalar currents is not large.
This is the primary reason for expressing the integrand 
in terms of scalar currents
instead of Feynman parameters themselves.

Initially, scalar currents 
were built by hands by two people working independently \cite{kl1}.
Comparison of resulting integrands enabled us to weed out human errors
quickly.
Recently we wrote programs in FORM (and MAPLE) which generate
$B_{ij}$ for each chain diagrams of Fig. \ref{chaindgrm}
and show further that they satisfy all loop laws and junction laws.

Around 1992 we re-derived the entire integrands by a computer program
in a form different from the original version.
In this manner we obtained two sets of integrands 
with different algebraic structures.
Their equivalence was checked by numerical {\it spot check}.
This process was repeated for each renormalization term.
Of course this comparison does not establish
the validity of all terms.
Since each diagram has several subtraction terms 
(as many as 20 in some cases), however,
checks repeated for all subtraction terms 
cover a much larger fraction of the integrand 
than that of a single subtraction term and enhance strongly the
confidence in the validity of the entire code.  

The procedure outlined above is especially potent for the diagrams that 
contain a second-order electron self-energy part.
When the integrand is expressed as a function of scalar currents,
individual terms of the integrand, except those that contain
the Feynman parameter  $z_{se}$ for the electron of the self-energy part
explicitly, are UV divergent.
This means that the corresponding self-energy subtraction term,
{\it when expressed in terms of its own scalar currents,}
has a form identical with that of the {\it mother integrand}
{\it minus} terms proportional to $z_{se}$. 
Their difference comes solely from different structure of
scalar currents for the mother and daughter integrands.

This argument applies equally well to diagrams which contain two
second-order electron self-energy parts.
In this case {\it all} terms of the mother integrand can be verified
by applying the above argument to each self-energy part,
verifying the correctness of the template completely and thus
all integrands obtained using the same template,
namely, the entire Group V diagrams.
In this manner we were able to confirm the validity
of all terms of Group V integrals of {\it Version A}.
Our very tight scheme {\it a., b., c.}, and {\it d}
leaves no room for further algebraic error.

\section{Two-Step Renormalization of Group V Integrals}
\label{renormalization} 

Following the discussion
in Refs.\cite{ck1,ck2,qedbook}, we adopted the two-step
renormalization scheme in the numerical evaluation of $g-2$.
The {\bf K} operation defined in Refs.\cite{ck1,qedbook} picks up
only the overall UV divergent part of the on-shell renormalization
constant.  It is  used  as the UV subtraction term in numerical 
calculation.  The rest of the renormalization constants are  
combined in the end with those of other diagrams, 
which gives a finite contribution,
referred to as the residual renormalization. 
The standard on-shell renormalization is
the sum of {\bf K} operation and residual renormalization. 
Only the renormalization scheme of Version A is given here since
Group V has not been treated in Version B.

\subsection{Renormalization Scheme for  Group V} 
\label{renormalization8V} 

The eighth-order Group V diagrams have very complicated
divergence structure.
Only the {\bf K} operation part was listed in Ref. \cite{qedbook}
because of page restriction.
For convenience of access and completeness we list here
{\bf K}-renormalization and standard renormalization together. 
Few typographic errors in the previous version are corrected.

We first list the  standard on-shell renormalized amplitude $a_{i}$
expressed in terms of the unrenormalized integral $M_{i}$ 
and various renormalizing terms and then present
the intermediate renormalization scheme which
describes the relation between $M_{i}$
and $\Delta M_{i}$, the finite part of $M_{i}$ defined
by {\bf K} operation.
Substitution of $M_{i}$ from the second part into the equation for $a_{i}$ 
expresses it in terms of $\Delta M_{i}$ and pieces of renormalization
constants as well as (unrenormalized) amplitudes of lower orders.
Plugging them in (\ref{renormalizeda8})
and using relations among lower order terms given in Appendix \ref{renormalization6} we obtain  Eq.({\ref{a8form}).

$M_{na}$ is the unrenormalized amplitude of the $n$th-order diagram $a$,
where $n \leq 6$.
A quantity with prefactor $\Delta$ implies that it is finite.
The infrared subtraction terms are denoted as $I_{na}$. 

Quantities $L_{n}$, $B_{n}$, and $\delta m_{n}$ denote  vertex
renormalization constant, wave-function renormalization constant, 
and mass renormalization constant of the $n$th-order,
of the standard on-shell renormalization, respectively.
In our approach by numerical integration, they are expressed as
integrals over $3n$, $3n-1$, and $3n-1$ Feynman parameters,
which enable us to perform cancellation of divergence
at every point in the domain of integration.
In the parametric form the divergent terms of an integrand can be readily identified
by means of a power counting rule for each divergence.
The sum of these terms are denoted as ${\hat L}_{n}$,
${\hat B}_{n}$, and $\delta {\hat m}_{n}$, and the remainder as
${\tilde L}_{n}$, etc.
For instance, 
$  L_2 = \hat{L}_2 + \tilde{L}_2 $, 
where $ \hat{L}_2 $ is UV-divergent part picked up by the {\bf K} operation and $\tilde{L}_2$
is the residual UV-finite term. 

Within each subdiagram the electron lines are assumed to be consecutively
labeled as [1, 2, ..., $n_m$], where $n_m$ = $2n$ for $L_{n}$ and $2n-1$ 
for $B_{n}$ and $\delta m_{n}$, and the line 1 is at the tip of the arrow
attached to the electron line.

The additional suffix such as $c$ in $L_{4c}$ indicates that it comes from
the so-called {\it corner} diagram.
Here we stick to the notation introduced previously.
See Ref. \cite{qedbook}, p.235, Fig. 8.
The suffix $(3^{*})$ means the insertion of two-point vertex (such as mass-vertex) in the fermion line $3$.
The primed quantity $L_{4x(1^\prime )}$  is the derivative amplitude
obtained by applying $-z_1 \frac{\partial}{\partial z_1} $ 
operation on the integrand, where $z_1$ is
the Feynman parameter assigned to the fermion line $1$. Note
that  $L_{4x(1^\prime)}$ is equal to $L_{4x}$, but its UV-divergent
part $\hat{L}_{4x(1^\prime)}$ is not equal to $\hat{L}_{4x}$.
Since second order quantities such as $B_2$ and $\delta m_2$
have only one electron line,
it is not really necessary to distinguish different electron lines.
We therefore use somewhat sloppy notations $B_{2^{*}}$ and $B_{2^{\prime}}$ 
instead of $B_{2(1^{*})}$ and $B_{2(1^{\prime})}$.
Similarly, double insertions of two-point vertices are denoted as $B_{2^{**}}$,
$B_{2^{\prime \ast}}$, $B_{2^{\prime\prime}}$, etc.
Since $\delta m_2$ has no UV-finite part, $\delta {\hat m}_2$
is replaced by $\delta m_2$ everywhere.
For $L_2$, which contains electron lines 1 and 2, it is sometimes
necessary to distinguish lines in which insertion is made.
$L_{2^{\ast \ast \dagger}}$  implies that two two-point vertices are inserted into the fermion 
line 1 of $L_{2}$, while $L_{2^{\ast \dagger \ast}}$ 
means that one two-point vertex is inserted into
the line 1 and another  into the line 2. 
$M_{4a}$ contains three electron lines 1,2,3 and $M_{4a(1^{**})}$ means that two-point vertex 
insertion has been made twice in the electron line 1, and so on.

The validity of all equations has been confirmed algebraically
by FORM.  They are then translated into LATEX format listed below \cite{aoyama}.


\vspace*{6mm}
\noindent
{\bf On-Shell Renormalization for the Eighth-order Moments}

\begin{align*}
a_{01}&=
M_{01}-2 M_{6F} L_{2}-2 L_{6F1} M_{2}-2 L_{4c} M_{4a}+6 L_{4c} L_{2} M_{2} \\
& +3 M_{4a} (L_{2})^2-4 (L_{2})^3 M_{2}
\end{align*}
\begin{align*}
a_{02}&=
M_{02}-M_{6D} L_{2}-M_{6F} B_{2}-M_{6F(1^\ast)} \delta m_{2}-L_{6D5} M_{2} \\
& +L_{4c(3^\ast)} \delta m_{2} M_{2}-L_{4c} M_{4b}+L_{4c} \delta m_{2} M_{2^\ast}+2 L_{4c} B_{2} M_{2}-L_{4s} M_{4a} \\
& +2 L_{4s} L_{2} M_{2}+M_{4a(1^\ast)} \delta m_{2} L_{2}+M_{4b} (L_{2})^2+M_{4a} \delta m_{2} L_{2^\ast}+2 M_{4a} B_{2} L_{2} \\
& -2 \delta m_{2} L_{2} L_{2^\ast} M_{2}-\delta m_{2} (L_{2})^2 M_{2^\ast}-3 B_{2} (L_{2})^2 M_{2}
\end{align*}
\begin{align*}
a_{03}&=
M_{03}-2 M_{6D} L_{2}-M_{6F} B_{2}-M_{6F(3^\ast)} \delta m_{2}-2 L_{6D1} M_{2} \\
& +2 L_{4c(1^\ast)} \delta m_{2} M_{2}+2 L_{4c} B_{2} M_{2}+2 L_{4s} L_{2} M_{2}+2 M_{4a(1^\ast)} \delta m_{2} L_{2}+M_{4b} (L_{2})^2 \\
& +2 M_{4a} B_{2} L_{2}-2 \delta m_{2} L_{2} L_{2^\ast} M_{2}-\delta m_{2} (L_{2})^2 M_{2^\ast}-3 B_{2} (L_{2})^2 M_{2}
\end{align*}
\begin{align*}
a_{04}&=
M_{04}-M_{6A} L_{2}-2 M_{6D} B_{2}-M_{6D(1^\ast)} \delta m_{2}-M_{6D(3^\ast)} \delta m_{2} \\
& -L_{6A1} M_{2}+2 L_{4s(1^\ast)} \delta m_{2} M_{2}+2 L_{4s} B_{2} M_{2}+2 M_{4a(1^\ast)} \delta m_{2} B_{2}+2 M_{4b(1^\ast)} \delta m_{2} L_{2} \\
& +2 M_{4b} B_{2} L_{2}+M_{4a} (B_{2})^2+M_{4a(1^\ast1^\ast)} (\delta m_{2})^2-2 \delta m_{2} B_{2} L_{2} M_{2^\ast}-2 \delta m_{2} B_{2} L_{2^\ast} M_{2} \\
& -(\delta m_{2})^2 L_{2} M_{2^{\ast \ast}}-(\delta m_{2})^2 L_{2^{\ast \ast}\dagger} M_{2}-2 (B_{2})^2 L_{2} M_{2}
\end{align*}
\begin{align*}
a_{05}&=
M_{05}-M_{6H} L_{2}-L_{6H1} M_{2}-L_{6F2} M_{2}-L_{4x} M_{4a} \\
& +3 L_{4x} L_{2} M_{2}
\end{align*}
\begin{align*}
a_{06}&=
M_{06}-M_{6F} L_{2}-M_{6G} L_{2}-L_{6F3} M_{2}-L_{6G5} M_{2} \\
& +3 L_{4c} L_{2} M_{2}-L_{4l} M_{4a}+2 L_{4l} L_{2} M_{2}+2 M_{4a} (L_{2})^2-3 (L_{2})^3 M_{2}
\end{align*}
\begin{align*}
a_{07}&=
M_{07}-M_{6F} L_{2}-M_{6G} L_{2}-L_{6G1} M_{2}-L_{6D2} M_{2} \\
& -L_{4c} M_{4a}+4 L_{4c} L_{2} M_{2}+L_{4l} L_{2} M_{2}+2 M_{4a} (L_{2})^2-3 (L_{2})^3 M_{2}
\end{align*}
\begin{align*}
a_{08}&=
M_{08}-M_{6C} L_{2}-2 M_{6D} L_{2}-L_{6C1} M_{2}-\delta m_{4a} M_{4a(1^\ast)} \\
& +\delta m_{4a} L_{2} M_{2^\ast}+\delta m_{4a} L_{2^\ast} M_{2}-B_{4a} M_{4a}+2 B_{4a} L_{2} M_{2}+2 L_{4s} L_{2} M_{2} \\
& +2 M_{4a(1^\ast)} \delta m_{2} L_{2}+2 M_{4b} (L_{2})^2+2 M_{4a} B_{2} L_{2}-2 \delta m_{2} L_{2} L_{2^\ast} M_{2}-2 \delta m_{2} (L_{2})^2 M_{2^\ast} \\
& -4 B_{2} (L_{2})^2 M_{2}
\end{align*}
\begin{align*}
a_{09}&=
M_{09}-M_{6E} L_{2}-M_{6F} B_{2}-M_{6F(2^\ast)} \delta m_{2}-L_{6E1} M_{2} \\
& -L_{6D3} M_{2}+L_{4c(1^\ast)} \delta m_{2} M_{2}+L_{4c(2^\ast)} \delta m_{2} M_{2}+2 L_{4c} B_{2} M_{2}-L_{4s} M_{4a} \\
& +3 L_{4s} L_{2} M_{2}+M_{4a(2^\ast)} \delta m_{2} L_{2}+M_{4a} \delta m_{2} L_{2^\ast}+2 M_{4a} B_{2} L_{2}-3 \delta m_{2} L_{2} L_{2^\ast} M_{2} \\
& -3 B_{2} (L_{2})^2 M_{2}
\end{align*}
\begin{align*}
a_{10}&=
M_{10}-M_{6B} L_{2}-M_{6D} B_{2}-M_{6D(2^\ast)} \delta m_{2}-L_{6B1} M_{2} \\
& +L_{4s(2^\ast)} \delta m_{2} M_{2}-\delta m_{4b} M_{4a(1^\ast)}+\delta m_{4b} L_{2} M_{2^\ast}+\delta m_{4b} L_{2^\ast} M_{2}-B_{4b} M_{4a} \\
& +2 B_{4b} L_{2} M_{2}+L_{4s} B_{2} M_{2}+M_{4a(1^\ast)} \delta m_{2} \delta m_{2^\ast}+M_{4a(1^\ast)} \delta m_{2} B_{2}+M_{4b(2^\ast)} \delta m_{2} L_{2} \\
& +M_{4b} B_{2} L_{2}+M_{4a} \delta m_{2} B_{2^\ast}+M_{4a} (B_{2})^2-\delta m_{2} \delta m_{2^\ast} L_{2} M_{2^\ast}-\delta m_{2} \delta m_{2^\ast} L_{2^\ast} M_{2} \\
& -\delta m_{2} B_{2} L_{2} M_{2^\ast}-\delta m_{2} B_{2} L_{2^\ast} M_{2}-2 \delta m_{2} B_{2^\ast} L_{2} M_{2}-2 (B_{2})^2 L_{2} M_{2}
\end{align*}
\begin{align*}
a_{11}&=
M_{11}-2 M_{6D} B_{2}-2 M_{6D(5^\ast)} \delta m_{2}-2 L_{4s} M_{4b}+2 L_{4s} \delta m_{2} M_{2^\ast} \\
& +2 L_{4s} B_{2} M_{2}+2 M_{4a(1^\ast)} \delta m_{2} B_{2}+2 M_{4b} \delta m_{2} L_{2^\ast}+2 M_{4b} B_{2} L_{2}+M_{4a} (B_{2})^2 \\
& +M_{4a(1^\ast3^\ast)} (\delta m_{2})^2-2 \delta m_{2} B_{2} L_{2} M_{2^\ast}-2 \delta m_{2} B_{2} L_{2^\ast} M_{2}-2 (\delta m_{2})^2 L_{2^\ast} M_{2^\ast}-2 (B_{2})^2 L_{2} M_{2}
\end{align*}
\begin{align*}
a_{12}&=
M_{12}-3 M_{6A} B_{2}-2 M_{6A(1^\ast)} \delta m_{2}-M_{6A(3^\ast)} \delta m_{2}+6 M_{4b(1^\ast)} \delta m_{2} B_{2} \\
& +3 M_{4b} (B_{2})^2+2 M_{4b(1^\ast1^\ast)} (\delta m_{2})^2+M_{4b(1^\ast3^\ast)} (\delta m_{2})^2-3 \delta m_{2} (B_{2})^2 M_{2^\ast}-3 (\delta m_{2})^2 B_{2} M_{2^{\ast \ast}} \\
& -(\delta m_{2})^3 M_{2^{\ast \ast \ast}}-(B_{2})^3 M_{2}
\end{align*}
\begin{align*}
a_{13}&=
M_{13}-M_{6H} B_{2}-M_{6H(1^\ast)} \delta m_{2}-L_{6D4} M_{2}+L_{4x(1^\ast)} \delta m_{2} M_{2} \\
& -L_{4x} M_{4b}+L_{4x} \delta m_{2} M_{2^\ast}+2 L_{4x} B_{2} M_{2}
\end{align*}
\begin{align*}
a_{14}&=
M_{14}-M_{6D} L_{2}-M_{6G} B_{2}-M_{6G(5^\ast)} \delta m_{2}-L_{6D3} M_{2} \\
& +L_{4c(1^\ast)} \delta m_{2} M_{2}+L_{4c} B_{2} M_{2}+L_{4s} L_{2} M_{2}-L_{4l} M_{4b}+L_{4l} \delta m_{2} M_{2^\ast} \\
& +L_{4l} B_{2} M_{2}+M_{4a(1^\ast)} \delta m_{2} L_{2}+M_{4b} (L_{2})^2+M_{4a} B_{2} L_{2}-\delta m_{2} L_{2} L_{2^\ast} M_{2} \\
& -\delta m_{2} (L_{2})^2 M_{2^\ast}-2 B_{2} (L_{2})^2 M_{2}
\end{align*}
\begin{align*}
a_{15}&=
M_{15}-M_{6D} L_{2}-M_{6G} B_{2}-M_{6G(1^\ast)} \delta m_{2}-L_{6A2} M_{2} \\
& +L_{4l(1^\ast)} \delta m_{2} M_{2}-L_{4c} M_{4b}+L_{4c} \delta m_{2} M_{2^\ast}+L_{4c} B_{2} M_{2}+L_{4s} L_{2} M_{2} \\
& +L_{4l} B_{2} M_{2}+M_{4a(1^\ast)} \delta m_{2} L_{2}+M_{4b} (L_{2})^2+M_{4a} B_{2} L_{2}-\delta m_{2} L_{2} L_{2^\ast} M_{2} \\
& -\delta m_{2} (L_{2})^2 M_{2^\ast}-2 B_{2} (L_{2})^2 M_{2}
\end{align*}
\begin{align*}
a_{16}&=
M_{16}-2 M_{6A} L_{2}-M_{6C} B_{2}-M_{6C(1^\ast)} \delta m_{2}-\delta m_{4a} M_{4b(1^\ast)} \\
& +\delta m_{4a} \delta m_{2} M_{2^{\ast \ast}}+\delta m_{4a} B_{2} M_{2^\ast}-B_{4a} M_{4b}+B_{4a} \delta m_{2} M_{2^\ast}+B_{4a} B_{2} M_{2} \\
& +4 M_{4b(1^\ast)} \delta m_{2} L_{2}+4 M_{4b} B_{2} L_{2}-4 \delta m_{2} B_{2} L_{2} M_{2^\ast}-2 (\delta m_{2})^2 L_{2} M_{2^{\ast \ast}}-2 (B_{2})^2 L_{2} M_{2}
\end{align*}
\begin{align*}
a_{17}&=
M_{17}-M_{6D} B_{2}-M_{6E} B_{2}-M_{6E(1^\ast)} \delta m_{2}-M_{6D(4^\ast)} \delta m_{2} \\
& -L_{6A3} M_{2}+2 L_{4s(1^\ast)} \delta m_{2} M_{2}-L_{4s} M_{4b}+L_{4s} \delta m_{2} M_{2^\ast}+3 L_{4s} B_{2} M_{2} \\
& +M_{4a(1^\ast)} \delta m_{2} B_{2}+M_{4a(2^\ast)} \delta m_{2} B_{2}+M_{4b} \delta m_{2} L_{2^\ast}+M_{4b} B_{2} L_{2}+M_{4a} (B_{2})^2 \\
& +M_{4a(1^\ast2^\ast)} (\delta m_{2})^2-\delta m_{2} B_{2} L_{2} M_{2^\ast}-3 \delta m_{2} B_{2} L_{2^\ast} M_{2}-(\delta m_{2})^2 L_{2^\ast} M_{2^\ast}-(\delta m_{2})^2 L_{2^\ast\dagger^\ast} M_{2} \\
& -2 (B_{2})^2 L_{2} M_{2}
\end{align*}
\begin{align*}
a_{18}&=
M_{18}-M_{6A} B_{2}-M_{6B} B_{2}-M_{6B(1^\ast)} \delta m_{2}-M_{6A(2^\ast)} \delta m_{2} \\
& -\delta m_{4b} M_{4b(1^\ast)}+\delta m_{4b} \delta m_{2} M_{2^{\ast \ast}}+\delta m_{4b} B_{2} M_{2^\ast}-B_{4b} M_{4b}+B_{4b} \delta m_{2} M_{2^\ast} \\
& +B_{4b} B_{2} M_{2}+M_{4b(1^\ast)} \delta m_{2} \delta m_{2^\ast}+2 M_{4b(1^\ast)} \delta m_{2} B_{2}+M_{4b(2^\ast)} \delta m_{2} B_{2}+M_{4b} \delta m_{2} B_{2^\ast} \\
& +2 M_{4b} (B_{2})^2+M_{4b(1^\ast2^\ast)} (\delta m_{2})^2-\delta m_{2} \delta m_{2^\ast} B_{2} M_{2^\ast}-\delta m_{2} B_{2} B_{2^\ast} M_{2}-2 \delta m_{2} (B_{2})^2 M_{2^\ast} \\
& -(\delta m_{2})^2 \delta m_{2^\ast} M_{2^{\ast \ast}}-(\delta m_{2})^2 B_{2} M_{2^{\ast \ast}}-(\delta m_{2})^2 B_{2^\ast} M_{2^\ast}-(B_{2})^3 M_{2}
\end{align*}
\begin{align*}
a_{19}&=
M_{19}-2 L_{6H2} M_{2}
\end{align*}
\begin{align*}
a_{20}&=
M_{20}-M_{6H} L_{2}-L_{6F2} M_{2}-L_{6G4} M_{2}+2 L_{4x} L_{2} M_{2}
\end{align*}
\begin{align*}
a_{21}&=
M_{21}-2 L_{6G2} M_{2}
\end{align*}
\begin{align*}
a_{22}&=
M_{22}-M_{6G} L_{2}-L_{6F1} M_{2}-L_{6C2} M_{2}-L_{4c} M_{4a} \\
& +3 L_{4c} L_{2} M_{2}+L_{4l} L_{2} M_{2}+M_{4a} (L_{2})^2-2 (L_{2})^3 M_{2}
\end{align*}
\begin{align*}
a_{23}&=
M_{23}-M_{6H} B_{2}-M_{6H(2^\ast)} \delta m_{2}-L_{6E2} M_{2}-L_{6D4} M_{2} \\
& +L_{4x(1^\ast)} \delta m_{2} M_{2}+L_{4x(2^\ast)} \delta m_{2} M_{2}+2 L_{4x} B_{2} M_{2}
\end{align*}
\begin{align*}
a_{24}&=
M_{24}-M_{6G} B_{2}-M_{6G(2^\ast)} \delta m_{2}-L_{6B2} M_{2}-L_{6D5} M_{2} \\
& +L_{4c(1^\ast)} \delta m_{2} M_{2}+L_{4l(2^\ast)} \delta m_{2} M_{2}+L_{4c} B_{2} M_{2}-L_{4s} M_{4a}+2 L_{4s} L_{2} M_{2} \\
& +L_{4l} B_{2} M_{2}+M_{4a} \delta m_{2} L_{2^\ast}+M_{4a} B_{2} L_{2}-2 \delta m_{2} L_{2} L_{2^\ast} M_{2}-2 B_{2} (L_{2})^2 M_{2}
\end{align*}
\begin{align*}
a_{25}&=
M_{25}-2 M_{6G} L_{2}-2 L_{6D2} M_{2}+2 L_{4c} L_{2} M_{2}+2 L_{4l} L_{2} M_{2} \\
& +M_{4a} (L_{2})^2-2 (L_{2})^3 M_{2}
\end{align*}
\begin{align*}
a_{26}&=
M_{26}-2 M_{6C} L_{2}-\delta m_{6F} M_{2^\ast}-B_{6F} M_{2}+2 \delta m_{4a} L_{2} M_{2^\ast} \\
& +2 B_{4a} L_{2} M_{2}-2 L_{4c} M_{4b}+2 L_{4c} \delta m_{2} M_{2^\ast}+2 L_{4c} B_{2} M_{2}+3 M_{4b} (L_{2})^2 \\
& -3 \delta m_{2} (L_{2})^2 M_{2^\ast}-3 B_{2} (L_{2})^2 M_{2}
\end{align*}
\begin{align*}
a_{27}&=
M_{27}-M_{6E} L_{2}-M_{6G} B_{2}-M_{6G(4^\ast)} \delta m_{2}-L_{6D1} M_{2} \\
& -L_{6A2} M_{2}+L_{4c(1^\ast)} \delta m_{2} M_{2}+L_{4l(1^\ast)} \delta m_{2} M_{2}+L_{4c} B_{2} M_{2}+2 L_{4s} L_{2} M_{2} \\
& +L_{4l} B_{2} M_{2}+M_{4a(2^\ast)} \delta m_{2} L_{2}+M_{4a} B_{2} L_{2}-2 \delta m_{2} L_{2} L_{2^\ast} M_{2}-2 B_{2} (L_{2})^2 M_{2}
\end{align*}
\begin{align*}
a_{28}&=
M_{28}-M_{6B} L_{2}-M_{6C} B_{2}-M_{6C(2^\ast)} \delta m_{2}-\delta m_{6D} M_{2^\ast} \\
& -B_{6D} M_{2}+\delta m_{4a(1^\ast)} \delta m_{2} M_{2^\ast}+B_{4a(1^\ast)} \delta m_{2} M_{2}+\delta m_{4a} B_{2} M_{2^\ast}+\delta m_{4b} L_{2} M_{2^\ast} \\
& +B_{4a} B_{2} M_{2}+B_{4b} L_{2} M_{2}-L_{4s} M_{4b}+L_{4s} \delta m_{2} M_{2^\ast}+L_{4s} B_{2} M_{2} \\
& +M_{4b(2^\ast)} \delta m_{2} L_{2}+M_{4b} \delta m_{2} L_{2^\ast}+2 M_{4b} B_{2} L_{2}-\delta m_{2} \delta m_{2^\ast} L_{2} M_{2^\ast}-2 \delta m_{2} B_{2} L_{2} M_{2^\ast} \\
& -\delta m_{2} B_{2} L_{2^\ast} M_{2}-\delta m_{2} B_{2^\ast} L_{2} M_{2}-(\delta m_{2})^2 L_{2^\ast} M_{2^\ast}-2 (B_{2})^2 L_{2} M_{2}
\end{align*}
\begin{align*}
a_{29}&=
M_{29}-2 M_{6E} B_{2}-2 M_{6E(2^\ast)} \delta m_{2}-2 L_{6A1} M_{2}+4 L_{4s(1^\ast)} \delta m_{2} M_{2} \\
& +4 L_{4s} B_{2} M_{2}+2 M_{4a(2^\ast)} \delta m_{2} B_{2}+M_{4a} (B_{2})^2+M_{4a(2^\ast2^\ast)} (\delta m_{2})^2-4 \delta m_{2} B_{2} L_{2^\ast} M_{2} \\
& -2 (\delta m_{2})^2 L_{2^{\ast \ast}\dagger} M_{2}-2 (B_{2})^2 L_{2} M_{2}
\end{align*}
\begin{align*}
a_{30}&=
M_{30}-2 M_{6B} B_{2}-2 M_{6B(2^\ast)} \delta m_{2}-\delta m_{6A} M_{2^\ast}-B_{6A} M_{2} \\
& +2 \delta m_{4b(1^\ast)} \delta m_{2} M_{2^\ast}+2 B_{4b(1^\ast)} \delta m_{2} M_{2}+2 \delta m_{4b} B_{2} M_{2^\ast}+2 B_{4b} B_{2} M_{2}+2 M_{4b(2^\ast)} \delta m_{2} B_{2} \\
& +M_{4b} (B_{2})^2+M_{4b(2^\ast2^\ast)} (\delta m_{2})^2-2 \delta m_{2} \delta m_{2^\ast} B_{2} M_{2^\ast}-2 \delta m_{2} B_{2} B_{2^\ast} M_{2}-\delta m_{2} (B_{2})^2 M_{2^\ast} \\
& -(\delta m_{2})^2 \delta m_{2^{\ast \ast}} M_{2^\ast}-(\delta m_{2})^2 B_{2^{\ast \ast}} M_{2}-(B_{2})^3 M_{2}
\end{align*}
\begin{align*}
a_{31}&=
M_{31}-2 L_{6H3} M_{2}
\end{align*}
\begin{align*}
a_{32}&=
M_{32}-L_{6H2} M_{2}-L_{6G3} M_{2}
\end{align*}
\begin{align*}
a_{33}&=
M_{33}-2 L_{6G3} M_{2}
\end{align*}
\begin{align*}
a_{34}&=
M_{34}-L_{6H1} M_{2}-L_{6C3} M_{2}-L_{4x} M_{4a}+2 L_{4x} L_{2} M_{2}
\end{align*}
\begin{align*}
a_{35}&=
M_{35}-M_{6H} L_{2}-L_{6E3} M_{2}-L_{6G4} M_{2}+2 L_{4x} L_{2} M_{2}
\end{align*}
\begin{align*}
a_{36}&=
M_{36}-M_{6G} L_{2}-L_{6B3} M_{2}-L_{6G5} M_{2}+L_{4c} L_{2} M_{2} \\
& -L_{4l} M_{4a}+3 L_{4l} L_{2} M_{2}+M_{4a} (L_{2})^2-2 (L_{2})^3 M_{2}
\end{align*}
\begin{align*}
a_{37}&=
M_{37}-2 L_{6G2} M_{2}
\end{align*}
\begin{align*}
a_{38}&=
M_{38}-\delta m_{6H} M_{2^\ast}-B_{6H} M_{2}-2 L_{4x} M_{4b}+2 L_{4x} \delta m_{2} M_{2^\ast} 
       +2 L_{4x} B_{2} M_{2}
\end{align*}
\begin{align*}
a_{39}&=
M_{39}-M_{6G} L_{2}-L_{6G1} M_{2}-L_{6C2} M_{2}-L_{4c} M_{4a} \\
& +3 L_{4c} L_{2} M_{2}+L_{4l} L_{2} M_{2}+M_{4a} (L_{2})^2-2 (L_{2})^3 M_{2}
\end{align*}
\begin{align*}
a_{40}&=
M_{40}-M_{6C} L_{2}-\delta m_{6G} M_{2^\ast}-B_{6G} M_{2}+\delta m_{4a} L_{2} M_{2^\ast} \\
& +B_{4a} L_{2} M_{2}-L_{4c} M_{4b}+L_{4c} \delta m_{2} M_{2^\ast}+L_{4c} B_{2} M_{2}-L_{4l} M_{4b} \\
& +L_{4l} \delta m_{2} M_{2^\ast}+L_{4l} B_{2} M_{2}+2 M_{4b} (L_{2})^2-2 \delta m_{2} (L_{2})^2 M_{2^\ast}-2 B_{2} (L_{2})^2 M_{2}
\end{align*}
\begin{align*}
a_{41}&=
M_{41}-2 M_{6E} L_{2}-2 L_{6C1} M_{2}-\delta m_{4a} M_{4a(2^\ast)}+2 \delta m_{4a} L_{2^\ast} M_{2} \\
& -B_{4a} M_{4a}+2 B_{4a} L_{2} M_{2}+4 L_{4s} L_{2} M_{2}+2 M_{4a(2^\ast)} \delta m_{2} L_{2}+2 M_{4a} B_{2} L_{2} \\
& -4 \delta m_{2} L_{2} L_{2^\ast} M_{2}-4 B_{2} (L_{2})^2 M_{2}
\end{align*}
\begin{align*}
a_{42}&=
M_{42}-2 M_{6B} L_{2}-\delta m_{6C} M_{2^\ast}-B_{6C} M_{2}-\delta m_{4a} M_{4b(2^\ast)} \\
& +\delta m_{4a} \delta m_{2^\ast} M_{2^\ast}+\delta m_{4a} B_{2^\ast} M_{2}+2 \delta m_{4b} L_{2} M_{2^\ast}-B_{4a} M_{4b}+B_{4a} \delta m_{2} M_{2^\ast} \\
& +B_{4a} B_{2} M_{2}+2 B_{4b} L_{2} M_{2}+2 M_{4b(2^\ast)} \delta m_{2} L_{2}+2 M_{4b} B_{2} L_{2}-2 \delta m_{2} \delta m_{2^\ast} L_{2} M_{2^\ast} \\
& -2 \delta m_{2} B_{2} L_{2} M_{2^\ast}-2 \delta m_{2} B_{2^\ast} L_{2} M_{2}-2 (B_{2})^2 L_{2} M_{2}
\end{align*}
\begin{align*}
a_{43}&=
M_{43}-M_{6H} B_{2}-M_{6H(3^\ast)} \delta m_{2}-2 L_{6E2} M_{2}+2 L_{4x(2^\ast)} \delta m_{2} M_{2} \\
& +2 L_{4x} B_{2} M_{2}
\end{align*}
\begin{align*}
a_{44}&=
M_{44}-M_{6G} B_{2}-M_{6G(3^\ast)} \delta m_{2}-L_{6E1} M_{2}-L_{6B2} M_{2} \\
& +L_{4c(2^\ast)} \delta m_{2} M_{2}+L_{4l(2^\ast)} \delta m_{2} M_{2}+L_{4c} B_{2} M_{2}-L_{4s} M_{4a}+2 L_{4s} L_{2} M_{2} \\
& +L_{4l} B_{2} M_{2}+M_{4a} \delta m_{2} L_{2^\ast}+M_{4a} B_{2} L_{2}-2 \delta m_{2} L_{2} L_{2^\ast} M_{2}-2 B_{2} (L_{2})^2 M_{2}
\end{align*}
\begin{align*}
a_{45}&=
M_{45}-M_{6C} B_{2}-M_{6C(3^\ast)} \delta m_{2}-\delta m_{6E} M_{2^\ast}-B_{6E} M_{2} \\
& +\delta m_{4a(2^\ast)} \delta m_{2} M_{2^\ast}+B_{4a(2^\ast)} \delta m_{2} M_{2}+\delta m_{4a} B_{2} M_{2^\ast}+B_{4a} B_{2} M_{2}-2 L_{4s} M_{4b} \\
& +2 L_{4s} \delta m_{2} M_{2^\ast}+2 L_{4s} B_{2} M_{2}+2 M_{4b} \delta m_{2} L_{2^\ast}+2 M_{4b} B_{2} L_{2}-2 \delta m_{2} B_{2} L_{2} M_{2^\ast} \\
& -2 \delta m_{2} B_{2} L_{2^\ast} M_{2}-2 (\delta m_{2})^2 L_{2^\ast} M_{2^\ast}-2 (B_{2})^2 L_{2} M_{2}
\end{align*}
\begin{align*}
a_{46}&=
M_{46}-M_{6E} B_{2}-M_{6E(3^\ast)} \delta m_{2}-2 L_{6B1} M_{2}+2 L_{4s(2^\ast)} \delta m_{2} M_{2} \\
& -\delta m_{4b} M_{4a(2^\ast)}+2 \delta m_{4b} L_{2^\ast} M_{2}-B_{4b} M_{4a}+2 B_{4b} L_{2} M_{2}+2 L_{4s} B_{2} M_{2} \\
& +M_{4a(2^\ast)} \delta m_{2} \delta m_{2^\ast}+M_{4a(2^\ast)} \delta m_{2} B_{2}+M_{4a} \delta m_{2} B_{2^\ast}+M_{4a} (B_{2})^2-2 \delta m_{2} \delta m_{2^\ast} L_{2^\ast} M_{2} \\
& -2 \delta m_{2} B_{2} L_{2^\ast} M_{2}-2 \delta m_{2} B_{2^\ast} L_{2} M_{2}-2 (B_{2})^2 L_{2} M_{2}
\end{align*}
\begin{align*}
a_{47}&=
M_{47}-M_{6B} B_{2}-M_{6B(3^\ast)} \delta m_{2}-\delta m_{6B} M_{2^\ast}-B_{6B} M_{2} \\
& +\delta m_{4b(2^\ast)} \delta m_{2} M_{2^\ast}+B_{4b(2^\ast)} \delta m_{2} M_{2}-\delta m_{4b} M_{4b(2^\ast)}+\delta m_{4b} \delta m_{2^\ast} M_{2^\ast}+\delta m_{4b} B_{2} M_{2^\ast} \\
& +\delta m_{4b} B_{2^\ast} M_{2}-B_{4b} M_{4b}+B_{4b} \delta m_{2} M_{2^\ast}+2 B_{4b} B_{2} M_{2}+M_{4b(2^\ast)} \delta m_{2} \delta m_{2^\ast} \\
& +M_{4b(2^\ast)} \delta m_{2} B_{2}+M_{4b} \delta m_{2} B_{2^\ast}+M_{4b} (B_{2})^2-\delta m_{2} \delta m_{2^\ast} B_{2} M_{2^\ast}-\delta m_{2} \delta m_{2^\ast} B_{2^\ast} M_{2} \\
& -\delta m_{2} (\delta m_{2^\ast})^2 M_{2^\ast}-2 \delta m_{2} B_{2} B_{2^\ast} M_{2}-\delta m_{2} (B_{2})^2 M_{2^\ast}-(\delta m_{2})^2 B_{2^\ast} M_{2^\ast}-(B_{2})^3 M_{2}
\end{align*}

\vspace*{6mm}
\noindent
{\bf Intermidiate Renormalization for the Eighth-order Moments}

\begin{align*}
M_{01}&=
\Delta M_{01}+2 M_{6F} \hat{L}_{2}+2 \hat{L}_{6F1} M_{2}+2 \hat{L}_{4c} M_{4a}-6 \hat{L}_{4c} \hat{L}_{2} M_{2} \\
& -3 M_{4a} (\hat{L}_{2})^2+4 (\hat{L}_{2})^3 M_{2}
\end{align*}
\begin{align*}
M_{02}&=
\Delta M_{02}+M_{6D} \hat{L}_{2}+M_{6F} \hat{B}_{2}+M_{6F(1^\ast)} \delta m_{2}+I_{6F1} M_{2} \\
& +\hat{L}_{6D5} M_{2}+\hat{L}_{4c} M_{4b}-\hat{L}_{4c} \delta m_{2} M_{2^\ast}-\hat{L}_{4c} \hat{B}_{2} M_{2}+\hat{L}_{4s} M_{4a} \\
& -2 \hat{L}_{4s} \hat{L}_{2} M_{2}-\hat{L}_{4c(3^\prime )} \hat{B}_{2} M_{2}+\Delta L_{4c} M_{2} I_{2}-M_{4a(1^\ast)} \delta m_{2} \hat{L}_{2}-M_{4b} (\hat{L}_{2})^2 \\
& -M_{4a} \hat{B}_{2} \hat{L}_{2}-M_{4a} \hat{B}_{2} \hat{L}_{2^\prime}+\delta m_{2} (\hat{L}_{2})^2 M_{2^\ast}+2 \hat{B}_{2} \hat{L}_{2} M_{2} \hat{L}_{2^\prime}+\hat{B}_{2} (\hat{L}_{2})^2 M_{2}
\end{align*}
\begin{align*}
M_{03}&=
\Delta M_{03}+2 M_{6D} \hat{L}_{2}+M_{6F} \hat{B}_{2}+M_{6F(3^\ast)} \delta m_{2}+2 \hat{L}_{6D1} M_{2} \\
& +I_{6F3} M_{2}-2 \hat{L}_{4s} \hat{L}_{2} M_{2}-2 \hat{L}_{4c((1^\prime )^\prime )} \hat{B}_{2} M_{2}-2 M_{4a(1^\ast)} \delta m_{2} \hat{L}_{2}-M_{4b} (\hat{L}_{2})^2 \\
& -2 M_{4a} \hat{B}_{2} \hat{L}_{2}+\delta m_{2} (\hat{L}_{2})^2 M_{2^\ast}+2 \hat{B}_{2} \hat{L}_{2} M_{2} \hat{L}_{2^\prime}+\hat{B}_{2} (\hat{L}_{2})^2 M_{2}
\end{align*}
\begin{align*}
M_{04}&=
\Delta M_{04}+M_{6A} \hat{L}_{2}+2 M_{6D} \hat{B}_{2}+M_{6D(1^\ast)} \delta m_{2}+M_{6D(3^\ast)} \delta m_{2} \\
& +\hat{L}_{6A1} M_{2}+I_{6D1} M_{2}+I_{6D3} M_{2}-\hat{L}_{4s((1^\prime )^\prime )} \hat{B}_{2} M_{2}-\hat{L}_{4s(3^\prime )} \hat{B}_{2} M_{2} \\
& -2 M_{4a(1^\ast)} \delta m_{2} \hat{B}_{2}-2 M_{4b(1^\ast)} \delta m_{2} \hat{L}_{2}-2 M_{4b} \hat{B}_{2} \hat{L}_{2}-M_{4a} (\hat{B}_{2})^2-M_{4a(1^\ast1^\ast)} (\delta m_{2})^2 \\
& +\hat{L}_{2^{\prime \prime}} (\hat{B}_{2})^2 M_{2}+2 \delta m_{2} \hat{B}_{2} \hat{L}_{2} M_{2^\ast}+(\delta m_{2})^2 \hat{L}_{2} M_{2^{\ast \ast}}+(\hat{B}_{2})^2 \hat{L}_{2} M_{2}
\end{align*}
\begin{align*}
M_{05}&=
\Delta M_{05}+M_{6H} \hat{L}_{2}+\hat{L}_{6H1} M_{2}+\hat{L}_{6F2} M_{2}+\hat{L}_{4x} M_{4a} \\
& -3 \hat{L}_{4x} \hat{L}_{2} M_{2}
\end{align*}
\begin{align*}
M_{06}&=
\Delta M_{06}+M_{6F} \hat{L}_{2}+M_{6G} \hat{L}_{2}+\hat{L}_{6F3} M_{2}+\hat{L}_{6G5} M_{2} \\
& -3 \hat{L}_{4c} \hat{L}_{2} M_{2}+\hat{L}_{4l} M_{4a}-2 \hat{L}_{4l} \hat{L}_{2} M_{2}-2 M_{4a} (\hat{L}_{2})^2+3 (\hat{L}_{2})^3 M_{2}
\end{align*}
\begin{align*}
M_{07}&=
\Delta M_{07}+M_{6F} \hat{L}_{2}+M_{6G} \hat{L}_{2}+\hat{L}_{6G1} M_{2}+\hat{L}_{6D2} M_{2} \\
& +\hat{L}_{4c} M_{4a}-4 \hat{L}_{4c} \hat{L}_{2} M_{2}-\hat{L}_{4l} \hat{L}_{2} M_{2}-2 M_{4a} (\hat{L}_{2})^2+3 (\hat{L}_{2})^3 M_{2}
\end{align*}
\begin{align*}
M_{08}&=
\Delta M_{08}+M_{6C} \hat{L}_{2}+2 M_{6D} \hat{L}_{2}+\hat{L}_{6C1} M_{2}+\delta\hat{m}_{4a} M_{4a(1^\ast)} \\
& -\delta\hat{m}_{4a} \hat{L}_{2} M_{2^\ast}+\hat{B}_{4a} M_{4a}-\hat{B}_{4a} \hat{L}_{2} M_{2}-\hat{B}_{4a} M_{2} \hat{L}_{2^\prime}-2 \hat{L}_{4s} \hat{L}_{2} M_{2} \\
& +\Delta \delta m_{4a} I_{4a(1^\ast)}-2 M_{4a(1^\ast)} \delta m_{2} \hat{L}_{2}-2 M_{4b} (\hat{L}_{2})^2-2 M_{4a} \hat{B}_{2} \hat{L}_{2}+\Delta M_{4a} I_{4c} \\
& +2 \delta m_{2} (\hat{L}_{2})^2 M_{2^\ast}+2 \hat{B}_{2} \hat{L}_{2} M_{2} \hat{L}_{2^\prime}+2 \hat{B}_{2} (\hat{L}_{2})^2 M_{2}
\end{align*}
\begin{align*}
M_{09}&=
\Delta M_{09}+M_{6E} \hat{L}_{2}+M_{6F} \hat{B}_{2}+M_{6F(2^\ast)} \delta m_{2}+\hat{L}_{6E1} M_{2} \\
& +I_{6F2} M_{2}+\hat{L}_{6D3} M_{2}+\hat{L}_{4s} M_{4a}-3 \hat{L}_{4s} \hat{L}_{2} M_{2}-\hat{L}_{4c(1^\prime )} \hat{B}_{2} M_{2} \\
& -\hat{L}_{4c(2^\prime )} \hat{B}_{2} M_{2}-M_{4a(2^\ast)} \delta m_{2} \hat{L}_{2}-M_{4a} \hat{B}_{2} \hat{L}_{2}-M_{4a} \hat{B}_{2} \hat{L}_{2^\prime}+3 \hat{B}_{2} \hat{L}_{2} M_{2} \hat{L}_{2^\prime}
\end{align*}
\begin{align*}
M_{10}&=
\Delta M_{10}+M_{6B} \hat{L}_{2}+M_{6D} \hat{B}_{2}+M_{6D(2^\ast)} \delta m_{2}+\hat{L}_{6B1} M_{2} \\
& +I_{6D2} M_{2}+\delta\hat{m}_{4b} M_{4a(1^\ast)}-\delta\hat{m}_{4b} \hat{L}_{2} M_{2^\ast}+\hat{B}_{4b} M_{4a}-\hat{B}_{4b} \hat{L}_{2} M_{2} \\
& -\hat{B}_{4b} M_{2} \hat{L}_{2^\prime}+\Delta \delta m_{4b} I_{4a(1^\ast)}-\hat{L}_{4s(2^\prime )} \hat{B}_{2} M_{2}-M_{4a(1^\ast)} \delta m_{2} \delta\hat{m}_{2^\ast}-M_{4a(1^\ast)} \hat{B}_{2} \delta\hat{m}_{2^\prime} \\
& -M_{4b(2^\ast)} \delta m_{2} \hat{L}_{2}-M_{4b} \hat{B}_{2} \hat{L}_{2}-M_{4a} \hat{B}_{2} \hat{B}_{2^\prime}+\Delta M_{4b} I_{4c}+I_{4c} M_{2} I_{2} \\
& +\delta m_{2} \hat{L}_{2} \delta\hat{m}_{2^\ast} M_{2^\ast}+\hat{B}_{2} \hat{L}_{2} M_{2^\ast} \delta\hat{m}_{2^\prime}+\hat{B}_{2} \hat{L}_{2} M_{2} \hat{B}_{2^\prime}+\hat{B}_{2} M_{2} \hat{B}_{2^\prime} \hat{L}_{2^\prime}
\end{align*}
\begin{align*}
M_{11}&=
\Delta M_{11}+2 M_{6D} \hat{B}_{2}+2 M_{6D(5^\ast)} \delta m_{2}+2 I_{6D5} M_{2}+2 \hat{L}_{4s} M_{4b} \\
& -2 \hat{L}_{4s} \delta m_{2} M_{2^\ast}-2 \hat{L}_{4s} \hat{B}_{2} M_{2}+2 \Delta L_{4s} M_{2} I_{2}-2 M_{4a(1^\ast)} \delta m_{2} \hat{B}_{2}-2 M_{4b} \hat{B}_{2} \hat{L}_{2^\prime} \\
& -M_{4a} (\hat{B}_{2})^2-M_{4a(1^\ast3^\ast)} (\delta m_{2})^2+2 \delta m_{2} \hat{B}_{2} M_{2^\ast} \hat{L}_{2^\prime}+2 (\hat{B}_{2})^2 M_{2} \hat{L}_{2^\prime}
\end{align*}
\begin{align*}
M_{12}&=
\Delta M_{12}+3 M_{6A} \hat{B}_{2}+2 M_{6A(1^\ast)} \delta m_{2}+M_{6A(3^\ast)} \delta m_{2}+2 I_{6A1} M_{2} \\
& +I_{6A3} M_{2}-6 M_{4b(1^\ast)} \delta m_{2} \hat{B}_{2}-3 M_{4b} (\hat{B}_{2})^2-2 M_{4b(1^\ast1^\ast)} (\delta m_{2})^2-M_{4b(1^\ast3^\ast)} (\delta m_{2})^2 \\
& +3 \delta m_{2} (\hat{B}_{2})^2 M_{2^\ast}+3 (\delta m_{2})^2 \hat{B}_{2} M_{2^{\ast \ast}}+(\delta m_{2})^3 M_{2^{\ast \ast \ast}}+(\hat{B}_{2})^3 M_{2}
\end{align*}
\begin{align*}
M_{13}&=
\Delta M_{13}+M_{6H} \hat{B}_{2}+M_{6H(1^\ast)} \delta m_{2}+I_{6H1} M_{2}+\hat{L}_{6D4} M_{2} \\
& +\hat{L}_{4x} M_{4b}-\hat{L}_{4x} \delta m_{2} M_{2^\ast}-\hat{L}_{4x} \hat{B}_{2} M_{2}+\Delta L_{4x} M_{2} I_{2}-\hat{L}_{4x(1^\prime )} \hat{B}_{2} M_{2}
\end{align*}
\begin{align*}
M_{14}&=
\Delta M_{14}+M_{6D} \hat{L}_{2}+M_{6G} \hat{B}_{2}+M_{6G(5^\ast)} \delta m_{2}+\hat{L}_{6D3} M_{2} \\
& +I_{6G5} M_{2}-\hat{L}_{4s} \hat{L}_{2} M_{2}+\hat{L}_{4l} M_{4b}-\hat{L}_{4l} \delta m_{2} M_{2^\ast}-\hat{L}_{4l} \hat{B}_{2} M_{2} \\
& +\Delta L_{4l} M_{2} I_{2}-\hat{L}_{4c(1^\prime )} \hat{B}_{2} M_{2}-M_{4a(1^\ast)} \delta m_{2} \hat{L}_{2}-M_{4b} (\hat{L}_{2})^2-M_{4a} \hat{B}_{2} \hat{L}_{2} \\
& +\delta m_{2} (\hat{L}_{2})^2 M_{2^\ast}+\hat{B}_{2} \hat{L}_{2} M_{2} \hat{L}_{2^\prime}+\hat{B}_{2} (\hat{L}_{2})^2 M_{2}
\end{align*}
\begin{align*}
M_{15}&=
\Delta M_{15}+M_{6D} \hat{L}_{2}+M_{6G} \hat{B}_{2}+M_{6G(1^\ast)} \delta m_{2}+I_{6G1} M_{2} \\
& +\hat{L}_{6A2} M_{2}+\hat{L}_{4c} M_{4b}-\hat{L}_{4c} \delta m_{2} M_{2^\ast}-\hat{L}_{4c} \hat{B}_{2} M_{2}-\hat{L}_{4s} \hat{L}_{2} M_{2} \\
& +\Delta L_{4c} M_{2} I_{2}-\hat{L}_{4l(1^\prime )} \hat{B}_{2} M_{2}-M_{4a(1^\ast)} \delta m_{2} \hat{L}_{2}-M_{4b} (\hat{L}_{2})^2-M_{4a} \hat{B}_{2} \hat{L}_{2} \\
& +\delta m_{2} (\hat{L}_{2})^2 M_{2^\ast}+\hat{B}_{2} \hat{L}_{2} M_{2} \hat{L}_{2^\prime}+\hat{B}_{2} (\hat{L}_{2})^2 M_{2}
\end{align*}
\begin{align*}
M_{16}&=
\Delta M_{16}+2 M_{6A} \hat{L}_{2}+M_{6C} \hat{B}_{2}+M_{6C(1^\ast)} \delta m_{2}+I_{6C1} M_{2} \\
& +1/2 J_{6C} M_{2}+\delta\hat{m}_{4a} M_{4b(1^\ast)}-\delta\hat{m}_{4a} \delta m_{2} M_{2^{\ast \ast}}-\delta\hat{m}_{4a} \hat{B}_{2} M_{2^\ast}+\hat{B}_{4a} M_{4b} \\
& -\hat{B}_{4a} \delta m_{2} M_{2^\ast}-\hat{B}_{4a} \hat{B}_{2} M_{2}+\Delta \delta m_{4a} I_{4b(1^\ast)}-\Delta \delta m_{4a} M_{2} I_{2^\ast}-4 M_{4b(1^\ast)} \delta m_{2} \hat{L}_{2} \\
& -4 M_{4b} \hat{B}_{2} \hat{L}_{2}+\Delta M_{4a} I_{4s}+4 \delta m_{2} \hat{B}_{2} \hat{L}_{2} M_{2^\ast}+2 (\delta m_{2})^2 \hat{L}_{2} M_{2^{\ast \ast}}+2 (\hat{B}_{2})^2 \hat{L}_{2} M_{2}
\end{align*}
\begin{align*}
M_{17}&=
\Delta M_{17}+M_{6D} \hat{B}_{2}+M_{6E} \hat{B}_{2}+M_{6E(1^\ast)} \delta m_{2}+M_{6D(4^\ast)} \delta m_{2} \\
& +I_{6E1} M_{2}+\hat{L}_{6A3} M_{2}+I_{6D4} M_{2}+\hat{L}_{4s} M_{4b}-\hat{L}_{4s} \delta m_{2} M_{2^\ast} \\
& -\hat{L}_{4s} \hat{B}_{2} M_{2}+\Delta L_{4s} M_{2} I_{2}-2 \hat{L}_{4s(1^\prime )} \hat{B}_{2} M_{2}-M_{4a(1^\ast)} \delta m_{2} \hat{B}_{2}-M_{4a(2^\ast)} \delta m_{2} \hat{B}_{2} \\
& -M_{4b} \hat{B}_{2} \hat{L}_{2^\prime}-M_{4a} (\hat{B}_{2})^2-M_{4a(1^\ast2^\ast)} (\delta m_{2})^2+\hat{L}_{2^{\prime \prime}} (\hat{B}_{2})^2 M_{2}+\delta m_{2} \hat{B}_{2} M_{2^\ast} \hat{L}_{2^\prime} \\
& +(\hat{B}_{2})^2 M_{2} \hat{L}_{2^\prime}
\end{align*}
\begin{align*}
M_{18}&=
\Delta M_{18}+M_{6A} \hat{B}_{2}+M_{6B} \hat{B}_{2}+M_{6B(1^\ast)} \delta m_{2}+M_{6A(2^\ast)} \delta m_{2} \\
& +I_{6B1} M_{2}+I_{6A2} M_{2}+1/2 J_{6B} M_{2}+\delta\hat{m}_{4b} M_{4b(1^\ast)}-\delta\hat{m}_{4b} \delta m_{2} M_{2^{\ast \ast}} \\
& -\delta\hat{m}_{4b} \hat{B}_{2} M_{2^\ast}+\hat{B}_{4b} M_{4b}-\hat{B}_{4b} \delta m_{2} M_{2^\ast}-\hat{B}_{4b} \hat{B}_{2} M_{2}+\Delta \delta m_{4b} I_{4b(1^\ast)} \\
& -\Delta \delta m_{4b} M_{2} I_{2^\ast}-M_{4b(1^\ast)} \delta m_{2} \hat{B}_{2}-M_{4b(1^\ast)} \delta m_{2} \delta\hat{m}_{2^\ast}-M_{4b(1^\ast)} \hat{B}_{2} \delta\hat{m}_{2^\prime}-M_{4b(2^\ast)} \delta m_{2} \hat{B}_{2} \\
& -M_{4b} \hat{B}_{2} \hat{B}_{2^\prime}-M_{4b} (\hat{B}_{2})^2-M_{4b(1^\ast2^\ast)} (\delta m_{2})^2+\Delta M_{4b} I_{4s}+I_{4s} M_{2} I_{2} \\
& +\delta m_{2} \hat{B}_{2} \delta\hat{m}_{2^\ast} M_{2^\ast}+\delta m_{2} \hat{B}_{2} M_{2^\ast} \hat{B}_{2^\prime}+\delta m_{2} \hat{B}_{2} M_{2^{\ast \ast}} \delta\hat{m}_{2^\prime}+(\delta m_{2})^2 \delta\hat{m}_{2^\ast} M_{2^{\ast \ast}}+(\hat{B}_{2})^2 M_{2^\ast} \delta\hat{m}_{2^\prime} \\
& +(\hat{B}_{2})^2 M_{2} \hat{B}_{2^\prime}
\end{align*}
\begin{align*}
M_{19}&=
\Delta M_{19}+2 \hat{L}_{6H2} M_{2}
\end{align*}
\begin{align*}
M_{20}&=
\Delta M_{20}+M_{6H} \hat{L}_{2}+\hat{L}_{6F2} M_{2}+\hat{L}_{6G4} M_{2}-2 \hat{L}_{4x} \hat{L}_{2} M_{2}
\end{align*}
\begin{align*}
M_{21}&=
\Delta M_{21}+2 \hat{L}_{6G2} M_{2}
\end{align*}
\begin{align*}
M_{22}&=
\Delta M_{22}+M_{6G} \hat{L}_{2}+\hat{L}_{6F1} M_{2}+\hat{L}_{6C2} M_{2}+\hat{L}_{4c} M_{4a} \\
& -3 \hat{L}_{4c} \hat{L}_{2} M_{2}-\hat{L}_{4l} \hat{L}_{2} M_{2}-M_{4a} (\hat{L}_{2})^2+2 (\hat{L}_{2})^3 M_{2}
\end{align*}
\begin{align*}
M_{23}&=
\Delta M_{23}+M_{6H} \hat{B}_{2}+M_{6H(2^\ast)} \delta m_{2}+\hat{L}_{6E2} M_{2}+I_{6H2} M_{2} \\
& +\hat{L}_{6D4} M_{2}-\hat{L}_{4x(1^\prime )} \hat{B}_{2} M_{2}-\hat{L}_{4x(2^\prime )} \hat{B}_{2} M_{2}
\end{align*}
\begin{align*}
M_{24}&=
\Delta M_{24}+M_{6G} \hat{B}_{2}+M_{6G(2^\ast)} \delta m_{2}+\hat{L}_{6B2} M_{2}+I_{6G2} M_{2} \\
& +\hat{L}_{6D5} M_{2}+\hat{L}_{4s} M_{4a}-2 \hat{L}_{4s} \hat{L}_{2} M_{2}-\hat{L}_{4c(3^\prime )} \hat{B}_{2} M_{2}-\hat{L}_{4l(2^\prime )} \hat{B}_{2} M_{2} \\
& -M_{4a} \hat{B}_{2} \hat{L}_{2^\prime}+2 \hat{B}_{2} \hat{L}_{2} M_{2} \hat{L}_{2^\prime}
\end{align*}
\begin{align*}
M_{25}&=
\Delta M_{25}+2 M_{6G} \hat{L}_{2}+2 \hat{L}_{6D2} M_{2}-2 \hat{L}_{4c} \hat{L}_{2} M_{2}-2 \hat{L}_{4l} \hat{L}_{2} M_{2} \\
& -M_{4a} (\hat{L}_{2})^2+2 (\hat{L}_{2})^3 M_{2}
\end{align*}
\begin{align*}
M_{26}&=
\Delta M_{26}+\Delta M_{6F} I_{2}+2 M_{6C} \hat{L}_{2}+\delta\hat{m}_{6F} M_{2^\ast}+\hat{B}_{6F} M_{2} \\
& +\Delta \delta m_{6F} M_{2^\ast}[I]-2 \delta\hat{m}_{4a} \hat{L}_{2} M_{2^\ast}-2 \hat{B}_{4a} \hat{L}_{2} M_{2}+2 \hat{L}_{4c} M_{4b}-2 \hat{L}_{4c} \delta m_{2} M_{2^\ast} \\
& -2 \hat{L}_{4c} \hat{B}_{2} M_{2}-3 M_{4b} (\hat{L}_{2})^2+3 \delta m_{2} (\hat{L}_{2})^2 M_{2^\ast}+3 \hat{B}_{2} (\hat{L}_{2})^2 M_{2}
\end{align*}
\begin{align*}
M_{27}&=
\Delta M_{27}+M_{6E} \hat{L}_{2}+M_{6G} \hat{B}_{2}+M_{6G(4^\ast)} \delta m_{2}+\hat{L}_{6D1} M_{2} \\
& +\hat{L}_{6A2} M_{2}+I_{6G4} M_{2}-2 \hat{L}_{4s} \hat{L}_{2} M_{2}-\hat{L}_{4c((1^\prime )^\prime )} \hat{B}_{2} M_{2}-\hat{L}_{4l(1^\prime )} \hat{B}_{2} M_{2} \\
& -M_{4a(2^\ast)} \delta m_{2} \hat{L}_{2}-M_{4a} \hat{B}_{2} \hat{L}_{2}+2 \hat{B}_{2} \hat{L}_{2} M_{2} \hat{L}_{2^\prime}
\end{align*}
\begin{align*}
M_{28}&=
\Delta M_{28}+\Delta M_{6D} I_{2}+M_{6B} \hat{L}_{2}+M_{6C} \hat{B}_{2}+M_{6C(2^\ast)} \delta m_{2} \\
& +\delta\hat{m}_{6D} M_{2^\ast}+I_{6C2} M_{2}+\hat{B}_{6D} M_{2}+\Delta \delta m_{6D} M_{2^\ast}[I]-\delta\hat{m}_{4b} \hat{L}_{2} M_{2^\ast} \\
& -\hat{B}_{4b} \hat{L}_{2} M_{2}+\hat{L}_{4s} M_{4b}-\hat{L}_{4s} \delta m_{2} M_{2^\ast}-\hat{L}_{4s} \hat{B}_{2} M_{2}-\delta\hat{m}_{4a(1^\ast)} \delta m_{2} M_{2^\ast} \\
& -\delta\hat{m}_{4a(1^\prime )} \hat{B}_{2} M_{2^\ast}-\hat{B}_{4a(1^\prime )} \hat{B}_{2} M_{2}-M_{4b(2^\ast)} \delta m_{2} \hat{L}_{2}-M_{4b} \hat{B}_{2} \hat{L}_{2}-M_{4b} \hat{B}_{2} \hat{L}_{2^\prime} \\
& +I_{4c} M_{2} I_{2}+\delta m_{2} \hat{B}_{2} M_{2^\ast} \hat{L}_{2^\prime}+\delta m_{2} \hat{L}_{2} \delta\hat{m}_{2^\ast} M_{2^\ast}+\hat{B}_{2} \hat{L}_{2} M_{2^\ast} \delta\hat{m}_{2^\prime}+\hat{B}_{2} \hat{L}_{2} M_{2} \hat{B}_{2^\prime} \\
& +(\hat{B}_{2})^2 M_{2} \hat{L}_{2^\prime}
\end{align*}
\begin{align*}
M_{29}&=
\Delta M_{29}+2 M_{6E} \hat{B}_{2}+2 M_{6E(2^\ast)} \delta m_{2}+2 \hat{L}_{6A1} M_{2}+2 I_{6E2} M_{2} \\
& -2 \hat{L}_{4s((1^\prime )^\prime )} \hat{B}_{2} M_{2}-2 \hat{L}_{4s(3^\prime )} \hat{B}_{2} M_{2}-2 M_{4a(2^\ast)} \delta m_{2} \hat{B}_{2}-M_{4a} (\hat{B}_{2})^2-M_{4a(2^\ast2^\ast)} (\delta m_{2})^2 \\
& +2 \hat{L}_{2^{\prime \prime}} (\hat{B}_{2})^2 M_{2}
\end{align*}
\begin{align*}
M_{30}&=
\Delta M_{30}+\Delta M_{6A} I_{2}+2 M_{6B} \hat{B}_{2}+2 M_{6B(2^\ast)} \delta m_{2}+\delta\hat{m}_{6A} M_{2^\ast} \\
& +2 I_{6B2} M_{2}+\hat{B}_{6A} M_{2}+\Delta \delta m_{6A} M_{2^\ast}[I]-2 \delta\hat{m}_{4b(1^\ast)} \delta m_{2} M_{2^\ast}-2 \delta\hat{m}_{4b(1^\prime )} \hat{B}_{2} M_{2^\ast} \\
& -2 \hat{B}_{4b(1^\prime )} \hat{B}_{2} M_{2}-2 M_{4b(2^\ast)} \delta m_{2} \hat{B}_{2}-M_{4b} (\hat{B}_{2})^2-M_{4b(2^\ast2^\ast)} (\delta m_{2})^2+2 I_{4s} M_{2} I_{2} \\
& +2 \delta m_{2} \hat{B}_{2} \delta\hat{m}_{2^{\prime \ast}} M_{2^\ast}+(\hat{B}_{2})^2 M_{2^\ast} \delta\hat{m}_{2^{\prime \prime}}+(\hat{B}_{2})^2 M_{2} \hat{B}_{2^{\prime \prime}}
\end{align*}
\begin{align*}
M_{31}&=
\Delta M_{31}+2 \hat{L}_{6H3} M_{2}
\end{align*}
\begin{align*}
M_{32}&=
\Delta M_{32}+\hat{L}_{6H2} M_{2}+\hat{L}_{6G3} M_{2}
\end{align*}
\begin{align*}
M_{33}&=
\Delta M_{33}+2 \hat{L}_{6G3} M_{2}
\end{align*}
\begin{align*}
M_{34}&=
\Delta M_{34}+\hat{L}_{6H1} M_{2}+\hat{L}_{6C3} M_{2}+\hat{L}_{4x} M_{4a}-2 \hat{L}_{4x} \hat{L}_{2} M_{2}
\end{align*}
\begin{align*}
M_{35}&=
\Delta M_{35}+M_{6H} \hat{L}_{2}+\hat{L}_{6E3} M_{2}+\hat{L}_{6G4} M_{2}-2 \hat{L}_{4x} \hat{L}_{2} M_{2}
\end{align*}
\begin{align*}
M_{36}&=
\Delta M_{36}+M_{6G} \hat{L}_{2}+\hat{L}_{6B3} M_{2}+\hat{L}_{6G5} M_{2}-\hat{L}_{4c} \hat{L}_{2} M_{2} \\
& +\hat{L}_{4l} M_{4a}-3 \hat{L}_{4l} \hat{L}_{2} M_{2}-M_{4a} (\hat{L}_{2})^2+2 (\hat{L}_{2})^3 M_{2}
\end{align*}
\begin{align*}
M_{37}&=
\Delta M_{37}+2 \hat{L}_{6G2} M_{2}
\end{align*}
\begin{align*}
M_{38}&=
\Delta M_{38}+\Delta M_{6H} I_{2}+\delta\hat{m}_{6H} M_{2^\ast}+\hat{B}_{6H} M_{2}+\Delta \delta m_{6H} M_{2^\ast}[I] \\
& +2 \hat{L}_{4x} M_{4b}-2 \hat{L}_{4x} \delta m_{2} M_{2^\ast}-2 \hat{L}_{4x} \hat{B}_{2} M_{2}
\end{align*}
\begin{align*}
M_{39}&=
\Delta M_{39}+M_{6G} \hat{L}_{2}+\hat{L}_{6G1} M_{2}+\hat{L}_{6C2} M_{2}+\hat{L}_{4c} M_{4a} \\
& -3 \hat{L}_{4c} \hat{L}_{2} M_{2}-\hat{L}_{4l} \hat{L}_{2} M_{2}-M_{4a} (\hat{L}_{2})^2+2 (\hat{L}_{2})^3 M_{2}
\end{align*}
\begin{align*}
M_{40}&=
\Delta M_{40}+\Delta M_{6G} I_{2}+M_{6C} \hat{L}_{2}+\delta\hat{m}_{6G} M_{2^\ast}+\hat{B}_{6G} M_{2} \\
& +\Delta \delta m_{6G} M_{2^\ast}[I]-\delta\hat{m}_{4a} \hat{L}_{2} M_{2^\ast}-\hat{B}_{4a} \hat{L}_{2} M_{2}+\hat{L}_{4c} M_{4b}-\hat{L}_{4c} \delta m_{2} M_{2^\ast} \\
& -\hat{L}_{4c} \hat{B}_{2} M_{2}+\hat{L}_{4l} M_{4b}-\hat{L}_{4l} \delta m_{2} M_{2^\ast}-\hat{L}_{4l} \hat{B}_{2} M_{2}-2 M_{4b} (\hat{L}_{2})^2 \\
& +2 \delta m_{2} (\hat{L}_{2})^2 M_{2^\ast}+2 \hat{B}_{2} (\hat{L}_{2})^2 M_{2}
\end{align*}
\begin{align*}
M_{41}&=
\Delta M_{41}+2 M_{6E} \hat{L}_{2}+2 \hat{L}_{6C1} M_{2}+\delta\hat{m}_{4a} M_{4a(2^\ast)}+\hat{B}_{4a} M_{4a} \\
& -2 \hat{B}_{4a} M_{2} \hat{L}_{2^\prime}-4 \hat{L}_{4s} \hat{L}_{2} M_{2}+\Delta \delta m_{4a} I_{4a(2^\ast)}-2 M_{4a(2^\ast)} \delta m_{2} \hat{L}_{2}-2 M_{4a} \hat{B}_{2} \hat{L}_{2} \\
& +\Delta M_{4a} I_{4x}+4 \hat{B}_{2} \hat{L}_{2} M_{2} \hat{L}_{2^\prime}
\end{align*}
\begin{align*}
M_{42}&=
\Delta M_{42}+\Delta M_{6C} I_{2}+2 M_{6B} \hat{L}_{2}+\delta\hat{m}_{6C} M_{2^\ast}+\hat{B}_{6C} M_{2} \\
& +\Delta \delta m_{6C} M_{2^\ast}[I]+\delta\hat{m}_{4a} M_{4b(2^\ast)}-\delta\hat{m}_{4a} \delta\hat{m}_{2^\ast} M_{2^\ast}-2 \delta\hat{m}_{4b} \hat{L}_{2} M_{2^\ast}+\hat{B}_{4a} M_{4b} \\
& -\hat{B}_{4a} M_{2^\ast} \delta\hat{m}_{2^\prime}-\hat{B}_{4a} M_{2} \hat{B}_{2^\prime}-2 \hat{B}_{4b} \hat{L}_{2} M_{2}+\Delta \delta m_{4a} I_{4b(2^\ast)}-2 M_{4b(2^\ast)} \delta m_{2} \hat{L}_{2} \\
& -2 M_{4b} \hat{B}_{2} \hat{L}_{2}+\Delta M_{4a} I_{4l}+\Delta M_{4a} (I_{2})^2+2 \delta m_{2} \hat{L}_{2} \delta\hat{m}_{2^\ast} M_{2^\ast}+2 \hat{B}_{2} \hat{L}_{2} M_{2^\ast} \delta\hat{m}_{2^\prime} \\
& +2 \hat{B}_{2} \hat{L}_{2} M_{2} \hat{B}_{2^\prime}
\end{align*}
\begin{align*}
M_{43}&=
\Delta M_{43}+M_{6H} \hat{B}_{2}+M_{6H(3^\ast)} \delta m_{2}+2 \hat{L}_{6E2} M_{2}+I_{6H3} M_{2} \\
& -2 \hat{L}_{4x(2^\prime )} \hat{B}_{2} M_{2}
\end{align*}
\begin{align*}
M_{44}&=
\Delta M_{44}+M_{6G} \hat{B}_{2}+M_{6G(3^\ast)} \delta m_{2}+\hat{L}_{6E1} M_{2}+\hat{L}_{6B2} M_{2} \\
& +I_{6G3} M_{2}+\hat{L}_{4s} M_{4a}-2 \hat{L}_{4s} \hat{L}_{2} M_{2}-\hat{L}_{4c(2^\prime )} \hat{B}_{2} M_{2}-\hat{L}_{4l(2^\prime )} \hat{B}_{2} M_{2} \\
& -M_{4a} \hat{B}_{2} \hat{L}_{2^\prime}+2 \hat{B}_{2} \hat{L}_{2} M_{2} \hat{L}_{2^\prime}
\end{align*}
\begin{align*}
M_{45}&=
\Delta M_{45}+\Delta M_{6E} I_{2}+M_{6C} \hat{B}_{2}+M_{6C(3^\ast)} \delta m_{2}+\delta\hat{m}_{6E} M_{2^\ast} \\
& +I_{6C3} M_{2}+\hat{B}_{6E} M_{2}+\Delta \delta m_{6E} M_{2^\ast}[I]+2 \hat{L}_{4s} M_{4b}-2 \hat{L}_{4s} \delta m_{2} M_{2^\ast} \\
& -2 \hat{L}_{4s} \hat{B}_{2} M_{2}-\delta\hat{m}_{4a(2^\ast)} \delta m_{2} M_{2^\ast}-\delta\hat{m}_{4a(2^\prime )} \hat{B}_{2} M_{2^\ast}-\hat{B}_{4a(2^\prime )} \hat{B}_{2} M_{2}-2 M_{4b} \hat{B}_{2} \hat{L}_{2^\prime} \\
& +I_{4x} M_{2} I_{2}+2 \delta m_{2} \hat{B}_{2} M_{2^\ast} \hat{L}_{2^\prime}+2 (\hat{B}_{2})^2 M_{2} \hat{L}_{2^\prime}
\end{align*}
\begin{align*}
M_{46}&=
\Delta M_{46}+M_{6E} \hat{B}_{2}+M_{6E(3^\ast)} \delta m_{2}+2 \hat{L}_{6B1} M_{2}+I_{6E3} M_{2} \\
& +\delta\hat{m}_{4b} M_{4a(2^\ast)}+\hat{B}_{4b} M_{4a}-2 \hat{B}_{4b} M_{2} \hat{L}_{2^\prime}+\Delta \delta m_{4b} I_{4a(2^\ast)}-2 \hat{L}_{4s(2^\prime )} \hat{B}_{2} M_{2} \\
& -M_{4a(2^\ast)} \delta m_{2} \delta\hat{m}_{2^\ast}-M_{4a(2^\ast)} \hat{B}_{2} \delta\hat{m}_{2^\prime}-M_{4a} \hat{B}_{2} \hat{B}_{2^\prime}+\Delta M_{4b} I_{4x}+I_{4x} M_{2} I_{2} \\
& +2 \hat{B}_{2} M_{2} \hat{B}_{2^\prime} \hat{L}_{2^\prime}
\end{align*}
\begin{align*}
M_{47}&=
\Delta M_{47}+\Delta M_{6B} I_{2}+M_{6B} \hat{B}_{2}+M_{6B(3^\ast)} \delta m_{2}+\delta\hat{m}_{6B} M_{2^\ast} \\
& +I_{6B3} M_{2}+\hat{B}_{6B} M_{2}+\Delta \delta m_{6B} M_{2^\ast}[I]+\delta\hat{m}_{4b} M_{4b(2^\ast)}-\delta\hat{m}_{4b} \delta\hat{m}_{2^\ast} M_{2^\ast} \\
& +\hat{B}_{4b} M_{4b}-\hat{B}_{4b} M_{2^\ast} \delta\hat{m}_{2^\prime}-\hat{B}_{4b} M_{2} \hat{B}_{2^\prime}-\delta\hat{m}_{4b(2^\ast)} \delta m_{2} M_{2^\ast}-\delta\hat{m}_{4b(2^\prime )} \hat{B}_{2} M_{2^\ast} \\
& -\hat{B}_{4b(2^\prime )} \hat{B}_{2} M_{2}+\Delta \delta m_{4b} I_{4b(2^\ast)}-M_{4b(2^\ast)} \delta m_{2} \delta\hat{m}_{2^\ast}-M_{4b(2^\ast)} \hat{B}_{2} \delta\hat{m}_{2^\prime}-M_{4b} \hat{B}_{2} \hat{B}_{2^\prime} \\
& +\Delta M_{4b} I_{4l}+\Delta M_{4b} (I_{2})^2+2 I_{4l} M_{2} I_{2}+\delta m_{2} (\delta\hat{m}_{2^\ast})^2 M_{2^\ast}+\hat{B}_{2} \delta\hat{m}_{2^\ast} M_{2^\ast} \delta\hat{m}_{2^\prime} \\
& +\hat{B}_{2} M_{2^\ast} \delta\hat{m}_{2^\prime} \hat{B}_{2^\prime}+\hat{B}_{2} M_{2} (\hat{B}_{2^\prime})^2+M_{2} (I_{2})^3
\end{align*}

\subsection{Divergence Structure of Quantities of Sixth- and Lower-Orders } 
\label{renormalization6} 

Substitution of $M_i$ into $a_i$ expresses the latter in terms of
finite quantity $\Delta M_i$.  However, $a_i$ still contains
numerous (divergent) quantities of lower orders.
To obtain a finite result in the end we need the information
on the divergence structure of these quantities,
which are listed in the following.
The suffix $6\alpha$, ($\alpha$ = A,..., H), in $L_{6\alpha}$,
$B_{6\alpha}$, $\delta m_{6\alpha}$ refers to the diagrams
similar to those of Fig. \ref{vertex6}, before $\Pi_2$ insertion is made.

\vspace*{6mm}
\noindent
{\bf Vertex renormalization constants of sixth-order}
\begin{align*}
L_{6A1}&=
I_{6A1}+\Delta L_{6A1}+2 \delta m_{2} L_{4s(1^\ast)}+2 \hat{B}_{2} \tilde{L}_{4s(1^\prime )}-\delta m_{2} (\delta m_{2} L_{2^{\ast\ast\dagger}} \\
& +\hat{B}_{2} L_{2^{\prime \ast}})-\hat{B}_{2} (\delta m_{2} L_{2^{\prime \ast}}+\hat{B}_{2} \tilde{L}_{2^{\prime \prime}})+\hat{L}_{6A1}
\end{align*}
\begin{align*}
L_{6A2}&=
I_{6A2}+\Delta L_{6A2}+\hat{L}_{2} \tilde{L}_{4s}+\delta m_{2} L_{4l(1^\ast)}+\hat{B}_{2} \tilde{L}_{4l(1^\prime )} \\
& -\hat{L}_{2} (\delta m_{2} L_{2^\ast}+\hat{B}_{2} \tilde{L}_{2^{\prime}})+I_{4s} \tilde{L}_{2}+\hat{L}_{6A2}
\end{align*}
\begin{align*}
L_{6A3}&=
I_{6A3}+\Delta L_{6A3}+2 (\delta m_{2} L_{4s(1^\ast)}+\hat{B}_{2} \tilde{L}_{4s(1^\prime )})-\delta m_{2} (\delta m_{2} L_{2^{\ast\dagger\ast}} \\
& +\hat{B}_{2} L_{2^{\prime \ast}})-\hat{B}_{2} (\delta m_{2} L_{2^{\prime \ast}}+\hat{B}_{2} \tilde{L}_{2^{\prime \prime}})+\hat{L}_{6A3}
\end{align*}
\begin{align*}
L_{6B1}&=
I_{6B1}+\Delta L_{6B1}+\delta m_{2} L_{4s(2^\ast)}+\hat{B}_{2} \tilde{L}_{4s(2^\prime )}+\delta\hat{m}_{4b} L_{2^\ast} \\
& +\hat{B}_{4b} \tilde{L}_{2^{\prime}}-\delta m_{2} \delta\hat{m}_{2^\ast} L_{2^\ast}-\hat{B}_{2} (\delta\hat{m}_{2^{\prime}} L_{2^\ast}+\hat{B}_{2^{\prime}} \tilde{L}_{2^{\prime}})+1/2 J_{6B} 
 +\hat{L}_{6B1}
\end{align*}
\begin{align*}
L_{6B2}&=
I_{6B2}+\Delta L_{6B2}+\delta m_{2} L_{4l(2^\ast)}+\hat{B}_{2} \tilde{L}_{4l(2^\prime )}+\hat{L}_{4s} \tilde{L}_{2} \\
& -\hat{B}_{2} \hat{L}_{2^{\prime}} \tilde{L}_{2}+I_{2} \tilde{L}_{4s}-I_{2} (\delta m_{2} L_{2^\ast}+\hat{B}_{2} \tilde{L}_{2^{\prime}})+\hat{L}_{6B2}
\end{align*}
\begin{align*}
L_{6B3}&=
I_{6B3}+\Delta L_{6B3}+\hat{L}_{2} \tilde{L}_{4l}+\hat{L}_{4l} \tilde{L}_{2}-(\hat{L}_{2})^2 \tilde{L}_{2} \\
& +I_{2} \tilde{L}_{4l}-I_{2} \hat{L}_{2} \tilde{L}_{2}+I_{4l} \tilde{L}_{2}+\hat{L}_{6B3}
\end{align*}
\begin{align*}
L_{6C1}&=
I_{6C1}+\Delta L_{6C1}+2 \hat{L}_{2} \tilde{L}_{4s}+\delta\hat{m}_{4a} L_{2^\ast}+\hat{B}_{4a} \tilde{L}_{2^{\prime}} \\
& -2 \hat{L}_{2} (\delta m_{2} L_{2^\ast}+\hat{B}_{2} \tilde{L}_{2^{\prime}})+1/2 J_{6C}+\hat{L}_{6C1}
\end{align*}
\begin{align*}
L_{6C2}&=
I_{6C2}+\Delta L_{6C2}+\hat{L}_{2} \tilde{L}_{4l}+\hat{L}_{4c} \tilde{L}_{2}-(\hat{L}_{2})^2 \tilde{L}_{2} \\
& +I_{2} \tilde{L}_{4c}-I_{2} \hat{L}_{2} \tilde{L}_{2}+\hat{L}_{6C2}
\end{align*}
\begin{align*}
L_{6C3}&=
I_{6C3}+\Delta L_{6C3}+\hat{L}_{4x} \tilde{L}_{2}+I_{2} \tilde{L}_{4x}+\hat{L}_{6C3}
\end{align*}
\begin{align*}
L_{6D1}&=
I_{6D1}+\Delta L_{6D1}+\delta m_{2} L_{4c(1^\ast)}+\hat{B}_{2} \tilde{L}_{4c(1^\prime )}+\hat{L}_{2} \tilde{L}_{4s} \\
& -\hat{L}_{2} (\delta m_{2} L_{2^\ast}+\hat{B}_{2} \tilde{L}_{2^{\prime}})+\hat{L}_{6D1}
\end{align*}
\begin{align*}
L_{6D2}&=
I_{6D2}+\Delta L_{6D2}+\hat{L}_{2} \tilde{L}_{4c}+\hat{L}_{2} \tilde{L}_{4l}-(\hat{L}_{2})^2 \tilde{L}_{2} 
 +I_{4c} \tilde{L}_{2}+\hat{L}_{6D2}
\end{align*}
\begin{align*}
L_{6D3}&=
I_{6D3}+\Delta L_{6D3}+\delta m_{2} L_{4c(1^\ast)}+\hat{B}_{2} \tilde{L}_{4c(1^\prime )}+\hat{L}_{2} \tilde{L}_{4s} \\
& -\hat{L}_{2} (\delta m_{2} L_{2^\ast}+\hat{B}_{2} \tilde{L}_{2^{\prime}})+\hat{L}_{6D3}
\end{align*}
\begin{align*}
L_{6D4}&=
I_{6D4}+\Delta L_{6D4}+\delta m_{2} L_{4x(1^\ast)}+\hat{B}_{2} \tilde{L}_{4x(1^\prime )}+\hat{L}_{6D4}
\end{align*}
\begin{align*}
L_{6D5}&=
I_{6D5}+\Delta L_{6D5}+\delta m_{2} L_{4c(3^\ast)}+\hat{B}_{2} \tilde{L}_{4c(3^\prime )}+\hat{L}_{4s} \tilde{L}_{2} \\
& -\hat{B}_{2} \hat{L}_{2^{\prime}} \tilde{L}_{2}+I_{2} \Delta L_{4s}+\hat{L}_{6D5}
\end{align*}
\begin{align*}
L_{6E1}&=
I_{6E1}+\Delta L_{6E1}+\delta m_{2} L_{4c(2^\ast)}+\hat{B}_{2} \tilde{L}_{4c(2^\prime )}+\hat{L}_{4s} \tilde{L}_{2} \\
& -\hat{B}_{2} \hat{L}_{2^{\prime}} \tilde{L}_{2}+I_{2} \Delta L_{4s}+\hat{L}_{6E1}
\end{align*}
\begin{align*}
L_{6E2}&=
I_{6E2}+\Delta L_{6E2}+\delta m_{2} L_{4x(2^\ast)}+\hat{B}_{2} \tilde{L}_{4x(2^\prime )}+\hat{L}_{6E2}
\end{align*}
\begin{align*}
L_{6E3}&=
I_{6E3}+\Delta L_{6E3}+I_{4x} \tilde{L}_{2}+\hat{L}_{2} \tilde{L}_{4x}+\hat{L}_{6E3}
\end{align*}
\begin{align*}
L_{6F1}&=
I_{6F1}+\Delta L_{6F1}+\hat{L}_{2} \tilde{L}_{4c}+\hat{L}_{4c} \tilde{L}_{2}-(\hat{L}_{2})^2 \tilde{L}_{2} 
 +I_{2} \Delta L_{4c}+\hat{L}_{6F1}
\end{align*}
\begin{align*}
L_{6F2}&=
I_{6F2}+\Delta L_{6F2}+\hat{L}_{2} \tilde{L}_{4x}+\hat{L}_{6F2}
\end{align*}
\begin{align*}
L_{6F3}&=
I_{6F3}+\Delta L_{6F3}+2 \hat{L}_{2} \tilde{L}_{4c}-(\hat{L}_{2})^2 \tilde{L}_{2}+\hat{L}_{6F3}
\end{align*}
\begin{align*}
L_{6G1}&=
I_{6G1}+\Delta L_{6G1}+\hat{L}_{2} \tilde{L}_{4c}+\hat{L}_{4c} \tilde{L}_{2}-(\hat{L}_{2})^2 \tilde{L}_{2} 
 +I_{2} \Delta L_{4c}+\hat{L}_{6G1}
\end{align*}
\begin{align*}
L_{6G2}&=
I_{6G2}+\Delta L_{6G2}+\hat{L}_{6G2}
\end{align*}
\begin{align*}
L_{6G3}&=
I_{6G3}+\Delta L_{6G3}+\hat{L}_{6G3}
\end{align*}
\begin{align*}
L_{6G4}&=
I_{6G4}+\Delta L_{6G4}+\hat{L}_{2} \tilde{L}_{4x}+\hat{L}_{6G4}
\end{align*}
\begin{align*}
L_{6G5}&=
I_{6G5}+\Delta L_{6G5}+\hat{L}_{2} \tilde{L}_{4c}+\hat{L}_{4l} \tilde{L}_{2}-(\hat{L}_{2})^2 \tilde{L}_{2} 
 +I_{2} \Delta L_{4l}+\hat{L}_{6G5}
\end{align*}
\begin{align*}
L_{6H1}&=
I_{6H1}+\Delta L_{6H1}+\hat{L}_{4x} \tilde{L}_{2}+I_{2} \Delta L_{4x}+\hat{L}_{6H1}
\end{align*}
\begin{align*}
L_{6H2}&=
I_{6H2}+\Delta L_{6H2}+\hat{L}_{6H2}
\end{align*}
\begin{align*}
L_{6H3}&=
I_{6H3}+\Delta L_{6H3}+\hat{L}_{6H3}
\end{align*}

\vspace*{6mm}
\noindent
{\bf Wave-function renormalization constants of sixth-order}
\begin{align*}
B_{6A}&=
-\sum_{\alpha=1}^{5} I_{6A\alpha}+\Delta B_{6A}+2 (\delta m_{2} B_{4b(1^\ast)}+\hat{B}_{2} \tilde{B}_{4b(1^\prime )}) \\
& -\delta m_{2} (\delta m_{2} B_{2^{\ast\ast}}+\hat{B}_{2} B_{2^{\prime \ast}})-\hat{B}_{2} (\delta m_{2} B_{2^{\prime \ast}}+\hat{B}_{2} \tilde{B}_{2^{\prime \prime}})+2 I_{4s} \tilde{B}_{2} 
 +\hat{B}_{6A}
\end{align*}
\begin{align*}
B_{6B}&=
-\sum_{\alpha=1}^5 I_{6B\alpha}+\Delta B_{6B}+\delta m_{2} B_{4b(2^\ast)}+\hat{B}_{2} \tilde{B}_{4b(2^\prime )} \\
& +\delta\hat{m}_{4b} B_{2^\ast}+\hat{B}_{4b} \tilde{B}_{2^{\prime}}-\delta m_{2} \delta\hat{m}_{2^\ast} B_{2^\ast}-\hat{B}_{2} (\delta\hat{m}_{2^{\prime}} B_{2^\ast}+\hat{B}_{2^{\prime}} \tilde{B}_{2^{\prime}}) \\
& +I_{2} \tilde{B}_{4b}-I_{2} (\delta m_{2} B_{2^\ast}+\hat{B}_{2} \tilde{B}_{2^{\prime}})+I_{4l} \tilde{B}_{2}+B_{2^\ast}[I] \Delta \delta m_{4b} 
 -J_{6B}+\hat{B}_{6B}
\end{align*}
\begin{align*}
B_{6C}&=
-\sum_{\alpha=1}^5 I_{6C\alpha}+\Delta B_{6C}+2 \hat{L}_{2} \tilde{B}_{4b}+\delta\hat{m}_{4a} B_{2^\ast} 
 +\hat{B}_{4a} \tilde{B}_{2^{\prime}}
\\
&-2 \hat{L}_{2} (\delta m_{2} B_{2^\ast}+\hat{B}_{2} \tilde{B}_{2^{\prime}})+I_{2} \tilde{B}_{4a}-2 I_{2} \hat{L}_{2} \tilde{B}_{2} 
 +B_{2^\ast}[I] \Delta \delta m_{4a}-J_{6C}+\hat{B}_{6C}
\end{align*}
\begin{align*}
B_{6D}&=
-\sum_{\alpha=1}^5 I_{6D\alpha}+\Delta B_{6D}+\delta m_{2} B_{4a(1^\ast)}+\hat{B}_{2} \tilde{B}_{4a(1^\prime )} 
 +\hat{L}_{2} \tilde{B}_{4b} +\hat{L}_{4s} \tilde{B}_{2} \\
& -\hat{L}_{2} (\delta m_{2} B_{2^\ast}+\hat{B}_{2} \tilde{B}_{2^{\prime}})-\hat{B}_{2} \hat{L}_{2^{\prime}} \tilde{B}_{2} 
 +I_{4c} \tilde{B}_{2}+\tilde{B}_{2} \Delta L_{4s}+\hat{B}_{6D}
\end{align*}
\begin{align*}
B_{6E}&=
-\sum_{\alpha=1}^5 I_{6E\alpha}+\Delta B_{6E}+\delta m_{2} B_{4a(2^\ast)}+\hat{B}_{2} \tilde{B}_{4a(2^\prime )} \\
& +2 \hat{L}_{4s} \tilde{B}_{2}-2 \hat{B}_{2} \hat{L}_{2^{\prime}} \tilde{B}_{2}+I_{4x} \tilde{B}_{2}+2 \tilde{B}_{2} \Delta L_{4s}+\hat{B}_{6E}
\end{align*}
\begin{align*}
B_{6F}&=
-\sum_{\alpha=1}^5 I_{6F\alpha}+\Delta B_{6F}+2 \hat{L}_{2} \tilde{B}_{4a}+2 \hat{L}_{4c} \tilde{B}_{2} 
 -3 (\hat{L}_{2})^2 \tilde{B}_{2}+2 \tilde{B}_{2} \Delta L_{4c}+\hat{B}_{6F}
\end{align*}
\begin{align*}
B_{6G}&=
-\sum_{\alpha=1}^5 I_{6G\alpha}+\Delta B_{6G}+\hat{L}_{2} \tilde{B}_{4a}+\hat{L}_{4c} \tilde{B}_{2} \\
& +\hat{L}_{4l} \tilde{B}_{2}-2 (\hat{L}_{2})^2 \tilde{B}_{2}+\tilde{B}_{2} (\Delta L_{4l}+\Delta L_{4c})+\hat{B}_{6G}
\end{align*}
\begin{align*}
B_{6H}&=
-\sum_{\alpha=1}^5 I_{6H\alpha}+\Delta B_{6H}+2 \hat{L}_{4x} \tilde{B}_{2}+2 \tilde{B}_{2} \Delta L_{4x} 
 +\hat{B}_{6H}
\end{align*}

\vspace*{6mm}
\noindent
{\bf Mass renormalization constants of sixth-order}
%
\begin{align*}
\delta m_{6A}&=
\Delta \delta m_{6A}+2 (\delta m_{2} \delta\tilde{m}_{4b(1^\ast)}+\hat{B}_{2} \delta\tilde{m}_{4b(1^\prime )})-\delta m_{2} (\delta m_{2} \delta m_{2^{\ast\ast}}+\hat{B}_{2} \delta\tilde{m}_{2^{\prime \ast}}) \\
& -\hat{B}_{2} (\delta m_{2} \delta\tilde{m}_{2^{\prime \ast}}+\hat{B}_{2} \delta\tilde{m}_{2^{\prime \prime}})+\delta\hat{m}_{6A}
\end{align*}
\begin{align*}
\delta m_{6B}&=
\Delta \delta m_{6B}+\delta m_{2} \delta\tilde{m}_{4b(2^\ast)}+\hat{B}_{2} \delta\tilde{m}_{4b(2^\prime )}+\delta\hat{m}_{4b} \delta\tilde{m}_{2^\ast}+\hat{B}_{4b} \delta\tilde{m}_{2^{\prime}} \\
& -\delta m_{2} \delta\hat{m}_{2^\ast} \delta\tilde{m}_{2^\ast}-\hat{B}_{2} (\delta\hat{m}_{2^{\prime}} \delta\tilde{m}_{2^\ast}+\hat{B}_{2^{\prime}} \delta\tilde{m}_{2^{\prime}})+I_{2} \delta\tilde{m}_{4b}-I_{2} (\delta m_{2} \delta\tilde{m}_{2^\ast} \\
& +\hat{B}_{2} \delta\tilde{m}_{2^{\prime}})+\delta\hat{m}_{6B}
\end{align*}
\begin{align*}
\delta m_{6C}&=
\Delta \delta m_{6C}+2 \hat{L}_{2} \delta\tilde{m}_{4b}+\delta\hat{m}_{4a} \delta\tilde{m}_{2^\ast}+\hat{B}_{4a} \delta\tilde{m}_{2^{\prime}}-2 \hat{L}_{2} (\delta m_{2} \delta\tilde{m}_{2^\ast} \\
& +\hat{B}_{2} \delta\tilde{m}_{2^{\prime}})+I_{2} \delta\tilde{m}_{4a}+\delta\hat{m}_{6C}
\end{align*}
\begin{align*}
\delta m_{6D}&=
\Delta \delta m_{6D}+\delta m_{2} \delta\tilde{m}_{4a(1^\ast)}+\hat{B}_{2} \delta\tilde{m}_{4a(1^\prime )}+\hat{L}_{2} \delta\tilde{m}_{4b}-\hat{L}_{2} (\delta m_{2} \delta\tilde{m}_{2^\ast} \\
& +\hat{B}_{2} \delta\tilde{m}_{2^{\prime}})+\delta\hat{m}_{6D}
\end{align*}
\begin{align*}
\delta m_{6E}&=
\Delta \delta m_{6E}+\delta m_{2} \delta\tilde{m}_{4a(2^\ast)}+\hat{B}_{2} \delta\tilde{m}_{4a(2^\prime )}+\delta\hat{m}_{6E}
\end{align*}
\begin{align*}
\delta m_{6F}&=
\Delta \delta m_{6F}+2 \hat{L}_{2} \delta\tilde{m}_{4a}+\delta\hat{m}_{6F}
\end{align*}
\begin{align*}
\delta m_{6G}&=
\Delta \delta m_{6G}+\hat{L}_{2} \delta\tilde{m}_{4a}+\delta\hat{m}_{6G}
\end{align*}
\begin{align*}
\delta m_{6H}&=
\Delta \delta m_{6H}+\delta\hat{m}_{6H}
\end{align*}
%

\vspace*{6mm}
\noindent
{\bf Renormalization Scheme for Sixth-order Moment}
\begin{align*}
a_{6A}&=
M_{6A}-2 \delta m_{2} M_{4b(1^\ast)}-2 B_{2} M_{4b}+\delta m_{2} (\delta m_{2} M_{2^{\ast\ast}}+B_{2} M_{2^\ast}) \\
& +B_{2} (\delta m_{2} M_{2^\ast}+B_{2} M_{2})
\end{align*}
\begin{align*}
a_{6B}&=
M_{6B}-\delta m_{2} M_{4b(2^\ast)}-B_{2} M_{4b}-\delta m_{4b} M_{2^\ast}-B_{4b} M_{2} \\
& +\delta m_{2} (\delta m_{2^\ast} M_{2^\ast}+B_{2^\ast} M_{2})+B_{2} (\delta m_{2} M_{2^\ast}+B_{2} M_{2})
\end{align*}
\begin{align*}
a_{6C}&=
M_{6C}-2 L_{2} M_{4b}-\delta m_{4a} M_{2^\ast}-B_{4a} M_{2}+2 L_{2} (\delta m_{2} M_{2^\ast} 
 +B_{2} M_{2})
\end{align*}
\begin{align*}
a_{6D}&=
M_{6D}-L_{4s} M_{2}-\delta m_{2} M_{4a(1^\ast)}-B_{2} M_{4a}-L_{2} M_{4b} \\
& +B_{2} L_{2} M_{2}+L_{2} (\delta m_{2} M_{2^\ast}+B_{2} M_{2})+L_{2^\ast} \delta m_{2} M_{2}
\end{align*}
\begin{align*}
a_{6E}&=
M_{6E}-2 L_{4s} M_{2}-\delta m_{2} M_{4a(2^\ast)}-B_{2} M_{4a}+2 L_{2^\ast} \delta m_{2} M_{2} 
 +2 L_{2} B_{2} M_{2}
\end{align*}
\begin{align*}
a_{6F}&=
M_{6F}-2 L_{4c} M_{2}-2 L_{2} M_{4a}+3 L_{2} L_{2} M_{2}
\end{align*}
\begin{align*}
a_{6G}&=
M_{6G}-L_{4c} M_{2}-L_{4l} M_{2}-L_{2} M_{4a}+2 L_{2} L_{2} M_{2}
\end{align*}
\begin{align*}
a_{6H}&=
M_{6H}-2 L_{4x} M_{2}
\end{align*}
\begin{align*}
M_{6A}&=
\Delta M_{6A}+2 \delta m_{2} M_{4b(1^\ast)}+2 \hat{B}_{2} M_{4b}-\delta m_{2} (\delta m_{2} M_{2^{\ast\ast}}+\hat{B}_{2} M_{2^\ast}) \\
& -\hat{B}_{2} (\delta m_{2} M_{2^\ast}+\hat{B}_{2} M_{2})+2 I_{4s} M_{2}
\end{align*}
\begin{align*}
M_{6B}&=
\Delta M_{6B}+\delta m_{2} M_{4b(2^\ast)}+\hat{B}_{2} M_{4b}+\delta\hat{m}_{4b} M_{2^\ast}+\hat{B}_{4b} M_{2} \\
& -\delta m_{2} \delta\hat{m}_{2^\ast} M_{2^\ast}-\hat{B}_{2} (\delta\hat{m}_{2^{\prime}} M_{2^\ast}+\hat{B}_{2^{\prime}} M_{2})+I_{2} \Delta M_{4b}+M_{2^\ast}[I] \Delta \delta m_{4b} \\
& +\{ I_{4l}+(I_{2})^2 \} M_{2}
\end{align*}
\begin{align*}
M_{6C}&=
\Delta M_{6C}+2 \hat{L}_{2} M_{4b}+\delta\hat{m}_{4a} M_{2^\ast}+\hat{B}_{4a} M_{2}-2 \hat{L}_{2} (\delta m_{2} M_{2^\ast} \\
& +\hat{B}_{2} M_{2})+I_{2} M_{4a}-2 I_{2} \hat{L}_{2} M_{2}+M_{2^\ast}[I] \Delta \delta m_{4a}
\end{align*}
\begin{align*}
M_{6D}&=
\Delta M_{6D}+\hat{L}_{4s} M_{2}+\delta m_{2} M_{4a(1^\ast)}+\hat{B}_{2} M_{4a}+\hat{L}_{2} M_{4b} \\
& -\hat{B}_{2} \hat{L}_{2^{\prime}} M_{2}-\hat{L}_{2} (\delta m_{2} M_{2^\ast}+\hat{B}_{2} M_{2})+I_{4c} M_{2}
\end{align*}
\begin{align*}
M_{6E}&=
\Delta M_{6E}+2 \hat{L}_{4s} M_{2}+\delta m_{2} M_{4a(2^\ast)}+\hat{B}_{2} M_{4a}-2 \hat{L}_{2^{\prime}} \hat{B}_{2} M_{2}  +I_{4x} M_{2}
\end{align*}
\begin{align*}
M_{6F}&=
\Delta M_{6F}+2 \hat{L}_{4c} M_{2}+2 \hat{L}_{2} M_{4a}-3 \hat{L}_{2} \hat{L}_{2} M_{2}
\end{align*}
\begin{align*}
M_{6G}&=
\Delta M_{6G}+\hat{L}_{4c} M_{2}+\hat{L}_{4l} M_{2}+\hat{L}_{2} M_{4a}-2 \hat{L}_{2} \hat{L}_{2} M_{2}
\end{align*}
\begin{align*}
M_{6H}&=
\Delta M_{6H}+2 \hat{L}_{4x} M_{2}
\end{align*}

\vspace*{6mm}
\noindent
{\bf Various Fourth-Order Quantities}
%
\begin{align*}
\delta m_{4a}&=
\delta\hat{m}_{4a}+\Delta \delta m_{4a}
\end{align*}
\begin{align*}
\delta m_{4b}&=
\delta\hat{m}_{4b}+\Delta \delta m_{4b}+\delta m_{2} \delta\tilde{m}_{2^\ast}+\hat{B}_{2} \delta\tilde{m}_{2^{\prime}}
\end{align*}
\begin{align*}
\tilde{L}_{4c((1^\prime )^\prime )}&=
\tilde{L}_{4c(1^\prime )}
\end{align*}
\begin{align*}
\tilde{L}_{4s((1^\prime )^\prime )}&=
\tilde{L}_{4s(1^\prime )}
\end{align*}
\begin{align*}
L_{4x}&=
\hat{L}_{4x}+I_{4x}+\Delta L_{4x}
\end{align*}
\begin{align*}
\tilde{L}_{4x}&=
I_{4x}+\Delta L_{4x}
\end{align*}
\begin{align*}
L_{4c}&=
\hat{L}_{4c}+I_{4c}+\Delta L_{4c}+\hat{L}_{2} \tilde{L}_{2}
\end{align*}
\begin{align*}
\tilde{L}_{4c}&=
I_{4c}+\Delta L_{4c}+\hat{L}_{2} \tilde{L}_{2}
\end{align*}
\begin{align*}
B_{4a}&=
\hat{B}_{4a}-I_{4x}+\Delta B_{4a}+2 \hat{L}_{2} \tilde{B}_{2}-2 I_{4c}
\end{align*}
\begin{align*}
\tilde{B}_{4a}&=
-I_{4x}+\Delta B_{4a}+2 \hat{L}_{2} \tilde{B}_{2}-2 I_{4c}
\end{align*}
\begin{align*}
L_{4l}&=
\hat{L}_{4l}+I_{4l}+(I_{2})^2+\Delta L_{4l}+\tilde{L}_{2} \hat{L}_{2}
\end{align*}
\begin{align*}
\tilde{L}_{4l}&=
I_{4l}+(I_{2})^2+\Delta L_{4l}+\tilde{L}_{2} \hat{L}_{2}
\end{align*}
\begin{align*}
L_{4s}&=
\hat{L}_{4s}+I_{4s}+\Delta L_{4s}+\delta m_{2} L_{2^\ast}+\hat{B}_{2} \tilde{L}_{2^{\prime}}
\end{align*}
\begin{align*}
B_{4b}&=
\hat{B}_{4b}+\Delta B_{4b}+\delta m_{2} B_{2^\ast}+\hat{B}_{2} \tilde{B}_{2^{\prime}}+I_{2} \tilde{B}_{2} 
 -2 I_{4s}-I_{4l}
\end{align*}
\begin{align*}
M_{4a(1^\ast)}&=
\Delta M_{4a(1^\ast)}+\hat{L}_{2} M_{2^\ast}+I_{4a(1^\ast)}
\end{align*}
\begin{align*}
M_{4a(2^\ast)}&=
\Delta M_{4a(2^\ast)}+I_{4a(2^\ast)}
\end{align*}
\begin{align*}
M_{4b(1^\ast)}&=
\Delta M_{4b(1^\ast)}+(\delta m_{2} M_{2^{\ast\ast}}+\hat{B}_{2} M_{2^\ast})+I_{4b(1^\ast)}
\end{align*}
\begin{align*}
M_{4b(2^\ast)}&=
\Delta M_{4b(2^\ast)}+\delta\hat{m}_{2^\ast} M_{2^\ast}+I_{4b(2^\ast)}+(I_{2} M_{2^\ast}+M_{2^\ast}[I] \delta\tilde{m}_{2^\ast}) 
 -2 M_{2^\ast}[I] I_{2}
\end{align*}
\begin{align*}
M_{4a}&=
\Delta M_{4a}+2 \hat{L}_{2} M_{2}
\end{align*}
\begin{align*}
M_{4b}&=
\Delta M_{4b}+\hat{B}_{2} M_{2}+\delta m_{2} M_{2^\ast}+I_{2} M_{2}
\end{align*}
\begin{align*}
\Delta M_{4a(2^\ast)}&=
\Delta M_{4s}-2 \Delta M_{4a(1^\ast)}-2 \Delta M_{4b(1^\ast)}-\Delta M_{4b(2^\ast)}
\end{align*}
\begin{align*}
\Delta B_{4}&\equiv
\Delta B_{4a}+\Delta B_{4b}
\end{align*}
\begin{align*}
\Delta M_{4}&\equiv
\Delta M_{4a}+\Delta M_{4b}
\end{align*}
\begin{align*}
\Delta L_{4}&\equiv
\Delta L_{4x}+2 \Delta L_{4c}+2 \Delta L_{4s}+\Delta L_{4l}
\end{align*}
\begin{align*}
\Delta \delta m_{4}& \equiv
\Delta \delta m_{4a}+\Delta \delta m_{4b}
\end{align*}
\begin{align*}
2 B_{4a(1^\ast)}+ B_{4a(2^\ast)} 
 = &-  4 ( 3 L_{4c(1^\ast)} +   L_{4c(2^\ast)} )
   -  4 (  L_{4x(1^\ast)} +  L_{4x(2^\ast)} )
\end{align*}
\begin{align*}
2 B_{4b(1^\ast)}+ B_{4b(2^\ast)}
 = &-  4 ( 3 L_{4s(1^\ast)} +   L_{4s(2^\ast)}  )
   -  4 (  L_{4l(1^\ast)} +  L_{4l(2^\ast)}  )
\end{align*}

\vspace*{6mm}
\noindent
{\bf Various Relations Among Second-Order Quantities}
\begin{align*}
B_{2^{\prime}}&=
B_{2}
\end{align*}
\begin{align*}
L_{2^{\prime}}&=
L_{2}
\end{align*}
\begin{align*}
\delta\hat{m}_{2^{\prime}}&=
\delta m_{2}-\delta\tilde{m}_{2^{\prime}}
\end{align*}
\begin{align*}
\delta\hat{m}_{2^{\prime \prime}}&=
\delta m_{2}-\delta\tilde{m}_{2^{\prime \prime}}
\end{align*}
\begin{align*}
B_{2^{\ast\ast}}&=
-2 (2 L_{2^{\ast\ast\dagger}}+L_{2^{\ast\dagger\ast}})
\end{align*}
\begin{align*}
L_{2^{\prime \ast}}&=
L_{2^\ast}= I_{2^\ast}+\Delta L_{2^\ast}
\end{align*}
\begin{align*}
B_{2^{\prime \ast}}&=
B_{2^\ast} =\Delta B_{2^\ast}-2 I_{2^\ast}
\end{align*}
\begin{align*}
\delta\hat{m}_{2^{\prime \ast}}&=
\delta m_{2^\ast}-\delta\tilde{m}_{2^{\prime \ast}}
\end{align*}
\begin{align*}
\delta m_{2^\ast}&=
\delta\hat{m}_{2^\ast}+\delta\tilde{m}_{2^\ast}
\end{align*}
\begin{align*}
\delta\tilde{m}_{2^\ast}&=
I_{2}+\Delta \delta m_{2^\ast}
\end{align*}
\begin{align*}
B_{2}&=
-L_{2}
\end{align*}
\begin{align*}
\tilde{B}_{2}&=
-I_{2}+\Delta B_{2}
\end{align*}
\begin{align*}
\hat{L}_{2}&=
-\hat{B}_{2}-\Delta B_{2}
\end{align*}
\begin{align*}
\tilde{L}_{2}&=
I_{2}
\end{align*}


\end{document}